\documentclass{aa} 
\usepackage{natbib}
\usepackage{graphicx}
\usepackage{txfonts}

\newcommand{\jmst}{J.~Mol.~Struct.}  
\newcommand{\jms}{J.~Mol.~Spectrosc.}  

\newcommand{\prev}{Phys. Rev.}

\newcommand{\pccp}{Phys. Chem. Chem. Phys.}

\begin{document}

\title{Analysis of the isotopologues of CS, CCS, CCCS, HCS$^+$, HCCS$^+$, and H$_2$CS in TMC-1 with the QUIJOTE line survey\thanks{Based on 
observations carried out
with the Yebes 40m telescope (projects 19A003,
20A014, 20D023, 21A011, 21D005, 22A007, 22B029 and 23A024) and the Institut de Radioastronomie Millim\'etrique (IRAM) 30m telescope. The 40m
radio telescope at Yebes Observatory is operated by the Spanish Geographic Institute
(IGN, Ministerio de Transportes y Movilidad Sostenible). IRAM is supported by INSU/CNRS
(France), MPG (Germany) and IGN (Spain).}}

\author{
R.~Fuentetaja\inst{1},
C.~Cabezas\inst{1},
Y.~Endo\inst{2},
M.~Ag\'undez\inst{1},
A.~Godard Palluet\inst{3},
F.~Lique\inst{3},
B.~Tercero\inst{4,5},
N.~Marcelino\inst{4,5},
P.~de~Vicente\inst{4},
J.~Cernicharo\inst{1}
}

\institute{Dept. de Astrof\'isica Molecular, Instituto de F\'isica Fundamental (IFF-CSIC),
C/ Serrano 121, 28006 Madrid, Spain. \newline \email r.fuentetaja@csic.es, jose.cernicharo@csic.es
\and Department of Applied Chemistry, Science Building II, National Yang Ming Chiao Tung University, 1001 Ta-Hsueh Rd., Hsinchu 300098, Taiwan
\and Université de Rennes, CNRS, IPR (Institut de Physique de Rennes) - UMR 6251, F-35000 Rennes, France.
\and Observatorio de Yebes (IGN), Cerro de la Palera s/n, 19141 Yebes, Guadalajara, Spain
\and Observatorio Astron\'omico Nacional (OAN, IGN), C/ Alfonso XII, 3, 28014, Madrid, Spain.
}

\date{Received; accepted}

\abstract 
{We performed a detailed analysis of the isotopologues with $^{13}$C, $^{34}$S, $^{33}$S, and $^{36}$S of
the sulphur-bearing molecules CS, CCS, CCCS, HCS$^+$, HCCS$^+$, and H$_2$CS towards the starless core TMC-1
using the QUIJOTE$^1$ line survey. The observations were obtained with the Yebes 40m radio telescope, and the sensitivity of the data 
varied between 0.08 and 0.2 mK in the 31-50 GHz range. Observations with the IRAM 30m radio telescope of the
most abundant isotopologues of these species are also presented and used to estimate volume densities and to
constrain the excitation conditions of these molecules. Among these species, 
we report the first detection in space of C$^{13}$C$^{34}$S, CC$^{33}$S, CCC$^{33}$S, HC$^{33}$S$^+$, and
HCC$^{34}$S$^+$. C$^{36}$S is also detected for the first time in a cold starless object.
These data were complemented with sensitive maps that provide the spatial distribution of 
most of these species.
Using the available collisional rate coefficients for each species,
we modeled the observed line intensities using the large velocity gradient method for the radiative transfer. The results allowed us to report
the most complete analysis of the column densities of the C$_n$S family and to compare the abundance ratios of all detected 
isotopologues. Adopting a kinetic temperature for TMC-1 of 9 K, we found that n(H$_2$)=
0.9-1.5$\times$10$^4$ cm$^{-3}$ can explain the observed decline in intensity with increasing rotational levels $J$ for
all observed molecules. 
We derived the rotational constants for the C$^{13}$C$^{34}$S, CC$^{33}$S, CCC$^{33}$S, HC$^{33}$S$^+$, and HCC$^{34}$S$^+$ isotopologues from new laboratory data and complemented them with the frequencies of the observed lines. 
We find that all sulphur isotopologues are consistent with solar isotopic abundance ratios. 
Accurate $^{12}$C/$^{13}$C abundances were derived and, as previously suggested, 
the $^{13}$C isotopologues of CCS and CCCS show strong abundance anomalies depending on the position of the substituted carbon. Nevertheless,
the $^{12}$C/$^{13}$C abundance ratio is practically identical to the solar value for CS, HCS$^+$, and H$_2$CS.
We also searched for the isotopologues of other S-bearing molecules in the 31-50 GHz domain (HCS, HSC, NCS, H$_2$CCS, HCSCN, HCCCS$^+$, C$_4$S, and C$_5$S). The expected intensities for their $^{34}$S and $^{13}$C isotopologues are too low to be detected with the present sensitivity of the QUIJOTE
line survey, however. The results presented in this work provide new insights into the molecular composition, isotopic abundances, and physical conditions 
of the cold starless core TMC-1. 
}


\titlerunning{CS, CCS and CCCS in TMC-1}
\authorrunning{Fuentetaja et al.}

\maketitle

\section{Introduction}

The cold dark cloud TMC-1 has emerged as a promising laboratory for the detection of a great number of molecules. This 
cloud has a 
carbon-rich chemistry that favours the formation of a high variety of molecules with a large unsaturated carbon chain. 
The recent discoveries made with the
QUIJOTE\footnote{\textbf{Q}-band \textbf{U}ltrasensitive \textbf{I}nspection \textbf{J}ourney to the \textbf{O}bscure 
\textbf{T}MC-1 \textbf{E}nvironment}
line survey of this source carried out with the Yebes 40m radio telescope \citep{Cernicharo2021a}, 
has shown an incredible unexpected chemical richness.

Sensitive line surveys are a powerful method with which to search for the chemical complexity of a cloud. The sensitivity 
reached by QUIJOTE has
permitted us to detect nearly 70 new molecules in TMC-1, including cations, anions, radicals, and cycles.
As the sensitivity increases, a
new issue appears in the interpretation of the data: All 
isotopologues of the most abundant 
species were detected in the data. This means that rare species containing D, $^{13}$C, $^{15}$N, 
$^{34}$S, $^{33}$S, and even
$^{36}$S or double substituted isotopologues, have to be identified in order to avoid possible pitfalls when searching for new molecular species. 
Many of these isotopologues have
been observed in the laboratory, and a search for them is easy. For other isotopologues, however, we lack information 
on their rotational spectroscopy,
and identifying their lines in the survey becomes as exciting as the detection of new species.

The study with QUIJOTE of the isotopologues present in the cloud is currently a work in progress. In recent years, new isotopologues were detected in TMC-1,
such as CH$_2$DC$_3$N \citep{Cabezas2021}, CH$_2$DC$_4$H \citep{Cabezas2022a}, singly substituted isotopologues of HCCNC and HNCCC \citep{Cernicharo2024a}, and doubly substituted isotopologues of HC$_3$N \citep{Tercero2024} . 
The isotopologues
of many species were previously studied at different galactocentric distances using different telescopes and sensitivities
\citep[see, e.g.,][and references therein]{Lucas1998,Milam2005,Sakai2007,Sakai2013,Yan2023}. Together with the observation
of a limited number of transitions for each species, this might introduce
some biases in the conclusions concerning their abundances. Moreover, the rate coefficients have been shown to be dependent on the isotopologues that are considered \citep{Flower2015, Navarro2023}.

The most abundant sulphur-bearing species in TMC-1, which allow us to measure the isotopic abundances of
sulphur and carbon, belong to the C$_n$S and HC$_n$S$^+$ families. In particular, CS, CCS, and CCCS are among the most abundant
molecules in cold dark clouds. The variety of S-bearing molecules is not as vast as those that contain nitrogen or oxygen, however. 
The abundance and number of sulphur-bearing species is limited, to a large extent, by the 
significant depletion of sulphur in the cloud \citep{Fuente2019,Navarro2020,Fuente2023}. 
In recent years, the models for the chemical composition of TMC-1 have 
been progressively refined with the discovery of new S-bearing molecules with the QUIJOTE line survey, such as C$_4$S, C$_5$S (previously detected 
in evolved stars), H$_2$CCS, H$_2$CCCS, HC$_4$S, NCS, HCSCN, HCSCCH, and NCCHCS
\citep{Cernicharo2021d,Cernicharo2021e,Fuentetaja2022,Cabezas2024}, HCCS$^+$ \citep{Cabezas2022b}, HCCCS$^+$ \citep{Cernicharo2021f}, HCNS \citep{Cernicharo2024b}, and CH$_3$CHS \citep{Agundez2024}.

In this work, we present an analysis of the isotopologues of CS, CCS, CCCS, HCS$^+$, HCCS$^+$, and H$_2$CS. 
The high abundance of CS, CCS, and CCCS in TMC-1, combined with the exceptional sensitivity of the QUIJOTE survey, allows 
us to detect singly and doubly substituted isotopologues. We provide a detailed study of isotopologues that are singly substituted with $^{33}$S, 
$^{34}$S, and $^{13}$C for these three abundant species, and of C$^{36}$S and the $^{13}$C$^{34}$S double 
isotopologue of CS. The lines of the different isotopologues are relatively close in frequency, which allows us to calculate column densities with high quality so that we avoid calibration errors, if there are any. We also derive the chemical fractionation enhancements of the $^{13}$C isotopologues of the different species. 

New laboratory rotational spectroscopy of CC$^{33}$S and CCC$^{33}$S have permitted us to detect these rare isotolopogues 
for the first time in space. In addition, using the hyperfine constants of C$^{13}$CS and the substitution structure of 
CCS derived from the spectroscopic information available for all its singly substituted isotopologues, we have found 
four lines in our data that we assign to C$^{13}$C$^{34}$S.  HC$^{33}$S$^+$ is also identified in our data. 
Its rotational constants have been derived from the measured line frequencies of seven hyperfine components
belonging to its $J$=1-0 and 2-1 rotational transitions. For the isotopologues of HCCS$^+$, we
assign four features of the QUIJOTE data to HCC$^{34}$S$^+$. Finally, the rotational and hyperfine constants
of C$^{13}$CS have been improved using the frequencies of the transitions observed in this study.

\section{Observations}
\subsection{Line surveys}
The observational data we used are part of QUIJOTE$^1$ \citep{Cernicharo2021a}, 
a spectral line survey of TMC-1 in the Q band carried out with the Yebes 40m telescope at 
the position $\alpha_{J2000}=4^{\rm h} 41^{\rm  m} 41.9^{\rm s}$ and $\delta_{J2000}=
+25^\circ 41' 27.0''$, which corresponds to the cyanopolyyne peak (CP) in TMC-1. The receiver 
was built within the Nanocosmos project\footnote{\texttt{https://nanocosmos.iff.csic.es/}}.
A detailed description of the system 
was given by \citet{Tercero2021}.

\begin{figure}
\centering
\includegraphics[angle=0,width=0.49\textwidth]{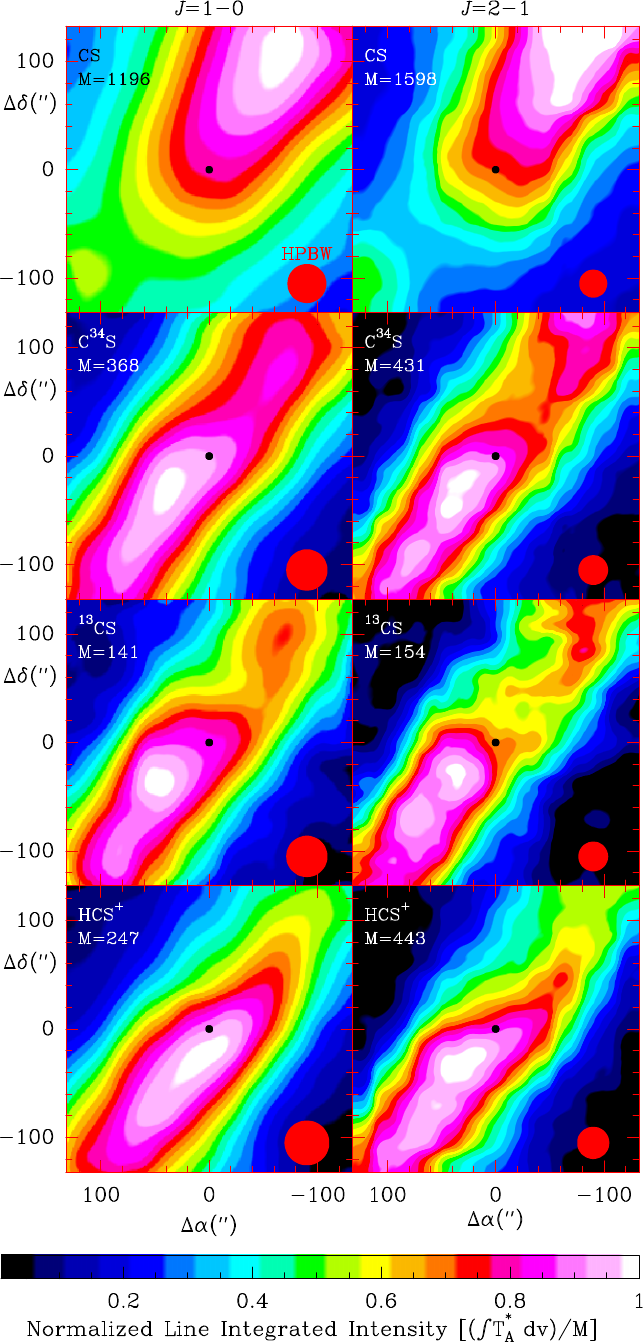}
\caption{Spatial distribution over a region of 264$''$$\times$264$''$ of the integrated-intensity emission of the $J$=1-0
and $J$=2-1 transitions (left and right panels, respectively) of CS, C$^{34}$S, $^{13}$CS, and HCS$^+$ (from top to bottom). For 
each map, the integrated-line emission has been normalized to its maximum value ($M$).
The value of $M$ in mK\,km\,s$^{-1}$ is indicated below the name of the molecule at the top left side of
each panel. The HPBW of the telescope for each transition is indicated by the red circle. The
central position of the map, corresponding to TMC-1(CP), is indicated by a black dot.
} 
\label{maps_CS}
\end{figure}

\begin{figure}
\centering
\includegraphics[angle=0,width=0.49\textwidth]{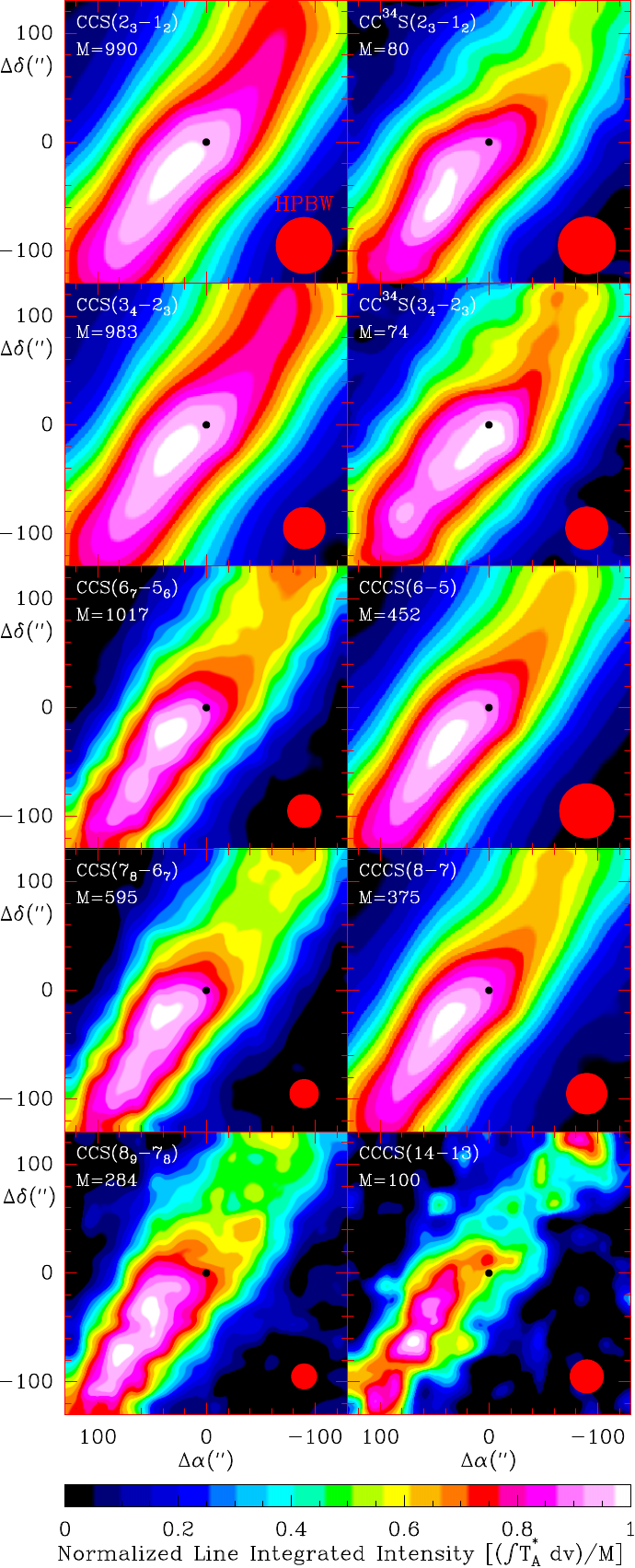}
\caption{Same as Fig. \ref{maps_CS}, but for the integrated-
intensity emission, normalized to M, of several transitions of CCS (left panels), CC$^{34}$S (two upper
right panels) and CCCS (remaining right panels).
The value of the maximum integrated intensity (in mK\,km\,s$^{-1}$) is shown at the top left side of
each panel. 
The HPBW of the telescope for each transition is indicated by the 
red circle. The
central position of the map, corresponding to TMC-1(CP), is indicated by a black dot.
} 
\label{maps_CCS_CCCS}
\end{figure}

The data of the QUIJOTE line survey presented here were gathered in several 
observing runs between November 2019 and July 2023.  
The measured sensitivity by July 2023 varied between 0.08 mK at 32 GHz and 0.2 mK at 49.5 GHz,
 and it is better by about 50 times than that of previous 
line surveys in the Q band of TMC-1 \citep{Kaifu2004}. A detailed description 
of the QUIJOTE line survey was provided by \citet{Cernicharo2021a}.

The main-beam efficiency measured during our observations in 2022 
varied from 0.66 at 32.4 GHz to 0.50 at 48.4 GHz \citep{Tercero2021} and is given across the Q band by
$B_{\rm eff}$=0.797 exp[$-$($\nu$(GHz)/71.1)$^2$]. The
forward telescope efficiency is 0.97.
The telescope beam size at half-power intensity is 54.4$''$ at 32.4 GHz and 36.4$''$ 
at 48.4 GHz. 

The data of TMC-1 taken with the IRAM 30m telescope consist of a 3mm line survey 
that covers the full available band at the telescope, between 71.6 GHz and 
117.6 GHz. The data were described by \citet{Cernicharo2012a}.
More recent high-sensitivity observations in 2021 were used to improve 
the signal-to-noise ratio (S/N) in several frequency windows \citep{Agundez2022,Cabezas2022b,Cernicharo2024b}.

The IRAM 30m beam 
varies between 34$''$ and 21$''$ at 72 GHz and 117 GHz, respectively, while the beam 
efficiency takes values of 0.83 and 0.78 at the same frequencies, following the 
relation B$_{eff}$= 0.871 exp${[-(\nu(GHz)/359)^2]}$. The forward efficiency at 3mm is 0.95.

The intensity scale used in this work for both telescopes, the antenna temperature
($T_A^*$), was calibrated using two absorbers at different temperatures and the ATM
atmospheric transmission model \citep{Cernicharo1985, Pardo2001}. Calibration uncertainties were adopted to be 10~\%. All data were analysed using the GILDAS package\footnote{\texttt{http://www.iram.fr/IRAMFR/GILDAS}}.

\begin{table}[h]
\caption{Density and column densities derived from the models$^*$.}
\label{columndensities}
\begin{centering}
\begin{tabular}{lccc}
\hline
 Molecule        & n(H$_2$)              &         N       & Notes \\
                 &  (cm$^{-3}$)          & (cm$^{-2}$)     &   \\
\hline				 
CS               & 2.0$\times$10$^4$   & 1.1$\times$10$^{14}$ &a\\
C$^{34}$S        & 0.9$\times$10$^4$   & 9.9$\times$10$^{12}$ &b\\
C$^{33}$S        & 0.9$\times$10$^4$   & 2.7$\times$10$^{12}$ & \\ 
C$^{36}$S        & 1.3$\times$10$^4$   & 2.8$\times$10$^{10}$ & \\
$^{13}$CS        & 0.7$\times$10$^4$   & 2.6$\times$10$^{12}$ & \\
$^{13}$C$^{34}$S & 1.3$\times$10$^4$   & 1.1$\times$10$^{11}$ & \\
$^{13}$C$^{33}$S & 0.9$\times$10$^4$   & 3.1 $\times$10$^{10}$ &c\\
\hline                                                         
HCS$^+$          & 0.9$\times$10$^4$   & 5.5$\times$10$^{12}$ & \\
H$^{13}$CS$^+$   & 1.0$\times$10$^4$   & 6.0$\times$10$^{10}$ & \\  
HC$^{34}$S$^+$   & 1.0$\times$10$^4$   & 2.5$\times$10$^{11}$ & \\
HC$^{33}$S$^+$   & 1.0$\times$10$^4$   & 8.0$\times$10$^{10}$ &d\\
\hline
CCS              & 1.3$\times$10$^4$   & 3.4$\times$10$^{13}$ & \\
CC$^{34}$S       & 1.3$\times$10$^4$   & 1.5$\times$10$^{12}$ & \\
CC$^{33}$S       & 1.3$\times$10$^4$   & 3.5$\times$10$^{11}$ & \\
$^{13}$CCS       & 1.3$\times$10$^4$   & 8.4$\times$10$^{10}$ & \\
C$^{13}$CS       & 1.3$\times$10$^4$   & 5.7$\times$10$^{11}$ & \\
C$^{13}$C$^{34}$S& 1.0$\times$10$^4$   & 1.8$\times$10$^{10}$ & \\
\hline
CCCS             & 1.3$\times$10$^4$   & 6.8$\times$10$^{12}$ & \\
CCC$^{34}$S      & 1.3$\times$10$^4$   & 2.7$\times$10$^{11}$ & \\
CCC$^{33}$S      & 1.3$\times$10$^4$   & 6.2$\times$10$^{10}$ & \\
$^{13}$CCCS      & 1.3$\times$10$^4$   & 2.2$\times$10$^{10}$ & \\
C$^{13}$CCS      & 1.3$\times$10$^4$   & 7.4$\times$10$^{10}$ & \\
\hline
HCCS$^+$         &                     & 1.1$\times$10$^{12}$ &e\\ 
HCC$^{34}$S$^+$  &                     & 4.7$\times$10$^{10}$ &f\\ 
\hline
H$_2$CS          & 1.5$\times$10$^4$   & 3.7$\times$10$^{13}$ & \\ 
H$_2$C$^{34}$S   & 1.5$\times$10$^4$   & 1.5$\times$10$^{12}$ & \\ 
H$_2$C$^{33}$S   & 1.5$\times$10$^4$   & 4.2$\times$10$^{11}$ &g\\ 
H$_2$$^{13}$CS   & 1.5$\times$10$^4$   & 4.5$\times$10$^{11}$ & \\ 
\hline
C$_4$S           & 4.0$\times$10$^4$   & 3.8$\times$10$^{10}$ &h\\ 
\hline 
C$_5$S           & 5.0$\times$10$^4$   & 3.0$\times$10$^{10}$ &i\\
\hline
\end{tabular}
\\
\tablefoot{
\tablefoottext{*}{The estimated absolute uncertainties for the column densities are 10\%. The relative uncertainties
for the column densities of the isotopologues of a given species are better than 5\%, however, i.e. the relative calibration 
of the observed lines in the QUIJOTE line survey.}
\tablefoottext{a}{Underestimated by the opacity of the $J$=1-0 and 2-1 transitions of CS. A
two-layer model is needed to explain the observed intensities (see section \ref{sec:CS}).}
\tablefoottext{b}{A small opacity effect in the $J$=2-1 line of C$^{34}$S may exist (see text).}
\tablefoottext{c}{The lines of $^{13}$C$^{33}$S are weak. The column density was
derived from the intensities of the two hyperfine components of the $J$=1-0 transition, 
which are detected above 5$\sigma$ (see Fig. \ref{fig:CS}).}
\tablefoottext{d}{Kinetic temperatures estimated from HC$^{34}$S$^+$ adopting the 
same value of T$_{K}$ for all
hyperfine components of HC$^{33}$S$^+$ (see section \ref{sec:HCS+}).}
\tablefoottext{e}{Column density from \citet{Cabezas2022b}.}
\tablefoottext{f}{The rotational temperatures were assumed to be identical to those of CC$^{34}$S (see text).}
\tablefoottext{g}{Derived adopting a rotational temperature of 8\,K and
an ortho/para ratio identical to that estimated for the other isotopologues
of H$_2$CS (see text).}
\tablefoottext{h}{Very uncertain determination of n(H$_2$). A similar column density is obtained assuming an uniform rotational temperature of 8\,K
(see section \ref{sec:C4S}).}
\tablefoottext{i}{A similar column density is obtained assuming a uniform 
kinetic temperature of 8.5\,K
(see section \ref{sec:C5S}).}
}\\
\end{centering}
\end{table}

\subsection{Maps}\label{obs_maps}
The observed emission in a frequency sweep is often modelled with very limited information on its spatial
extent. In the QUIJOTE line survey, the only available spatial information
is provided by the variation in the telescope half-power beam with the frequency across the
line survey. 
The spatial sizes 
of the observed molecules, together 
with the issues related to the line opacities and radiative transfer, can only be 
addressed through a spatial mapping of the molecular emission. 
To overcome these issues, the 
QUIJOTE line survey is being complemented with high-sensitivity maps obtained
with the Yebes 40m and IRAM 30m radio telescopes. 
At the 40m telescope, $\sim$100 hours of observing time have been devoted to cover
a region of 320$''$$\times$320$''$ around the CP position of TMC-1. The maps are
fully sampled and cover the whole Q band, as in QUIJOTE. We call these 
supplementary spatial data surveying the area of the neighbour TMC-1 cloud through heterodyne observations (SANCHO), and they are a faithful companion to the QUIJOTE line survey. Observation and data reduction details of the maps can be found in \citet{Cernicharo2023}.
The final goal of these maps is to permit the study of
the spatial distribution 
of any QUIJOTE line with an intensity $\geq$20 mK with a S/N $\geq$10, which means 
all lines of abundant species together with their
$^{13}$C, $^{34}$S, D, and $^{15}$N isotopologues (see \citet{Tercero2024}). 
The current sensitivity of SANCHO along the Q band
is $\sim$3 mK, which permits us
to obtain the spatial distribution of several of the molecules discovered with QUIJOTE, including 
cations, anions, radicals, and sulphur-bearing species. 
The maps at 3mm taken with the
IRAM 30m radio telescope 
cover the same spatial extent as those gathered with the Yebes
telescope, but they
are not as sensitive. Nevertheless, 
their sensitivity is enough to map the emission of the isotopologues
of the most abundant molecules. About 50 hours of observing time were 
used in four different frequency settings
with the IRAM 30m telescope. 

The spatial distribution of some of the molecules we studied are
shown in Fig. \ref{maps_CS} and \ref{maps_CCS_CCCS}. 
They are discussed in the next sections in the context of cloud
structure and possible line opacity effects for CS. A detailed study of the velocity
and physical structure of the cloud will be published elsewhere.

\section{Methods}\label{sec:methods}
The lines were identified using the MADEX catalogue \citep{Cernicharo2012b}, which contains the spectral information 
of 6632 species corresponding to 1853 different molecules, including their isotopologues and some of their vibrationally excited states. 
For some species, the CDMS \citep{Muller2005} and the JPL \citep{Pickett1998} catalogues were also used. The details of the
spectroscopic references are given for each molecule in the next sections.

The observed line parameters 
were obtained by means of a Gaussian fit using the CLASS package of GILDAS$^4$. The derived line-integrated 
intensities, their velocities, 
the antenna temperatures, and the full width at half intensity are given in Table B.1.
We considered a window of $\pm$15 km s$^{-1}$ around the v$_{LSR}$ (5.83 km s$^{-1}$) 
of the source for each transition to perform the fit after we removed the baseline. 

The emission of all observed transitions was modelled using the large velocity gradient (LVG) approach. The
basic formula and methods were described by \citet{Goldreich1974}. This approach has been implemented in
the MADEX code. In all cases, we assumed a kinetic 
temperature of 9 K \citep{Tercero2024} and a representative line width of 0.6 km\,s$^{-1}$ (see Table B.1). The densities and column densities we obtained for each species are listed in Table \ref{columndensities} (see details in the next sections).

For the source, we adopted a uniform brightness temperature over a disc with a diameter of 80$''$ \citep{Fosse2001,Cernicharo2023}. This is
a reasonable approximation to the spatial extent and structure of the source. The spatial structure of the source corresponds to a filament that is tilted south-east towards north-west, however, with an angle of
$\sim$60$^o$. The SANCHO maps \citep{Cernicharo2023} 
were used to estimate the effect of the spatial structure of the source and the opacity 
effect for the different observed lines. Figure \ref{maps_CS} shows the observed maps of the $J$=1-0 and 2-1 transitions
of CS, $^{13}$CS, C$^{34}$S, and HCS$^+$.

To determine the molecular abundances in TMC-1, we adopted a total gas column density of 10$^{22}$cm$^{-2}$ \citep[Av=10 mag,][]{Cernicharo1987}, which was derived from star counts as an average value over a beam of 2$'$$\times$2$'$. Values of the column density of H$_2$ derived from Herschel observations of the dust
emission with better angular resolution were obtained by \citet{Kirk2013} and \citet{Feher2016}.
The spatial distribution of N(H$_2$) around TMC-1 derived with Herschel have
values for A$_V$ that range from 5 mag at the
border of the cloud to 25 mag at the CP position \citep[see Fig.1 of][]{Navarro2020,Fuente2019}.
From these maps, the value derived by \citet{Cernicharo1987} appears to be a good compromise 
for the angular resolution of the observations we present here, which ranges from 25$''$ to 56$''$at the highest 
and lowest frequencies, respectively.

We also used the observed line-integrated intensities ($W$) of the same transition of two isotopologues, $A$ and $B$, to derive 
the $A$/$B$ abundance ratio. This method takes advantage of the fact that all data in the QUIJOTE line survey have a homogeneous calibration and
the same pointing uncertainty. For the 3mm line survey taken with the IRAM 30m radio telescope, the data were gathered for a 
significant number of runs with different frequency coverage. Consequently, different transitions observed at different
epochs can have different pointing and calibration errors. We therefore limited this method to the lines that were observed with QUIJOTE.
The results could be independent of the source structure if both isotopologues were assumed to have the same excitation conditions,
the same spatial structure, and optically thin emission in the considered lines. If these conditions are fulfilled, then the line integrated-intensity 
ratio $W_A/W_B$ is proportional to $N(A)/N(B)$ and to a function that depends smoothly on the excitation 
temperature, the energies of the levels involved in the transition, and the rotational constants of the two isotopologues
\citep[see Appendix A of][]{Cernicharo2021b}. Optically thin
emission is expected for the double isotopologues of CS and for all the singly substituted isotopologues of the other species we studied.

\begin{figure}
\centering
\includegraphics[angle=0,width=0.485\textwidth]{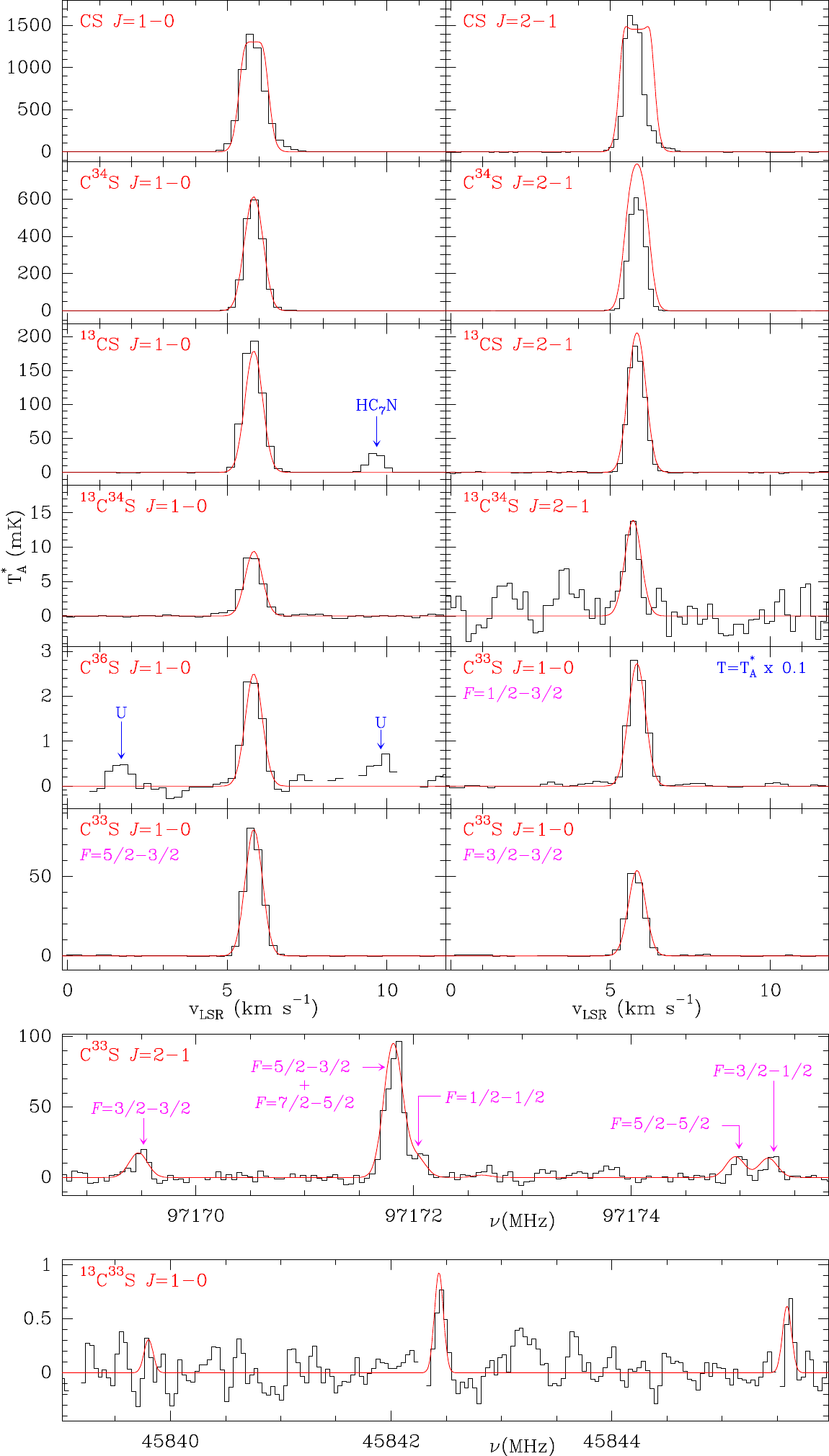}
\caption{Observed $J$ = 1-0 and 2-1 lines of CS and its isotopologues $^{13}$CS, C$^{34}$S, C$^{33}$S, and  $^{13}$C$^{34}$S toward TMC-1. For C$^{36}$S and $^{13}$C$^{33}$S, 
only the $J$=1-0 line was detected.
The abscissa 
corresponds to the local standard of rest velocity of the lines adopting the rest frequencies given in Table 
B.1. In the two bottom panel, however, the abscissa corresponds to the rest frequency. The 
ordinate corresponds to the antenna temperature corrected for 
atmospheric and telescope losses in mK. 
The derived line parameters for the observed lines are given in Table B.1. 
The synthetic spectra (red line) are derived for the column densities shown in Table 
\ref{columndensities} and the LVG model described in sections \ref{sec:methods} and
\ref{sec:CS}. For CS, a two-layer component model is used (see section \ref{sec:CS}).
Negative features appearing in the folding of the frequency switching data are blanked in all the
figures. The observed spectrum of the
hyperfine component $J$=1-0, $F$=1/2-3/2 of 
C$^{33}$S was multiplied by 0.1.} 
\label{fig:CS}
\end{figure}

\section{Results}\label{sec:results}

\subsection{CS}\label{sec:CS}
Carbon monosulfide (CS) is an important molecule in studies of the interstellar medium (ISM). It was the first confirmed 
sulphur-bearing species in the ISM and is commonly used as a tracer of the volume density in dense molecular clouds. It is also one of the most 
abundant molecules found in cold dark clouds such as TMC-1, which has permitted us to detect the lines of its five most 
abundant isotopologues with an unprecedented S/N.
The electronic ground state of CS is $^1\Sigma$, and its dipole moment is $\mu = 1.958 \pm 0.005 D$ 
\citep{Winnewisser1968}. A large set of laboratory rotational data was obtained for this molecule and its
isotopologues in different vibrational states \citep{Bogey1981,Bogey1982,Ahrens1999,Kim2003,Gottlieb2003}. 
Infrared data are also available, and they help us to
constrain the high-order distortion constants of the molecule \citep{Todd1977,Todd1979,Winkel1984,
Burkholder1987,Ram1995,Uehara2015}.
We used the measured frequencies in these references to obtain Dunham mass-independent parameters and implemented them in the MADEX
code. 
For the two transitions we considered, the difference between our frequencies and those in
the CDMS \citep{Muller2005} are below 1 kHz.

The observed intensities of the $J$=1-0 and $J$=2-1 transitions of CS and its isotopologues range from about 
1.4 K for CS to $\sim$ 1 mK for $^{13}$C$^{33}$S (see Fig. \ref{fig:CS} and Table B.1), that is, 
a dynamical range of 1400. We used the collision rate calculated by \citet{Denis-alpizar2018} for the CS/$p$-H$_2$ system to model the line profiles. 

The CS abundance is probably highly underestimated because the $J$ = 1-0 and 2-1 lines are expected to be opaque, and it cannot be used to derive isotopic abundances.
These line trapping problems are clearly seen in the maps of the integrated intensity of CS and its $^{13}$CS and C$^{34}$S isotopologues (see Fig. \ref{maps_CS}). 
The integrated intensity of the
two lines of CS clearly peaks towards the NW of the maps and is shifted along the main axis
of TMC-1 by more than 2$'$ with respect the emission of the isotopologues. Moreover, the emission of CS seems
to decline towards the zone in which the isotopologues are most intense. This effect
appears in the two transitions we studied. They were observed with different telescopes, the Yebes 40m telescope for $J$=1-0, and the IRAM 30m telescope for $J$=2-1.
The result is therefore not an artefact of the observations.

The situation for CS is similar to that analysed by \citet{Cernicharo1987} for HCO$^+$ and H$^{13}$CO$^+$. For these two species, it was found that the line profiles and the intensities reflected a huge opacity in the $J$=1-0 line of HCO$^+$. A simple two-layer model was 
sufficient to explain the observations qualitatively. A high-density core that in the maps of Fig. \ref{maps_CS}
corresponds to the narrow filament oriented SE-NW, causes the emission of H$^{13}$CO$^+$. A surrounding envelope with a
moderate density (a few 10$^3$ cm$^{-3}$) still has enough molecules
of HCO$^+$ (here CS) to absorb the photons from the core and to re-emit them over a
large volume. The line opacity in the envelope is so high that photons from the core
do not escape from the cloud. The column density of the cloud is not sufficent for H$^{13}$CO$^+$ (here, $^{13}$CS and C$^{34}$S) to produce significant intrinsic emission or notable
absorption of the photons from the dense filament, however. This situation is typical of cold
clouds with narrow line widths and small velocity gradients. As a consequence, all
points of the cloud with a high and low volume density are connected radiatively. 

The two-layer model is an oversimplification of the radiative transfer problem in TMC-1, and 
a more detailed analysis is required to implement density and velocity gradients from the external parts to the core. To achieve these goals, we require a multi-line study with a higher angular resolution than we used here. Nevertheless, we tested the reliabilty of the model
by comparing the predicted intensities for all the isotopologues of CS with the observations.

A direct determination of the $^{12}$C/$^{13}$C abundance ratio in CS can be derived from the line intensities of C$^{34}$S and $^{13}$C$^{34}$S. We estimated, however, that the $J$=1-0 and
$J$=2-1 lines of C$^{34}$S have opacities of $\sim$0.5 and 1.6, respectively. The ratios
we obtained from the direct comparison of the line intensities are therefore to be considered lower limits. From the $J$=1-0 line intensities, we obtained $^{12}$C/$^{13}$C=54.9$\pm$0.7.
From the $J$=2-1 transition, we derived a ratio of 44.0$\pm$6.5. Nevertheless, the column densities for C$^{34}$S and $^{13}$C$^{34}$S were determined from the LVG approximation used
for CS, and these opacity corrections are therefore taken into account in our estimation of the 
column densities in Table \ref{columndensities}. From the LVG calculations, we derived a $^{12}$C/$^{13}C$ abundance ratio
of 90$\pm$9.

Two of the three hyperfine components of the $J$=1-0 transition of $^{13}$C$^{33}$S are marginally detected in the QUIJOTE line survey. 
In the next months, improved QUIJOTE data might allow us to improve the data for this rare isotopologue of CS
and to derive a better isotopic $^{12}$C/$^{13}$C abundance ratio from C$^{33}$S and $^{13}$C$^{33}$S.

The derived column densities for the different isotopologues of CS are given in Table \ref{columndensities}. The derived isotopic abundances from these
column densities are given in Table \ref{isoabundances} and are discussed in Section \ref{sec:discussion}.

\begin{figure}
\centering
\includegraphics[angle=0,width=0.485\textwidth]{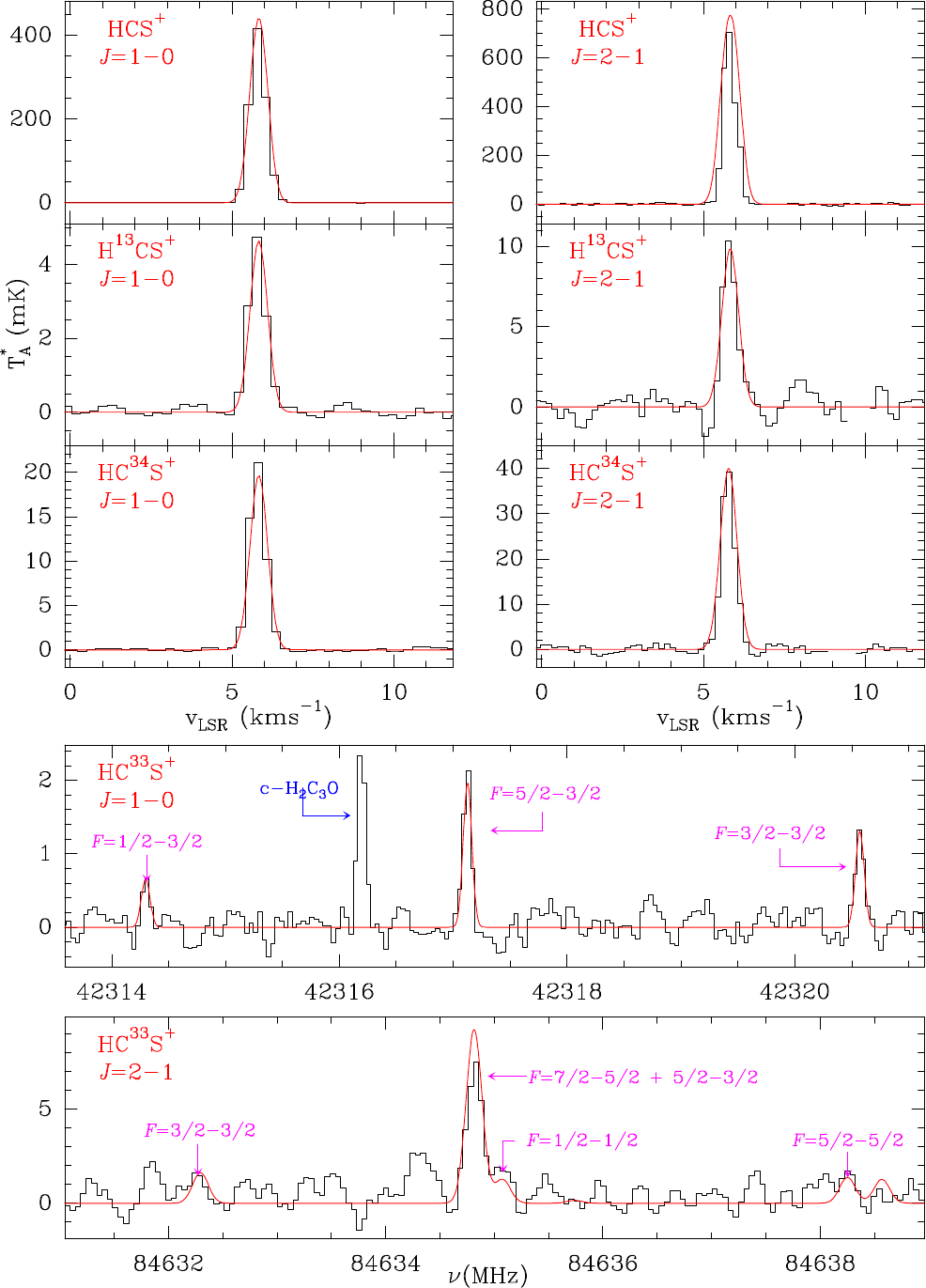}
\caption{Observed $J$ = 1-0 and 2-1 lines of HCS$^+$ and its isotopologues towards TMC-1. 
The abscissa 
corresponds to the local standard of rest velocity of the lines adopting the rest frequencies given in Table 
B.1. The bottom panels show the same transitions of HC$^{33}$S$^+$,
which exhibit several hyperfine components. The abscissa in this case is the rest frequency adopting a velocity for the source of 5.83 km\,s$^{-1}$ \citep{Cernicharo2020}. 
The ordinate corresponds to the antenna temperature corrected for 
atmospheric and telescope losses in mK.
The derived line parameters for the observed lines are given in Table B.1. 
The synthetic spectra (red line) are derived for the column densities shown in Table 
\ref{columndensities} and the LVG model described in sections \ref{sec:methods}
and \ref{sec:HCS+}.
} 
\label{fig:HCS+}
\end{figure}

\begin{figure*}
\centering
\includegraphics[angle=0,width=0.90\textwidth]{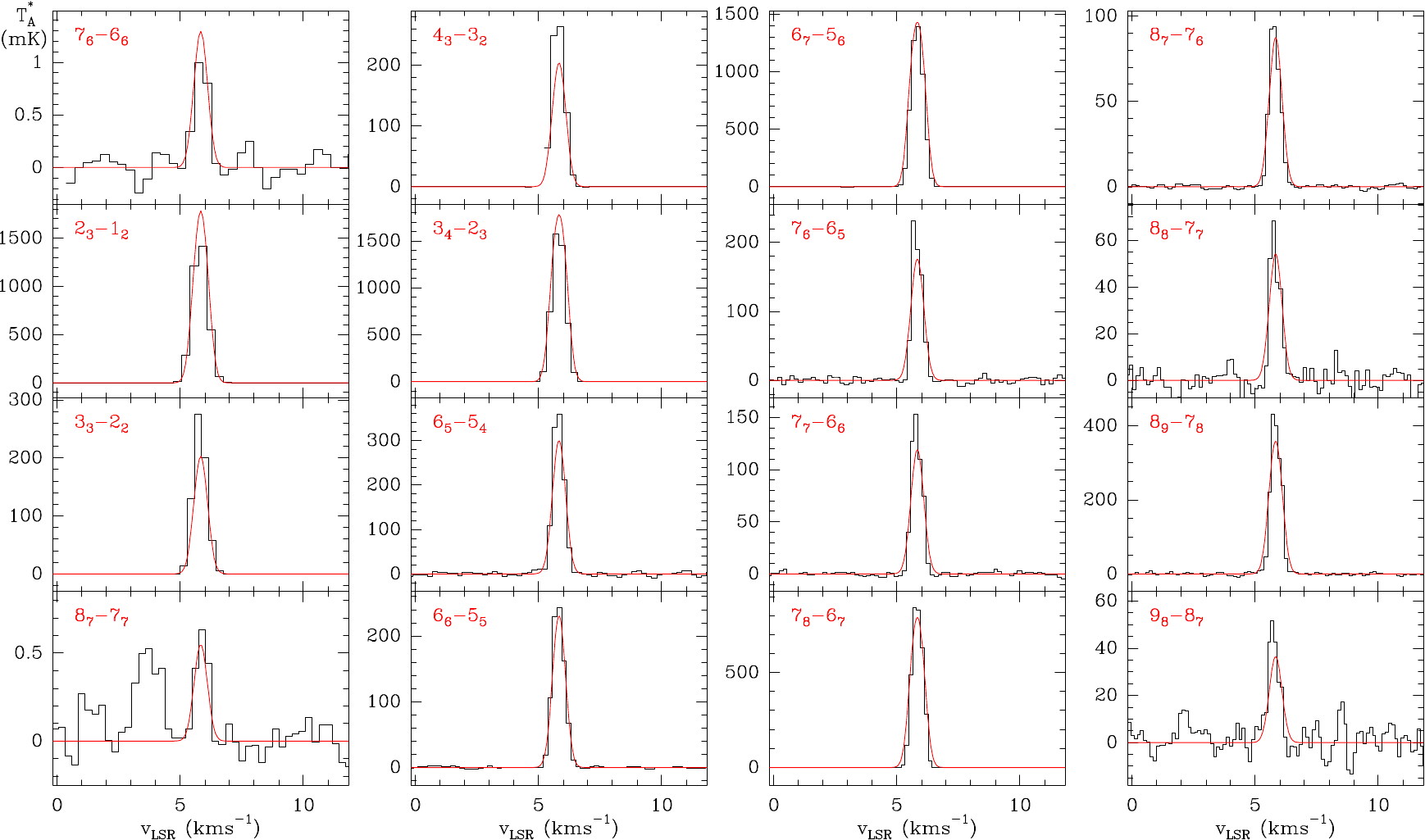}
\caption{Observed lines of CCS towards TMC-1. 
The abscissa 
corresponds to the local standard of rest velocity of the lines adopting the rest frequencies given in Table 
B.1.
The ordinate corresponds to the antenna temperature corrected for 
atmospheric and telescope losses in mK.
The derived line parameters are given in Table B.1. 
The synthetic spectra (red line) are derived for the column densities shown in Table 
\ref{columndensities} and the LVG model described in section \ref{sec:methods} 
with n(H$_2$)=1.3$\times$10$^4$ cm$^{-3}$ and N(CCS)=3.4$\times$10$^{13}$cm$^{-2}$ 
(see section \ref{sec:CCS}).} \label{fig:CCS}
\end{figure*}

\begin{figure}
\centering
\resizebox{9cm}{!}{
\includegraphics[angle=0,width=0.92\textwidth]{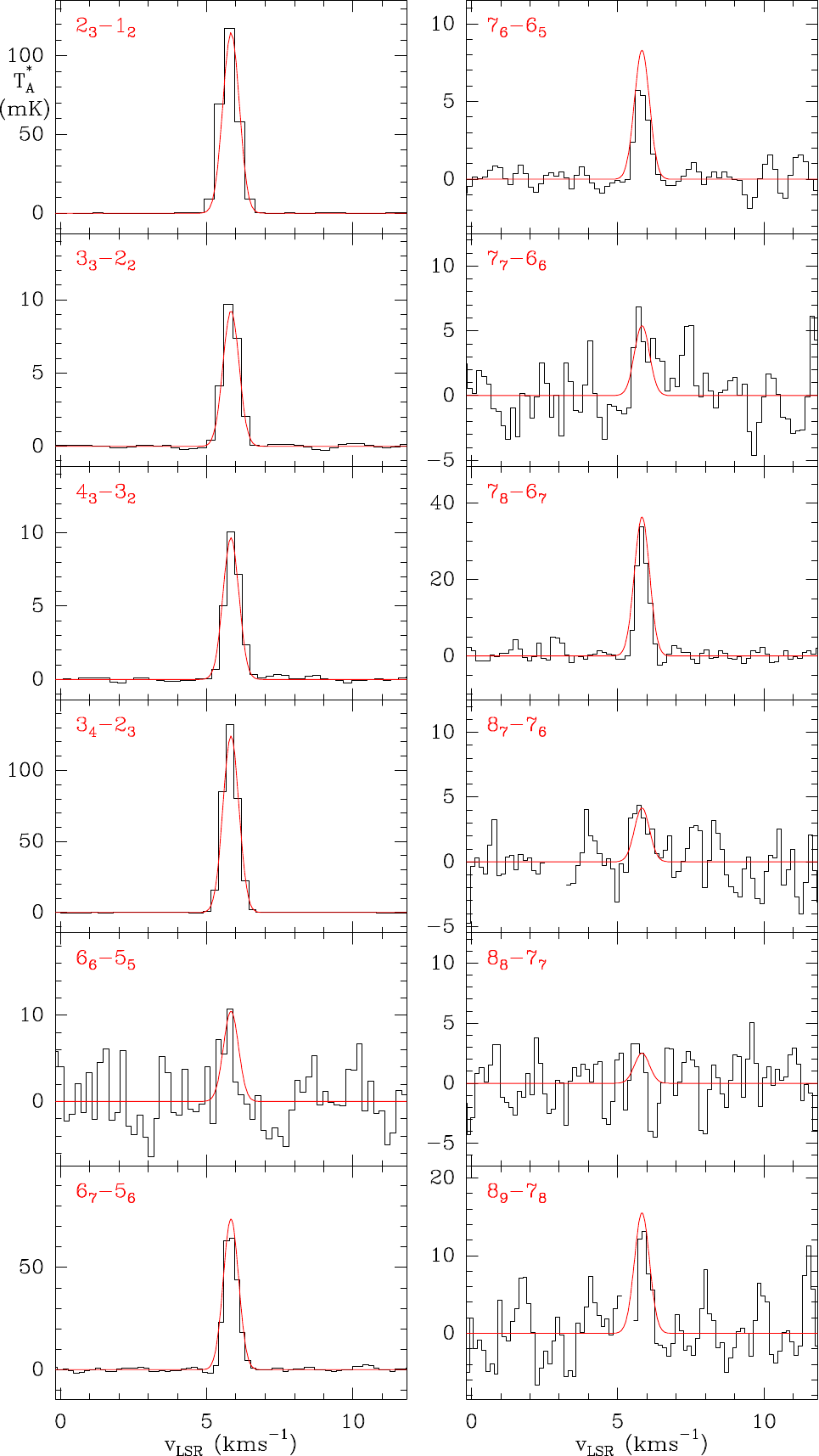}
}
\caption{Observed lines of CC$^{34}$S towards TMC-1. 
The abscissa 
corresponds to the local standard of rest velocity of the lines adopting the rest frequencies given in Table 
B.1.
The ordinate corresponds to the antenna temperature corrected for 
atmospheric and telescope losses in mK.
The derived line parameters are given in Table B.1.  
The synthetic spectra (red line) are derived from the LVG model described in section \ref{sec:methods}
with n(H$_2$)=1.3$\times$10$^4$ cm$^{-3}$ and N(CC$^{34}$S)=1.5x10$^{12}$cm$^{-2}$ (see section \ref{sec:CCS}).} 
\label{fig:CC34S}
\end{figure}

\begin{figure}
\includegraphics[angle=0,width=0.49\textwidth]{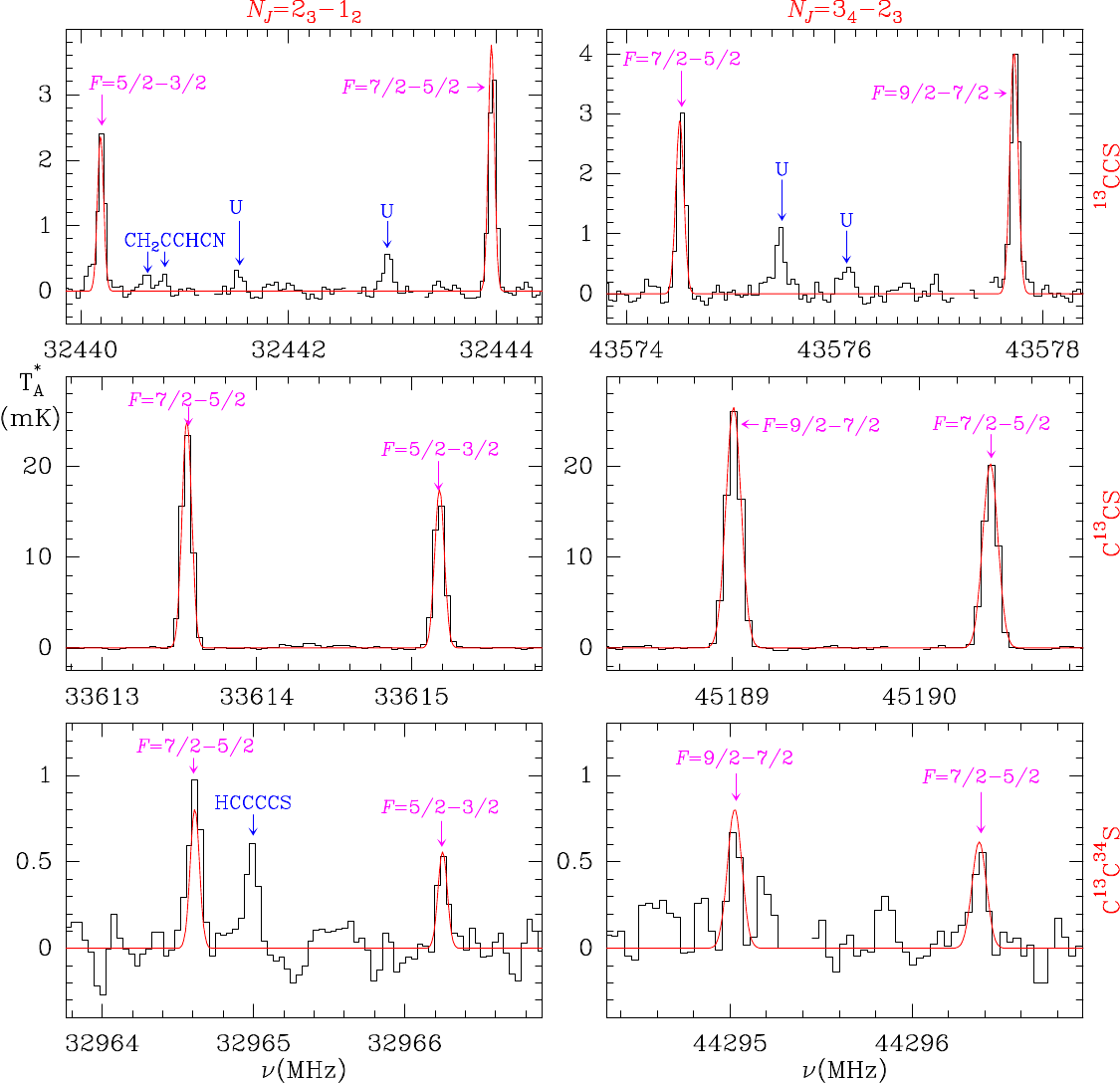}
\caption{Observed hyperfine components of the $N_J=2_3-1_2$ (left panels) and
$N_J=3_4-2_3$ (right panels) transitions of $^{13}$CCS, C$^{13}$CS, and
C$^{13}$C$^{34}$S towards TMC-1. 
The abscissa 
corresponds to the rest frequency in MHz.
The ordinate corresponds to the antenna temperature corrected for 
atmospheric and telescope losses in mK.
The derived line parameters are given in Table B.1.  
The synthetic spectra (red line) are derived from the LVG model described in section \ref{sec:methods}
with n(H$_2$)=1.3$\times$10$^4$ cm$^{-3}$ and the column densities given in Table \ref{columndensities} (see section \ref{sec:CCS}).} 
\label{fig:CCS_iso_13C}
\end{figure}

\begin{figure*}[h]
\centering
\includegraphics[angle=0,width=0.95\textwidth]{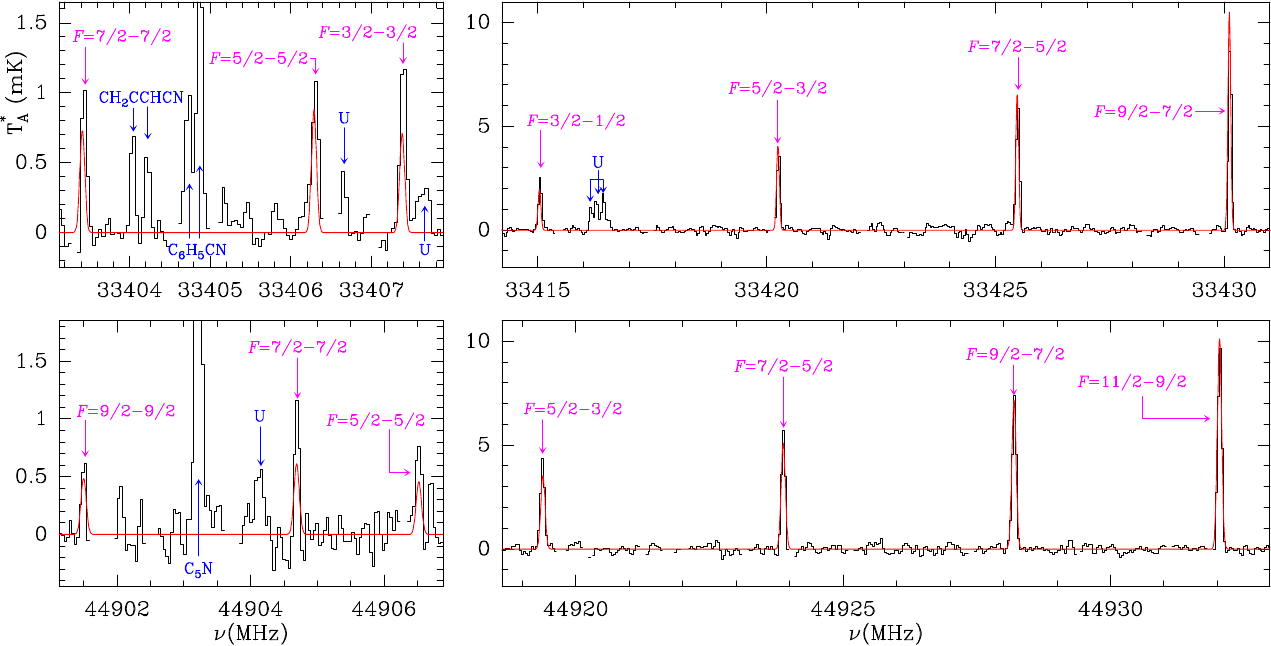}
\caption{Observed hyperfine components of the $N_J=2_3-1_2$ (upper panels) and 
$N_J=3_4-2_3$ (bottom panels) transitions of CC$^{33}$S towards TMC-1. 
The derived line parameters of the observed lines are given in Table B.1. 
The abscissa 
corresponds to the rest frequency in MHz.
The ordinate corresponds to the antenna temperature corrected for 
atmospheric and telescope losses in mK.
The synthetic spectra (red line) are derived from the LVG model described in section \ref{sec:methods}
with n(H$_2$)=1.3$\times$10$^4$ cm$^{-3}$ and $N$(CC$^{33}$S)=3.5$\times$10$^{11}$cm$^{-2}$ (see section \ref{sec:CCS}).} \label{C233S_primeros}

\end{figure*}

\subsection{HCS$^+$}\label{sec:HCS+}
This molecule was found in space in 1981 by \citet{Thaddeus1981} prior to any laboratory measurement through the
observation of four lines that were related harmonically. The identification was confirmed by the observation of the $J$=2-1
line in the laboratory by \citet{Gudeman1981}. Precise rotational constants for HCS$^+$ and several of its isotopologues 
were derived from laboratory measurements by different authors \citep{Bogey1984, Tang1995, Margules2003}. 
It is interesting to note that the dipole moment of this species is 1.958\,D \citep{Botschwina1985}, which is identical to that of CS.
Therefore, we assumed as a first approximation that CS and HCS$^{+}$ undergo the same excitation conditions, which depend on the Einstein coefficients and collisional rate coefficients, and derived their abundance ratios using the intensity ratios of optically thin lines. HCS$^+$ is chemically related to CS as it results from the protonation of this species with
H$_3$$^+$, HCO$^+$, and H$_3$O$^+$ as the main proton donators. Its principal destruction path is through electronic dissociative
recombination to produce CS.

We report the detection of the $J$=1-0 and 2-1 lines of HCS$^+$,
H$^{13}$CS$^+$, HC$^{34}$S$^+$ and, for the first time in space, of HC$^{33}$S$^+$. The rotational constants for this
isotopologue were derived from the observed line frequencies, as discussed in Appendix \ref{cons_HC33S+}. 
Figure \ref{fig:HCS+} shows the observed lines of all isotopologues of HCS$^+$. The spatial
distribution of the $J$=1-0 and $J$=2-1 lines of HCS$^+$ is shown in Fig. \ref{maps_CS}. 
The emission from the
two transitions appears to be very similar to that of C$^{34}$S and $^{13}$CS, which indicates that
CS and HCS$^+$ coexist spatially. Like for CS, the emission peak is shifted relative to our
central position. The change in intensity between the emission peak and the
central position is smaller than 15\%.

Using the same parameters for the source as for CS and the collisional rate coefficients calculated by \citet{Denis-alpizar2022}
for HCS$^+$ with $p$-H$_2$,
we derived a volume density from the observed lines of n(H$_2$)=1.0$\times$10$^4$ cm$^{-2}$. For the main isotopologue the best
fit to the line profiles was obtained for 
a slightly lower value of the volume density (0.9$\times$10$^4$ cm$^{-3}$). This fact, together with the observed line
intensity ratios between HCS$^+$ and its rare isotopologues, suggests minor opacity problems
for its lines. The computed LVG opacity values for HCS$^+$ are $\tau$($J$=1-0)$\sim$0.19 and $\tau$($J$=2-1)$\sim$0.76. The best fits to the column densities are 
given in Table \ref{columndensities}, and  the comparison between the modelled and observed
spectra is shown in Fig. \ref{fig:HCS+}. The opacity of the lines of all
isotopologues is much lower and does not affect our column density estimates.
For HCS$^+$, the column densities derived from the LVG analysis
take the line opacities into account as a first approximation.

The isotopic abundance ratios 
derived from HCS$^+$ and its isotopologues are given in Table \ref{isoabundances}.
The HCS$^+$/H$^{13}$CS$^+$ column density ratio is 91$\pm$9. Using the integrated line intensities of the $J$=1-0 of HCS$^+$ and H$^{13}$CS$^+$, we obtained 
$^{12}$C/$^{13}$C=78$\pm$3, and from the $J$=2-1 transition line, we derived 
an abundance ratio of 82$\pm$2. Both determinations are nearly
identical to the determination derived  from C$^{34}$S and $^{13}$C$^{34}$S. This value also agrees with the $^{12}$C/$^{13}$C 
averaged abundance ratio derived from the $^{13}$C isotopologues of C$_4$H \citep{Sakai2013},
of HC$_5$N \citep{Takano1998,Cernicharo2020} and of HCCCN, HNCCC, and HCCNC \citep{Cernicharo2024a,Tercero2024}. 

The abundance ratios of the protonated and neutral molecule depend on the degree of ionisation and on the formation and destruction rates of the cation. These ratios also increase with the proton affinity of the neutral species \citep{Agundez2015}. From the derived column densities for the isotopologues of CS and HCS$^+$, we derived 
N($^{13}$CS)/N(H$^{13}$CS$^+$)=43 $\pm$4,
N(C$^{34}$S)/N(HC$^{34}$S$^+$)=40 $\pm$4 and N(C$^{33}$S)/N(HC$^{33}$S$^+$)=34 $\pm$4. Based on this, the abundance ratio 
of CS and its protonated form in TMC-1 is $\sim$40.

\subsection{C$_2$S}\label{sec:CCS}

The linear molecule CCS (thioxoethenylidene) has a $^3\Sigma^{-}$ ground electronic state. 
It was first identified in the ISM 
towards TMC-1 and Sgr\,B2 \citep{Saito1987,Yamamoto1987,Kaifu1987} and towards the
carbon-rich evolved star IRC+10216 \citep{Cernicharo1987}. Nevertheless, one of
its rotational lines was previously reported by \citet{Suzuki1984} towards TMC-1,
but lack of laboratory spectroscopy at that time prevented the assignment of the line
to CCS. All the laboratory-measured frequencies of CCS 
\citep{Saito1987,Yamamoto1990,Lovas1992,McGuire2018}
were
used to fit a standard Hamiltonian for a $^3\Sigma$ molecule. The
resulting rotational constants were implemented in the code MADEX. The predicted frequencies for the transitions we observed are given in Table B.1.
The dipole moment of CCS was calculated to be 2.88\,D 
\citep{Pascoli1998,Lee1997}. In 
the Q band, we detected four lines for the CCS molecule and its most abundant isotopologue 
CC$^{34}$S 
(see Figures \ref{fig:CCS} and \ref{fig:CC34S}, and Table B.1), which correspond to the transitions 
with $N$=2 up to 4 ($J$=$N\pm$0,1). The intensities of two of these
transitions lie above 100\,mK. 

On the other hand, for the $^{13}$CCS,
C$^{13}$CS, and C$^{13}$C$^{34}$S
isotopologues, 
we observed the transitions $N_J$ = 2$_3$ - 1$_2$ and 
$N_J$ = 3$_4$ - 2$_3$ 
(see Fig. \ref{fig:CCS_iso_13C}), and all their hyperfine components were detected with
a good S/N.

We obtained the line parameters for thioxoethenylidene (CCS) using the same methods as for the 
previous molecules, and we report them in Table 
B.1. 
We also calculated the column densities using the set of collisional rate coefficients computed by 
\cite{Godard2023,Godard2024} and the 
LVG approach. These values are given in Table \ref{columndensities}.

Several unidentified lines were detected in our survey in the Q band with different patterns 
at 33.4 GHz and 44.9 GHz. Based on the CCS data and the expected abundances for the $^{33}$S 
isotopologue, we suspect that these lines correspond to the hyperfine structure of the 
CC$^{33}$S molecule. To confirm this assignment, we used ab initio theoretical 
calculations and laboratory data (see Appendix \ref{cons_cc33s}). These transitions correspond 
to $N_J$ = 2$_3$ - 1$_2$ and $N_J$ = 3$_4$ - 2$_3$ (see Fig. \ref{C233S_primeros}). We 
obtained the observed line parameters and report them in Table 
B.1.

The spatial distribution of CCS and its isotopologue CC$^{34}$S follows the same behaviour as observed for CS and its isotopologues (see Fig. \ref{maps_CCS_CCCS}). For the $J$ = 8$_9$ - 7$_8$ transition, however, the distribution is split into two peaks that follow the same SE-NW orientation.

\begin{figure}[h]
\centering
\includegraphics[angle=0,width=0.485\textwidth]{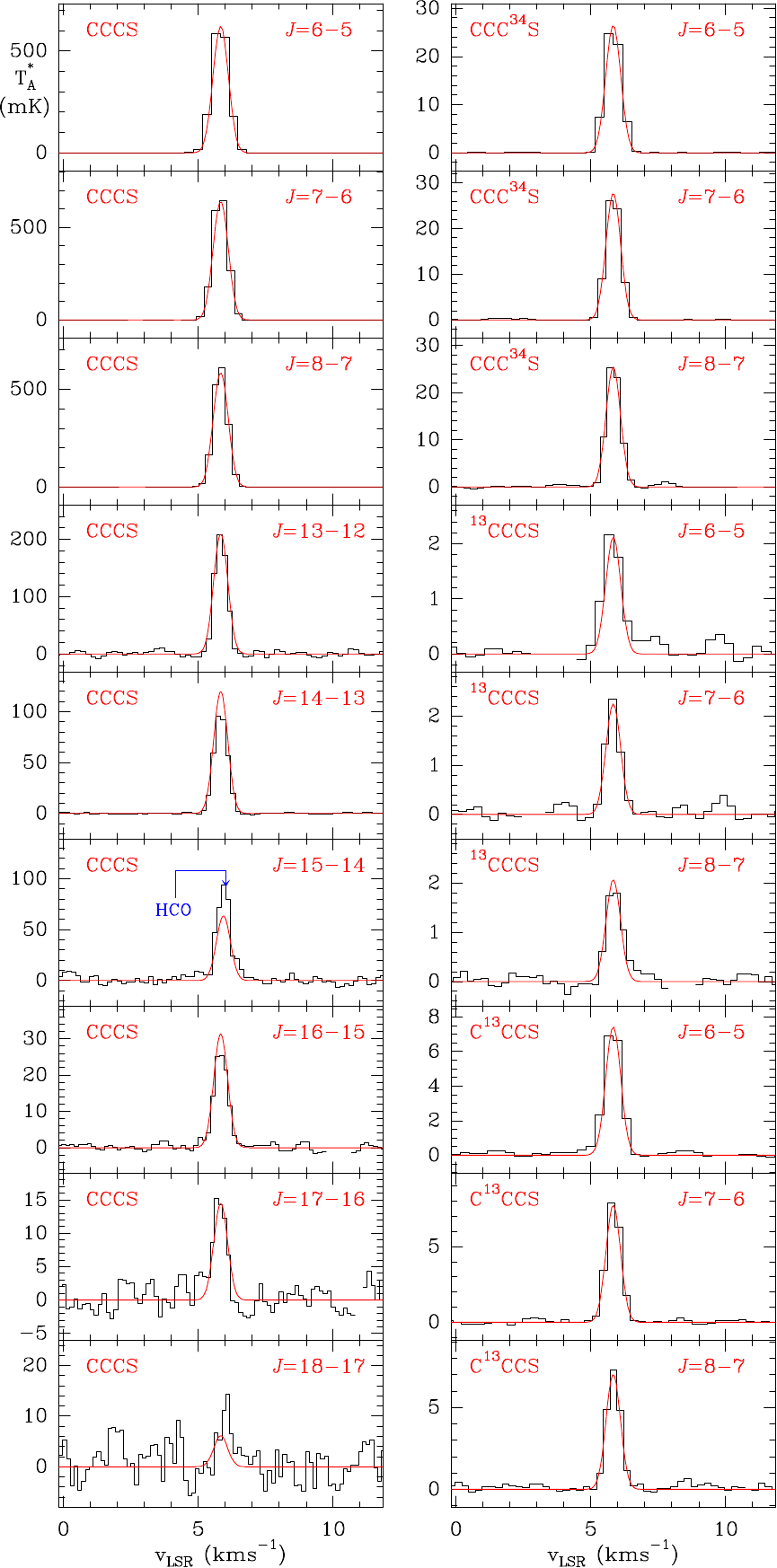}
\caption{Observed transitions of CCCS, CCC$^{34}$S, $^{13}$CCCS, and
C$^{13}$CCS towards TMC-1. All transitions of CCCS are shown in the left panels. The right
panels show $J$=6-5, 7-6, and 8-7 of the three isotopologues.
The abscissa 
corresponds to the local standard of rest velocity of the lines adopting the rest frequencies given in Table 
B.1.
The ordinate corresponds to the antenna temperature corrected for 
atmospheric and telescope losses in mK.
The derived line parameters are given in Table B.1. 
The synthetic spectra (red lines) have been calculated from the LVG model described in section \ref{sec:CCCS}. The
adopted volume density is n(H$_2$)=1.3$\times$10$^4$ cm$^{-3}$ for all isotopologues, and the
resulting column densities are given in Table \ref{columndensities}.} \label{fig:CCCS}
\end{figure}

Because the lines are opaque, we performed several calculations 
to obtain the most accurate column density for the molecules that lack a hyperfine structure (CCS and CC$^{34}$S) by varying the density 
between 10$^{2}$ and 10$^{8}$ cm$^{-3}$ for three values of the column 
densities (10$^{12}$, 10$^{13}$ and 10$^{14}$ cm$^{-2}$). These values 
represent abundances with respect to H$_2$ from 10$^{-10}$ to 10$^{-12}$. For both molecules, we found that the 
most intense transitions are 2$_3$ - 1$_2$ and 3$_4$ - 2$_3$, with the 
latter being the largest of them. These transitions thermalise at lower values than the other two transitions.
For these values, we found that the best-fit column density corresponds to $2.8\times10^{13}$ cm$^{-2}$ in the case of CCS and $1.0\times10^{12}$ cm$^{-2}$ for the CC$^{34}$S isotopologue. These two results correspond to a density of 1.3 $\times$10$^4$ cm$^{-3}$.

\subsection{$CCCS$}\label{sec:CCCS}

The linear molecule CCCS  has an electronic ground state $^1\Sigma$. Abundant laboratory data
of its rotational transitions and those of its isotopologues were reported \citep{Yamamoto1987,Lovas1992,Tang1995,Ohshima1992,Gordon2001,Sakai2013,McGuire2018}. 
Its dipole moment was measured to be $\mu=$ 3.704D 
\citep{Suenram1994}. The first detection of CCCS in space was achieved towards TMC-1 \citep{Kaifu1987,Yamamoto1987} and towards
the carbon-rich evolved star IRC+10216 \citep{Cernicharo1987}. We fitted the available laboratory data for each isotopologue of CCCS and implemented
them into the MADEX code. The frequencies we used for the observations are given in Table B.1. 

\begin{figure}
\centering
\includegraphics[angle=0,width=0.49\textwidth]{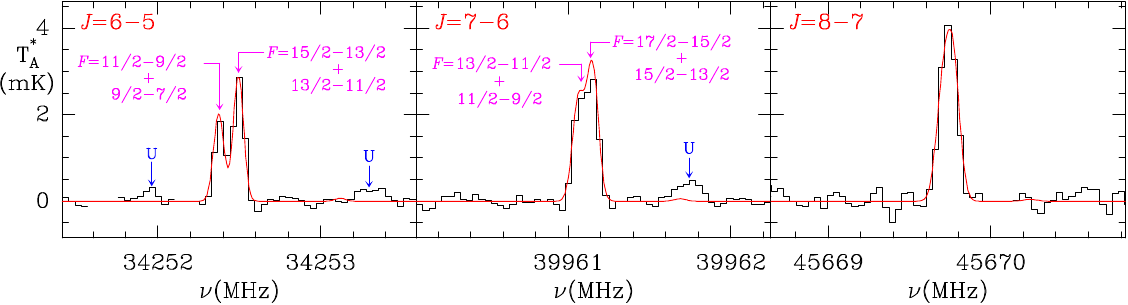}
\caption{Observed transitions of CCC$^{33}$S towards TMC-1. 
The abscissa 
corresponds to the rest frequency.
The ordinate corresponds to the antenna temperature corrected for 
atmospheric and telescope losses in mK.
The derived line parameters are given in Table B.1. 
The synthetic spectra (red line) are derived from a model with $N$=6.2x10$^{10}$cm$^{-2}$ and $T_{rot}$=8\, K (see section \ref{sec:CCCS}).} \label{fig:CCC33S}
\end{figure}

In our observations, we detected three transitions of CCCS, CCC$^{34}$S, $^{13}$CCCS, and C$^{13}$CCS in the Q band 
corresponding to $J = 8 - 7$, $J = 7 - 6$ and $J = 6 - 5$. In addition, CCCS lines from $J=13-12$ up to $J=18-17$  were detected
with the IRAM 30m telescope line survey in the 3mm domain. The lines of CCCS and its isotopologues are shown in Figs. \ref{fig:CCCS} and \ref{fig:CCC33S}.
The derived line parameters of the observed lines were obtained using the same methods as for the previous molecules and are given in Table 
B.1.
Because the rotational constants of CCCS and those of CC$^{13}$CS are nearly coincident, the latter isotopologue could not be detected
(its transitions differ by less than 0.1 MHz from those of CCCS and are not resolved in the QUIJOTE data). We detected several 
features at 34.2, 39.6, and 45.6 GHz that correspond to the isotopologue CCC$^{33}$S. Improved rotational constants from our measured
line frequencies in TMC-1 and those obtained from previous \citep{McGuire2018} and new laboratory data are given in Appendix \ref{cons_CCC33S}
and Table \ref{constants_CCC33S}.

The spatial distribution of CCCS is very similar to that of the isotopologues
of CS, HCS$^+$, CCS, and CC$^{34}$S as shown in Fig. \ref{maps_CCS_CCCS}, which can be compared
to the maps in Fig. \ref{maps_CS}. The emission peak of CCCS, like the other species in Figs. \ref{maps_CS} and \ref{maps_CCS_CCCS}, is shifted to $\Delta\alpha$=-30$''$ and $\Delta\delta$=-30$''$ with respect to the centre of the maps, but the intensity
variation between this position and the central one is smaller than 15\%. For $J$ = 14 - 13, similar to the CCS case, the distribution is split into two peaks, with a higher intensity located to the SE. Again, the
assumption of a source of uniform brightness over a diameter of 80$''$ seems to be
a reasonable hypothesis. 
\begin{figure*}
\centering
\includegraphics[angle=0,width=0.9\textwidth]{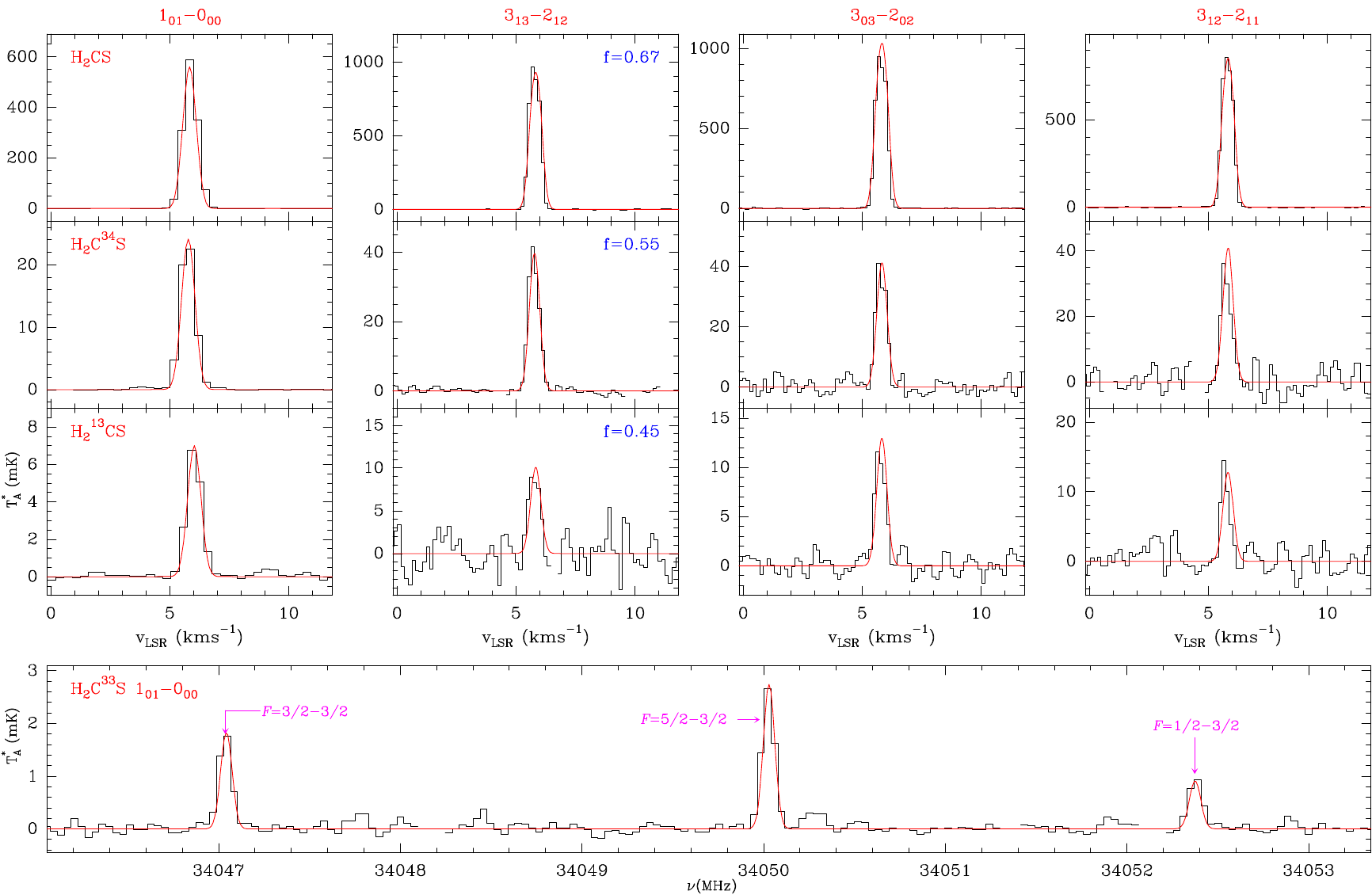}
\caption{Observed transitions of H$_2$CS and its isotopologues towards TMC-1. 
The abscissa 
corresponds to the local standard of rest velocity of the lines adopting the rest frequencies given in Table 
B.1, except for the bottom panel (H$_2$C$^{33}$S)
for which  it
corresponds to the rest frequency.
The ordinate corresponds to the antenna temperature corrected for 
atmospheric and telescope losses in mK.
The derived line parameters are given in Table B.1. 
The synthetic spectra (red line) are derived from the LVG model 
described in section \ref{sec:S-bearing}.
The modelled spectra for the $3_{13}-2_{12}$ transition of H$_2$CS, H$_2$C$^{34}$S and 
H$_2$$^{13}$CS
have been multiplied by a correction factor indicated in blue at the top right side
of the corresponding panels. For all the other transitions, no correction factors
are applied. This factor results from the adopted
collisional rate coefficients for H$_2$CS (see section \ref{sec:S-bearing}).} \label{fig:H2CS}
\end{figure*}

Collisional rate coefficients for CCCS with He were calculated by \citet{Sahnoun2020}. Only 11 levels were considered ($J_{max}$=10), however, that
only cover energies up to 15.3\,K. These rates cannot be used to model the nine observed lines of CCCS,
which go from $J$=6-5 up to $J$=18-17, with upper-level
energies up to 47.5\,K, which is well above the energies calculated by \citet{Sahnoun2020}. The detection of these high-$J$ lines of
CCCS provides a key constraint in the modelling of the observed intensities. An estimation
of the collisional rate coefficients above $J_u$=10 can be obtained  from the $\sigma$(0$\rightarrow$J) cross sections with the IOS approximation. 
This approach, however, introduces several uncertainties in the modelling of the high-$J$ lines. 
The available collisional rate coefficients between CCCS and He are similar to those of the HCCCN with $p$-H$_2$ \citep{Faure2016}.
We therefore adopted the HCCCN-$p$-H$_2$ rate coefficients to model the observed CCCS lines. 

In order to fit the lines of CCCS, we explored the predicted intensities and line intensity ratios for a wide range of volume densities. Surprisingly,
the best fit to the data was obtained for n(H$_2$)=(1.3$\pm$0.2)$\times$10$^4$, which is the same value as for CCS. Figure \ref{fig:CCCS} shows the
quality  of the synthetic line profiles obtained from our model for N(CCCS)=(6.8$\pm$0.7)$\times$10$^{12}$ cm$^{-2}$. The largest discrepancy is observed
for the $J$=15-14 line. It is due to the severe blending of this transition with a feature arising from HCO, however. The lines of the four isotopologues of CCCS,
$^{13}$CCCS, C$^{13}$CCS, CCC$^{34}$S, and CCC$^{33}$S are reproduced very well with the same model and for the column densities given in Table \ref{columndensities}. 

The column density ratio of CCCS and $^{13}$CCCS is 309$\pm$30, while for CCCS/C$^{13}$CCS, the value
is 92$\pm$9, which is nearly identical to the solar abundance $^{12}$C/$^{13}$C. These species were previously studied in TMC-1 by \citet{Sakai2013}, who derived values for these ratios of $\ge$209 and 48$\pm$15. Our more sensitive observations (at least a factor of 10) clearly indicate that C$^{13}$CCS has an 
isotopic abundance ratio slightly higher than that of the local ISM standard value of 60-70. 
The abundance of $^{13}$CCCS is depleted by a factor 3.4$\pm$0.3. The abundance ratio of CCCS and CCC$^{34}$S is 25.2$\pm$2.5, which is
identical to the solar abundance. The derived isotopic ratios are given in Table \ref{isoabundances} and are discussed in section \ref{sec:discussion}.

\subsection{H$_2$CS and other S-bearing species}\label{sec:S-bearing}
TMC-1 is known to be a factory of sulphur-bearing species \citep{Cernicharo2021c}. In addition to the species
considered in previous sections, other molecules such as NCS, HCCS, H$_2$CS, H$_2$CCS, H$_2$CCCS, HCCCCS, HCSCN, HCSCCH,
HCCS$^+$, HCCCS$^+$, C$_4$S, and C$_5$S were recently detected with the QUIJOTE line survey 
\citep{Cernicharo2021b,Cernicharo2021c,
Cernicharo2021d,Cabezas2022a,Fuentetaja2022}. 
When the observed line intensities of these species and their associated column densities are taken into account
\citep[see Table 1 of][]{Cernicharo2021c}, 
the abundances of H$_2$CS \citep{Cernicharo2021c}, HCCS$^+$ \citep{Cabezas2022a}, and HCSCN \citep{Cernicharo2021d} alone are high enough to allow the detection of their $^{34}$S and $^{13}$C isotopologues in our line survey.

For H$_2$CS our data from QUIJOTE and the IRAM 30m telescope observations cover two ortho and two para transitions.
They are detected for H$_2$CS, H$_2$C$^{34}$S, and H$_2$$^{13}$CS, and the observed lines are shown in
Fig.  \ref{fig:H2CS}. 
In addition, the three hyperfine components
of the 1$_{01}$-0$_{00}$ transition of H$_2$C$^{33}$S were also detected in space for the first
time. They are shown in Fig. \ref{fig:H2CS}, and
the derived line parameters are given in Table B.1. In spite of the high
observed intensities for the lines of H$_2$CS, the emission seems to be optically thin because the
intensity ratios of H$_2$CS and its isotopologues for the four observed lines
are near the solar abundance of $^{12}$C/$^{13}$C and $^{32}$S/$^{34}$S. 

No collisional rate coefficients are available
for H$_2$CS, but they can be inferred from those of H$_2$CO.
Two different sets of collisional rate coefficients can be employed for $o$-H$_2$CS, those of collisions of H$_2$CO
with He \citep{Green1991}, or those of collisions with $o$-H$_2$ and $p$-H$_2$ calculated by \citet{Troscompt2009}. 
For $p$-H$_2$CS, only the $p$-H$_2$CO-He rate coefficients \citep{Green1991} can be employed. The results of our LVG models 
using the rate coefficients of $o$/$p$-H$_2$CS with He
are shown in Fig. \ref{fig:H2CS}. It is surprising to find that the fitted line profiles reproduce
the two para and one of the ortho lines of the three isotopologues quite well. The volume density adopted in the
model is 1.5$\times$10$^4$ cm$^{-3}$, but
when the large uncertainty on the adopted rate coefficients is taken into account, this density has to be taken with caution.
In the modelling, we found an ortho-to-para ratio for all isotopologues
of 2.0$\pm$0.1, which is significantly different from the  expected value of 3.
The main discrepancy between observations and the model was found for the ortho $3_{13}-2_{12}$ transition, for which the 
predicted intensities have to be multiplied by a
factor $\sim$0.5 to match the observations (see Fig. \ref{fig:H2CS};
the correction factor for the $3_{13}-2_{12}$ transition is specified for each 
isotopologue). The computed excitation temperature for this 
transition is very sensitive to the adopted volume density. 
The computed line opacities for the two observed ortho transitions are identical, however. Hence, the
brightness temperature is always higher for the $3_{13}-2_{12}$ transition
than for the $3_{12}-2_{11}$ one.
We explored the effect of the volume density on the expected intensities. For higher values of
n(H$_2$), all lines are too strong. We also analysed the effect of the 
adopted collisional rate coefficients. Using the rate coefficients of $o$-H$_2$CO with $p$-H$_2$ of \citet{Troscompt2009}, we obtained an intensity
ratio for the two ortho lines, T$_B$($3_{13}-2_{12}$)/T$_B$($3_{12}-2_{11}$), of 1.34;
but for the collisional rate coefficients with He \citep{Green1991}, this ratio is
1.94. Consequently, the observed intensity discrepancy between the two ortho line
seems to be related to the set of collisional rate coefficients adopted for H$_2$CS. 
Unfortunately, the data of \citet{Troscompt2009} only cover ten energy levels of $o$-H$_2$CO (for H$_2$CS, this corresponds
to a maximum enery of 22\,K) and were not calculated for $p$-H$_2$CO. In the final model, we therefore adopted the collisional
data of H$_2$CO with He and applied the correction factor to the intensity of the $3_{13}-2_{12}$ transition. The
derived ortho/para ratio has to be taken with caution. 
Future calculations for the collisional rate coefficients of thioformaldehyde with He and/or H$_2$ will be very interesting. This
molecule is ubiquitous in space, and its lines are easily detected in cold, warm, and hot molecular clouds. It is also 
detected in evolved stars \citep{Agundez2008}.

The column density ratios given in Table \ref{isoabundances} indicate
that $^{12}$C/$^{13}$C and $^{32}$S/$^{34}$S are similar to the solar abundances.
These abundance ratios can be also obtained
from the line intensity ratios between the different isotopologues 
(see Table B.1 and Fig. \ref{fig:H2CS}). From the four lines of each of them, we derived
an averaged value for $^{12}$C/$^{13}$C and $^{32}$S/$^{34}$S of 84.7$\pm$3 and 24.9$\pm$1, respectively.
These isotopic abundance ratios are similar to those derived from S-bearing species
that only contain one carbon (CS and HCS$^+$).

\begin{figure}
\includegraphics[angle=0,width=0.50\textwidth]{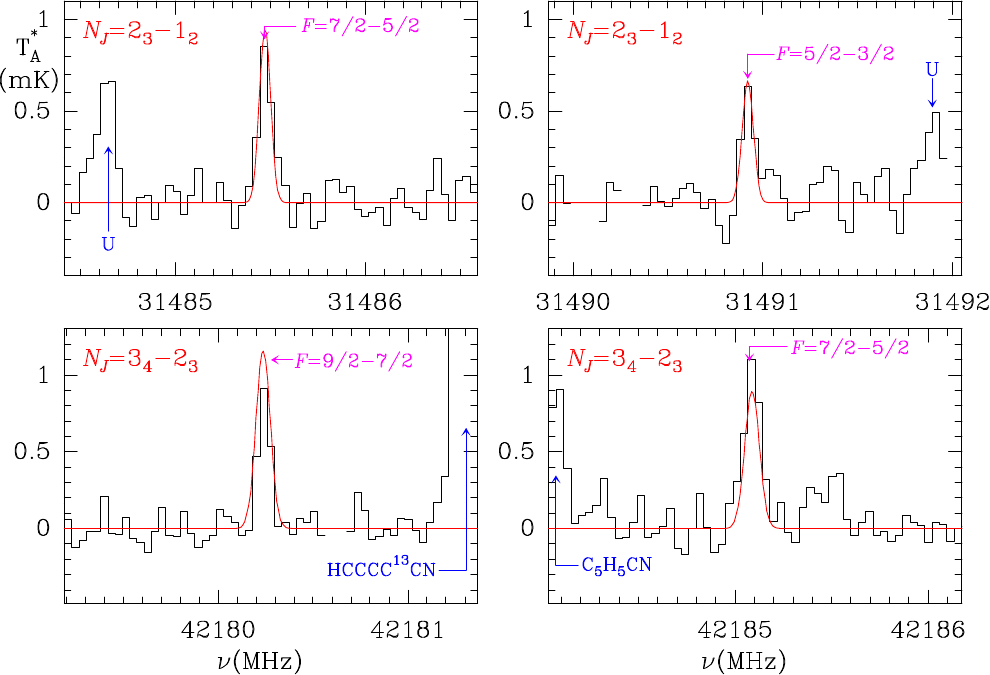}
\caption{Observed lines of HCC$^{34}$S$^+$ towards TMC-1. 
The abscissa 
corresponds to the rest frequency.
The ordinate corresponds to the antenna temperature corrected for 
atmospheric and telescope losses in mK.
The derived line parameters of the observed lines are given in Table B.1. 
The modelled spectra (red line) are derived for  $N$(HCC$^{34}$S$^+$)=4.7$\times$10$^{10}$cm$^{-2}$ adopting
the kinetic temperatures calculated for CC$^{34}$S as the rotational temperature for n(H$_2$)=1.3$\times$10$^4$ cm$^{-3}$.} \label{fig:HCC34S+}
\end{figure}

HCCS$^+$ has been detected in TMC-1 by \citet{Cabezas2022a}. 
The identification was based on ab initio calculations and the detection of 26 of its
rotational transitions, which contain fine and hyperfine structure.
The intensity of the lines of HCCS$^+$ is
high enough to allow the detection at least of the $^{34}$S isotopologue. Based
on the rotational constants predicted for HCC$^{34}$S$^+$, from our ab initio calculations
and those observed for HCCS$^+$, we predict the frequencies of the four strongest lines
of HCC$^{34}$S$^+$ in the QUIJOTE domain (see section \ref{cons_HCC34S+}). Four lines
are found at frequencies that differ from the predicted values by less than 0.1 MHz. They are
shown in Fig. \ref{fig:HCC34S+}, and their line parameters are given in Table B.1.
A fit to the observed frequencies provides accurate rotational constants of this species
(see section \ref{cons_HCC34S+}). To derive a column density for this isotopologue,
we adopted the rotational excitation temperature derived for CC$^{34}$S for all
hyperfine components of each rotational transition of HCC$^{34}$S$^+$ and derived a
column density of 4.7$\times$10$^{10}$ cm$^{-2}$. From the column density of HCCS$^+$
derived by \citet{Cabezas2022a}, we derived a $^{32}$S/$^{34}$S abundance ratio of 23.4$\pm$2.5,
which is similar to that obtained from other species. The expected intensities for
the $^{13}$C isotopologues, $\sim$0.15-0.3 mK, are below the current sensitivity of QUIJOTE.

The abundance ratios of the protonated and the neutral molecule are similar to those of the HCS$^+$/CS cases. We obtained a value of 30.9 for N(HCCS$^+$)/N(CCS) and 31.9 for the N(HCC$^{34}$S$^+$)/N(CC$^{34}$S) ratio.

The rotational constants for HC$^{34}$SCN were obtained by \citet{Cernicharo2021d}. 
From the intensities observed for the main isotopologue \citep{Cernicharo2021d}, the
expected intensities for the strongest transitions of HC$^{34}$SCN are $\sim$0.4 mK .
An exploration
of the QUIJOTE data provides the detection of only three lines at 3$\sigma$, while other lines
remain undetected. Future improved QUIJOTE data might allow us to detect this isotopologue
of HCSCN.

We also considered C$_4$S and C$_5$S. They are analysed in sections \ref{sec:C4S}
and \ref{sec:C5S}, respectively. Their observed intensities are too low, $\sim$1 mK, to permit the
detection of their isotopologues. Nevertheless,
improved rotational constants for both molecules are provided in Tables \ref{constants_C4S} and \ref{constants_C5S},
respectively.

\begin{table}[h]
\tiny
\caption{Isotopic ratio derived from S-bearing molecules in TMC-1$^\#$}
\label{isoabundances}

\centering
\begin{tabular}{|lccccc|}

\hline
Isotopes & Molecules & R$^*$ & SS$^+$${^c}$ & L483${^d}$ & Notes \\
\hline\\[-2ex]
$^{32}$S/$^{34}$S & CS/C$^{34}$S              & 11.1$\pm$1.1 & 22.6& &a,b \\
                  &  $^{13}$CS/$^{13}$C$^{34}$S & 23.6$\pm$2.4 &   & 31$\pm$9 & \\
                  & CCS/CC$^{34}$S            & 22.7$\pm$2.3 & 17$\pm$5    & 17$\pm$5 &  \\
                  & CCCS/CCC$^{34}$S          & 25.2$\pm$2.5 &     & &  \\
                  & HCS$^+$/HC$^{34}$S$^+$    & 22.0$\pm$2.2 &     & 21$\pm$6 &  \\
                  & HCCS$^+$/HCC$^{34}$S$^+$  & 23.4$\pm$2.3 &     & &  \\
                  & H$_2$CS/H$_2$C$^{34}$S    & 24.6$\pm$2.5 &     & 29$\pm$9 &  \\
                  &                           &      &     & &  \\
\hline	
                  &                           &      &     &  & \\
$^{32}$S/$^{33}$S & $^{13}$CS/$^{13}$C$^{33}$S& 83.9$\pm$8.4 &  127   & &  \\
                  & CCS/CC$^{33}$S            & 97.1$\pm$9.7 &    & &   \\
                  & CCCS/CCC$^{33}$S          & 109.7$\pm$10.9 &   &  &   \\
                  & HCS$^+$/HC$^{33}$S$^+$    & 68.8$\pm$6.9 &    & &   \\
                  & H$_2$CS/H$_2$C$^{33}$S    & 88.1$\pm$8.8 &    & &   \\
                  &                           &      &    & &   \\
\hline	
                  &                           &      &    & &   \\
$^{34}$S/$^{33}$S & C$^{34}$S/C$^{33}$S       & 3.7$\pm$0.4 &  5.6 & & b \\
         & $^{13}$C$^{34}$S/$^{13}$C$^{33}$S  & 3.6$\pm$0.4 &    & & \\
                  & CC$^{34}$S/CC$^{33}$S     & 4.3$\pm$0.4 &    & &\\
                  & CCC$^{34}$S/CCC$^{33}$S   & 4.4$\pm$0.4 &    & &\\
                &H$_2$C$^{34}$S/H$_2$C$^{33}$S& 3.6$\pm$0.4 &    & &   \\
                  &                           &      &    & &   \\
\hline				  
                  &                           &      &    & &   \\
$^{34}$S/$^{36}$S & C$^{34}$S/C$^{36}$S       & 353.4$\pm$35.3 & 425 & & b \\
                  &                           &      &    & &   \\
\hline				  
                  &                           &      &    & &   \\
$^{33}$S/$^{36}$S & C$^{33}$S/C$^{36}$S       & 96.4$\pm$9.6 & 75 &  &\\
                  &                           &      &    & &   \\
\hline				  
                  &                           &      &    & &   \\
$^{12}$C/$^{13}$C & CS/$^{13}$CS              & 68.8$\pm$6.9 & 89 & & a \\
                  & C$^{34}$S/$^{13}$C$^{34}$S& 90.0$\pm$9.0 &    & 58$\pm$18 & b \\
                  & C$^{33}$S/$^{13}$C$^{33}$S& 87.1$\pm$8.7 &    & &   \\
                  & CCS/$^{13}$CCS            &\textbf{404.7}$\pm$40.4 &   & >25 &   \\
                  & CCS/C$^{13}$CS            &\textbf{ 59.6}$\pm$6.0 &   & 28$\pm$8  &   \\
                  & CCCS/$^{13}$CCCS          &\textbf{309.1}$\pm$30.9 &   & &   \\
                  & CCCS/C$^{13}$CCS          &\textbf{ 91.9}$\pm$9.2 &   & &   \\
                  & H$_2$CS/H$_2$$^{13}$CS    & 82.2$\pm$8.2 &    & 113$\pm$34 &   \\
                  & HCS$^+$/H$^{13}$CS$^+$    & 91.0$\pm$9.1 \\
                  &                           &            &   &    &   \\
\hline				  
                  &                           &      &   &  &   \\
$^{13}$C$_2$/$^{13}$C$_1$ & C$^{13}$CS/$^{13}$CCS &\textbf{ 6.8}$\pm$0.7 & & & \\
                  & C$^{13}$CCS/$^{13}$CCCS   &\textbf{ 3.4}$\pm$0.3 &  & &\\
                  &                           &      &     &  & \\
\hline				  
                  &                          &      &     &  & \\
$^{13}$C/$^{34}$S & $^{13}$CS/C$^{34}$S      & 0.26$\pm$0.03 &  0.27 & & \\
                  & $^{13}$CCS/CC$^{34}$S    & 0.06$\pm$0.01 &      & & \\
                  & C$^{13}$CS/CC$^{34}$S    & 0.38$\pm$0.04 &      & & \\
                  & $^{13}$CCCS/CCC$^{34}$S  & 0.08$\pm$0.01 &      & & \\
                  & C$^{13}$CCS/CCC$^{34}$S  & 0.27$\pm$0.03 &      & & \\
               &H$_2$$^{13}$CS/H$_2$C$^{34}$S& 0.30$\pm$0.03 &    & &   \\
                  &                           &      &    & &   \\
\hline
\end{tabular}

\tablefoot{
\tablefoottext{\#}{The estimated uncertainties are 10\%. These ratios are obtained from the
column densities given in Table \ref{columndensities}.}\\
\tablefoottext{*}{Abundance ratio} \\
\tablefoottext{+}{Solar System} \\
\tablefoottext{a}{Underestimated by the opacity of the 1-0 of CS.}\\
\tablefoottext{b}{Some small correction of C$^{34}$S may exist.}\\
\tablefoottext{c}{Values derived by \citet{Anders1989} and \citet{Yan2023}.}\\
\tablefoottext{d}{Values derived by \citet{Agundez2019}.}}\\
\end{table}

\subsection{Discussion}\label{sec:discussion}

From the column densities we calculated for the various species as summarized in Table \ref{columndensities}, we derived the corresponding abundance ratios, which are given in Table \ref{isoabundances}, where they are compared with the values derived in the dark cloud L483 and in the Solar System. The abundance ratios CS/C$^{34}$S and CS/$^{13}$CS are not reliable because the $J = 1-0$ and $J = 1-0$ lines of CS are affected by opacity. This limitation can be overcome using the doubly substituted isotopologue $^{13}$C$^{34}$, which yields $^{32}$S/$^{34}$S and $^{12}$C/$^{13}$C ratios that agree with those derived using H$_2$CS and HCS$^+$. The
$^{32}$S/$^{34}$S ratios derived in TMC-1, in the range 22.0-25.2, are consistent with the Solar System value of 22.6 and with the values derived in the local ISM, 24.4$\pm$5.0 \citep{Chin1996}, 19$\pm$8  \citep{Lucas1998}, and $\backsim$22 \citep{Wilson1999}.
The values found in L483 are also consistent within the errors with the local ISM and Solar System values. On the other hand, the $^{13}$C$^{33}$S ratios in TMC-1, in the range 68.8-109.7, are somewhat lower than the Solar System value of 127 and the local ISM value of 153$\pm$40 \citep{Chin1996}. It therefore seems that there is minimal isotopic fractionation for $^{34}$S, while there could be a slight isotopic enrichment in $^{33}$S in TMC-1.
In the case of carbon, the $^{12}$C/$^{13}$C ratios derived from C$^{34}$S, C$^{33}$S, H$_2$CS, and HCS$^*$, which are in the range 82.2-91, are consistent with the Solar System value of 89 and somewhat above the local ISM values of 59$\pm$2 \citep{Lucas1998}, 69$\pm$6 \citep{Wilson1999}, 68$\pm$15 \citep{Milam2005}, 70$\pm$2 \citep{Sheffer2007}, 76$\pm$2 \citep{Stahl2008}, and 74,4$\pm$7.6 \citep{Ritchey2011}. The fact that the values found in the local ISM are systematically lower than those found in TMC-1 suggests that they might be affected by opacity, resulting in an underestimation of the true $^{12}$C/$^{13}$C interstellar ratio.
The $^{12}$C/$^{13}$C ratios derived from CCS and CCCS deserve special attention. Isotopic ratios are very different and depend on the position where $^{13}$C is substituted. In general, the isotopologue with a $^{13}$C at the terminal carbon exhibits a much lower abundance than the other observed isotopologue. These isotopic anomalies were first noted by \citet{Sakai2007} for the CCS isotopologues and later on by \citet{Sakai2013} for CCCS.
The explanation of these isotopic anomalies have been the subject of discussion. It was originally proposed that the formation pathway of the molecules themselves with non-equivalent carbon atoms was at the origin \citet{Sakai2007}. A different explanation was proposed by \citet{Furuya2011}, who suggested that the different abundances of the various isotopologues might be caused by isotopologue exchange reactions involving H atoms. These reactions would progressively transform the less stable isotopologues into the more stable one approaching abundance ratios determined by the difference in the zero-point energies (ZPE) of the different isotopologues.
The scenario proposed by \citet{Furuya2011} was validated by \citet{Talbi2018}, who studied the exchange reaction $^{13}$CCS + H $\rightarrow$ C$^{13}$CS + H theoretically and found it to be barrierless. If this reaction is rapid enough, it would indeed tend to drive the C$^{13}$CS/$^{13}$CCS abundance ratio to a value of exp($\Delta$$E/T$), where $T$ is the gas kinetic temperature, and $\Delta$$E$ is the difference between the ZPE of $^{13}$CCS and that of C$^{13}$CS. For a value of $\Delta$$E$ of 18.9 K \citep{Talbi2018}, the theoretically expected C$^{13}$CS/$^{13}$CCS would be 8.2 for a kinetic gas temperature of 9 K \citep{Agundez2023}, which is only slightly above the observed value in TMC-1 of 6.8. Isotopic fractionation of carbon was studied using chemical models \citep{Roueff2015, Colzi2020, Loison2020}. The implementation of the H exchange mechanism in the chemical model of \citet{Loison2020} resulted in a C$^{13}$CS/$^{13}$CCS ratio that was somewhat lower than observed, in the range 1-4, depending on the time and on the assumptions about the O + C$3$ reaction. It would be interesting to further explore which conditions allow us to reproduce the observed  C$^{13}$CS/$^{13}$CCS ratio and also to explore the isotopic anomalies found for CCCS.

\section{Conclusions}

We have presented a comprehensive analysis of the CS, CCS, CCCS, C$_4$S, C$_5$S, and H$_2$CS molecules and their isotopologues detected towards TMC-1 using the QUIJOTE line survey conducted with the Yebes 40m radio telescope. A total of 69 lines were observed, including the first detection of C$^{36}$S in this source. For the line-fitting procedure, we employed specific collisional rate coefficients and analysed the line intensity ratios for four of the studied species.
We also presented a laboratory study of rotational spectroscopy for the CC$^{33}$S and CCC$^{33}$S species, theoretical calculations for HCC$^{34}$S$^+$, and improved values of the rotational constants of other species detected based on the observations.

Our analysis of the abundance ratios derived from different isotopologues provides the most complete information to date on isotopic ratios for these molecules in TMC-1. These ratios were compared in detail with Solar System values, and we revealed a consistency and offered insights into isotopic fractionation processes within the cloud. Additionally, we analysed the abundance ratios of the protonated and neutral species, specifically, for CS and CCS, through their respective isotopologues.
We also commented on the anomaly observed in the $^{13}C$ isotopologues of CCS and CCCS and analysed its possible causes.
Furthermore, we investigated the spatial distribution of the most abundant molecules within TMC-1, where we observed a distribution pattern that is consistent with the expected pattern. This spatial analysis enhances our understanding of the molecular environment and the chemical processes governing the molecule formation and distribution in dark cold clouds.

\section*{Data availability}
Table B.1. are only available in electronic form at the CDS via anonymous ftp to cdsarc.u-strasbg.fr (130.79.128.5) or via http://cdsweb.u-strasbg.fr/cgi-bin/qcat?J/A+A

\begin{acknowledgements}
We thank Ministerio de Ciencia e Innovaci\'on of Spain (MICIU) for funding support through projects
PID2019-106110GB-I00, PID2019-107115GB-C21 / AEI / 10.13039/501100011033, and
PID2023-147545NB-I00. We also thank ERC for funding
through grant ERC-2013-Syg-610256-NANOCOSMOS.
C.C., Y.E., M.A, and J.C. thank Ministry of Science and Technology of Taiwan 
and Consejo Superior de Investigaciones Científicas for funding support under 
the MOST-CSIC Mobility Action 2021 (Grants 11-2927-I-A49-502 and OSTW200006).
AGP and FL acknowledge financial support from Rennes Metropole and the European Research Council (Consolidator Grant COLLEXISM, Grant Agreement No. 811363), the CEA/GENCI (Grand Equipement National de Calcul Intensif) for awarding them access to the TGCC (Très Grand Centre de Calcul) Joliot Curie/IRENE supercomputer within the A0110413001 project and from the Institut Universitaire de France.
\end{acknowledgements}

\begin{appendix}
\section{Spectroscopic data}
\label{sec:lab}
\subsection{New molecular constants for $^{13}$CCS and C$^{13}$CS species}\label{cons_13ccs_c13cs}

The observed lines of $^{13}$CCS and C$^{13}$CS species (see Table B.1) were merged with the laboratory data
\citep{Yamamoto1990,Ikeda1997} to provide a new set of molecular constants using the SPFIT code \citep{Pickett1991}. They are given in Tables 
\ref{constants_13CCS} and \ref{constants_C13CS}.

\begin{table}
\caption{New spectroscopic parameters of $^{13}$CCS}
\label{constants_13CCS}
\centering
\begin{tabular}{lc}
\hline \hline
\multicolumn{1}{c}{Parameter}  & \multicolumn{1}{c}{Lab + TMC-1}  \\
\hline
$B$ (MHz)                &  ~  6188.08703(58)\,$^a$ \\
$D$ (kHz)                &  ~   1.57299(68)   \\
$\lambda$ (MHz)          &  ~   97203.99(24)       \\
$\lambda_D$ (kHz)        &  ~   24.82(44)    \\
$\gamma$ (MHz)           &  ~   $-$14.07(14) \\
$\gamma_D$ (kHz)         &  ~      [35.8]\,$^b$  \\
$b_F$${(^{13}C)}$ (MHz)  &  ~      19.80(73)     \\
$c$${(^{13}C)}$ (MHz)    &  ~      $-$45.8(24) \\
$\sigma$ (kHz)           &  ~        28.9         \\
$N_{lines}$              &  ~        39          \\
\hline
\end{tabular}
\tablefoot{
\tablefoottext{a}{Numbers in parentheses are 1$\sigma$ uncertainties in units of the last digits.}\\
\tablefoottext{b}{Fixed to the value reported by \citet{Ikeda1997} for $^{13}$CCS.}\\
}
\end{table}

\begin{table}
\caption{New spectroscopic parameters of C$^{13}$CS.}
\label{constants_C13CS}
\centering
\begin{tabular}{lc}
\hline \hline
\multicolumn{1}{c}{Parameter}  & \multicolumn{1}{c}{Lab + TMC-1}  \\
\hline
$B$ (MHz)                &  ~  6446.96495(51)\,$^a$ \\
$D$ (kHz)                &  ~   1.71185(80)   \\
$\lambda$ (MHz)          &  ~   97226.77(27)       \\
$\lambda_D$ (kHz)        &  ~   26.79(64)    \\
$\gamma$ (MHz)           &  ~   $-$14.634(17) \\
$\gamma_D$ (kHz)         &  ~      [32]\,$^b$  \\
$b_F$${(^{13}C)}$ (MHz)  &  ~    $-$17.665(35)     \\
$c$${(^{13}C)}$ (MHz)    &  ~     $-$11.08(11) \\
$\sigma$ (kHz)           &  ~        30.9         \\
$N_{lines}$              &  ~        32          \\
\hline
\end{tabular}
\tablefoot{
\tablefoottext{a}{Numbers in parentheses are 1$\sigma$ uncertainties in units of the last digits.}\\
\tablefoottext{b}{Fixed to the value reported by \citet{Ikeda1997} for C$^{13}$CS.}\\
}
\end{table}

\subsection{Laboratory spectroscopy data for CC$^{33}$S}\label{cons_cc33s}

\begin{table}
\caption{Observed laboratory transition frequencies for CC$^{33}$S}
\label{lines_cc33s}
\centering
\begin{tabular}{cccccccc}
\hline
\hline
$N'$ & $J'$ & $F'$ & $N''$ &  $J''$ & $F''$ &   $\nu_{obs}$  &   Obs-Calc \\
\hline
2 & 1 & 1/2 & 1 & 0 & 3/2 &     11001.360 &  -0.003     \\
2 & 1 & 3/2 & 1 & 0 & 3/2 &     11008.341 &  -0.004     \\
2 & 1 & 5/2 & 1 & 0 & 3/2 &     11025.010 &  -0.005     \\
1 & 2 & 5/2 & 2 & 1 & 5/2 &     22118.069 &  -0.000     \\
1 & 2 & 1/2 & 2 & 1 & 1/2 &     22120.113 &   0.000     \\
1 & 2 & 3/2 & 2 & 1 & 3/2 &     22120.784 &   0.001     \\
1 & 2 & 3/2 & 2 & 1 & 1/2 &     22127.766 &   0.001     \\
1 & 2 & 5/2 & 2 & 1 & 3/2 &     22134.739 &  -0.000     \\
1 & 2 & 7/2 & 2 & 1 & 5/2 &     22140.130 &   0.001     \\
\hline
\hline
\end{tabular}
\end{table}

The rotational spectrum of the CC$^{33}$S radical was observed using a Balle-Flygare narrowband type
Fourier-transform microwave (FTMW) spectrometer operating in the frequency region of 4-40 GHz
\citep{Endo1994,Cabezas2016}. The short-lived species CC$^{33}$S was produced in a supersonic
expansion by a pulsed electric discharge of a gas mixture of CS$_2$ (0.3\%) and C$_2$H$_2$ (0.3\%)
diluted in Ar. This gas mixture was flowed through a pulsed-solenoid valve that is accommodated in
the backside of one of the cavity mirrors and aligned parallel to the optical axis of the resonator.
A pulse voltage of 1400 V with a duration of 450 $\mu$s was applied between stainless steel
electrodes attached to the exit of the pulsed discharge nozzle (PDN), resulting in an electric
discharge synchronized with the gas expansion. The resulting products generated in the discharge
were supersonically expanded, rapidly cooled to a rotational temperature of $\sim$2.5\,K between
the two mirrors of the Fabry-P\'erot resonator, and then probed by FTMW spectroscopy. For measurements
of the paramagnetic lines, the Earth’s magnetic field was cancelled by using three sets of Helmholtz
coils placed perpendicularly to one another. Since the PDN is arranged parallel to the cavity of the
spectrometer, it is possible to suppress the Doppler broadening of the spectral lines, allowing to
resolve small hyperfine splittings. The spectral resolution is 5\,kHz and the frequency measurements
have an estimated accuracy better than 3\,kHz.

Quantum chemical calculations were carried out to estimate the molecular parameters of CC$^{33}$S. Very
precise values for the rotational constant can be obtained using experimental/theoretical ratios derived
for CCS parent species. This is the most common method to predict the expected experimental rotational
constants for an isotopic species of a given molecule when the rotational constants for its parent species
are known. The calculations were done using the second order M{\o}ller-Plesset perturbation
(MP2; \citealt{Moller1934}) and Dunning's augmented correlation-consistent polarized quadruple-$\zeta$
basis sets (aug-cc-pVQZ; \citealt{Dunning1989}). This calculations were carried out using the Gaussian16
\citep{Frisch2016} program package. For the frequency predictions, other parameters determined for the
parent species CCS \citep{Yamamoto1990} were used.

\begin{figure}
\centering
\includegraphics[angle=0,width=0.5\textwidth]{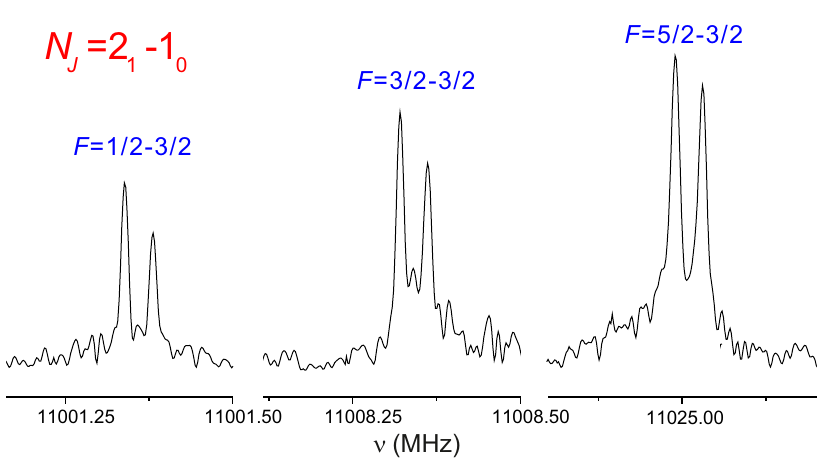}
\caption{Laboratory measurements of the three hyperfine components of the $N_J$=2$_1$-1$_0$ rotational transition of CC$^{33}$S we observed. The abscissa corresponds to the frequency of the lines in MHz. The spectra were achieved by 100-shots of accumulation. The coaxial arrangement of the adiabatic expansion and the resonator axis in the FTMW spectrometer produces an instrumental Doppler doubling.
The resonance frequencies are calculated as the average of the two Doppler components.} \label{ftmw_CCS}
\end{figure}

\begin{table}
\caption{Spectroscopic parameters of CC$^{33}$S}
\label{constants_CC33S}
\centering
\begin{tabular}{lc}
\hline \hline
\multicolumn{1}{c}{Parameter}  & \multicolumn{1}{c}{Lab + TMC-1}  \\
\hline
$B$ (MHz)                &  ~  6404.4551(34)\,$^a$   \\
$D$ (kHz)                &  ~   [1.728]\,$^b$     \\
$\lambda$ (MHz)          &  ~   97220.7(47)          \\
$\gamma$  (MHz)          &  ~   [$-$14.737]          \\
$\gamma_D$ (kHz)         &  ~      [27]           \\
$b_F$${(^{33}S)}$ (MHz)  &  ~     23.7(17)           \\
$c$${(^{33}S)}$ (MHz)    &  ~     $-$64.4(58)        \\
$eQq$${(^{33}S)}$ (MHz)  &  ~   $-$5.1471(79)        \\
$\sigma$ (kHz)           &  ~        13.1            \\
$N_{lines}$              &  ~        23              \\
\hline
\end{tabular}
\tablefoot{
\tablefoottext{a}{Numbers in parentheses are 1$\sigma$ uncertainties in units of the last digits.}\\
\tablefoottext{b}{Values in brackets were fixed to the theoretical values.}\\
}
\end{table}

We observed by FTMW spectroscopy, two rotational transitions with quantum numbers $N_J$=2$_1$-1$_0$ and
1$_2$-2$_1$ at 11 and 22 GHz, respectively. An example is shown in Figure \ref{ftmw_CCS}. Each transition
is split into several hyperfine components due to the presence of the $^{33}$S nucleus, which has a
non-zero nuclear spin. The frequencies for all these hyperfine components (see Table \ref{lines_cc33s})
and those observed in
TMC-1 were analysed with the SPFIT program \citep{Pickett1991} using a Hamiltonian for a linear
molecule in a $^3\Sigma$ electronic state. The employed Hamiltonian can be written as follows:
\begin{equation}
H = H_{rot} + H_{ss} + H_{sr} + H_{mhf}
,\end{equation}
where $H_{rot}$, $H_{ss}$, and $H_{sr}$ denote the rotational, spin-spin, and spin-rotation terms,
respectively, and $H_{mhf}$ represents the magnetic hyperfine coupling interaction term due to the
$^{33}$S nucleus. The coupling scheme used is \textbf{J}\,=\,\textbf{N}\,+\,\textbf{S},
\textbf{F}\,=\,\textbf{J}\,+\,\textbf{I}($^{33}$S).

The molecular constants determined from the fit are given in Table \ref{constants_CC33S}. We
determined the $B$ rotational constant, the spin-spin interaction constant, $\lambda$, and
the hyperfine constants for the $^{33}$S nucleus, named Fermi contact constant, $b_F$, the
dipole-dipole constant, $c$, and the nuclear quadrupole coupling constant, $eQq$. Other
parameters like the distortion constant, $D$, and the spin-rotation interaction constants,
$\gamma$ and $\gamma_D$, were kept fixed to the values determined for the parent species
CCS \citep{Yamamoto1990}.

\subsection{C$^{13}$C$^{34}$S}\label{cons_c13c34s}
Using similar theoretical and experimental methods than for CC$^{33}$S we explore
the rotational spectrum of C$^{13}$C$^{34}$S.
We detected four unidentified lines in our survey in the Q-band very close to the predicted
frequencies at 32.9 and 42.3 GHz. They were attributed to the most intense hyperfine components
of the rotational transitions with quantum numbers $N_J$=2$_3$-1$_2$ and 3$_4$-2$_3$,
respectively. Each transition is split in two intense hyperfine components due to the
presence of the $^{13}$C nucleus, which has a non-zero nuclear spin. The frequencies
for these four hyperfine components were analysed with the SPFIT program \citep{Pickett1991}
using the Hamiltonian for a linear molecule in a $^3\Sigma$ electronic state described above. 
The molecular constants determined from the fit are given in Table \ref{constants_C13C34S}. We
determined the $B$ rotational constant and  the spin-spin interaction constant, $\lambda$.
Other parameters were kept fixed to the values determined for CCS, and CC$^{34}$S
\citep{Yamamoto1990,Ikeda1997}, and for C$^{13}$CS determined here (see \ref{cons_13ccs_c13cs}).

\begin{table}
\caption{Spectroscopic parameters of C$^{13}$C$^{34}$S}
\label{constants_C13C34S}
\centering
\begin{tabular}{lc}
\hline \hline
\multicolumn{1}{c}{Parameter}  & \multicolumn{1}{c}{Lab + TMC-1}  \\
\hline
$B$ (MHz)                     &  ~  6303.213(14)\,$^a$ \\
$D$ (kHz)                     &  ~   [1.654]\,$^b$  \\
$\lambda$  (MHz)              &  ~   97226.7(16)       \\
$\lambda_D$  (kHz)            &  ~   [26.76]\,$^b$   \\
$\gamma$  (MHz)               &  ~   [$-$14.338]\,$^b$ \\
$\gamma_D$ (kHz)              &  ~      [32]\,$^c$  \\
$b_F$${(^{13}C)}$ (MHz)       &  ~    $-$17.74(14)     \\
$c$${(^{13}C)}$   (MHz)       &  ~     [$-$11.08]\,$^d$ \\
$\sigma$ (kHz)                &  ~        10.5         \\
$N_{lines}$                   &  ~        4            \\
\hline
\end{tabular}
\tablefoot{
\tablefoottext{a}{Numbers in parentheses are 1$\sigma$ uncertainties in units of the last digits.}\\
\tablefoottext{b}{Fixed to the value reported by \citet{Yamamoto1990} for CC$^{34}$S.}\\
\tablefoottext{c}{Fixed to the value reported by \citet{Ikeda1997} for C$^{13}$CS.}\\
\tablefoottext{d}{Fixed to the value obtained in this work for C$^{13}$CS.}\\
}
\end{table}

\begin{table}
\caption{Spectroscopic parameters of CCC$^{33}$S}
\label{constants_CCC33S}
\centering
\begin{tabular}{lcc}
\hline \hline
\multicolumn{1}{c}{Parameter}  & \multicolumn{1}{c}{Lab + TMC-1} & \multicolumn{1}{c}{\citet{McGuire2018}}  \\
\hline
$B$ (MHz)                &  ~  2854.386789(89)\,$^a$ &  ~  2854.3868(2)   \\
$D$ (kHz)                &  ~   0.2169(16)           &  ~         0.222(4)   \\
$eQq$${(^{33}S)}$        &  ~   $-$10.6029(29)       &  ~        $-$10.593(6)$^b$     \\
$\sigma$ (kHz)           &  ~        2.5             &  ~        -        \\
$N_{lines}$              &  ~        32              &  ~        9        \\
\hline
\end{tabular}
\tablefoot{
\tablefoottext{a}{Numbers in parentheses are 1$\sigma$ uncertainties in units of the last digits.}\\
\tablefoottext{b}{A value of -15.889(9) MHz is provided by \citet{McGuire2018}. We assume that it corresponds to 3/2$\times$$eQq$.}
}
\end{table}

\subsection{CCC$^{33}$S}\label{cons_CCC33S}
The rotational spectrum for the CCC$^{33}$S species has been observed in the laboratory by \citep{McGuire2018},
who reported the observation of nine hyperfine components. In this work, we measured the rotational
transition for CCC$^{33}$S in the 10-34 GHz frequency region and we observed a total of 23 hyperfine
components. The measured frequencies are given in Table \ref{lines_CCC33S}. The experimental setup and experimental conditions are the same that those used in this
work for CC$^{33}$S.

Towards TMC-1 we observe two well resolved hyperfine components for the $J$=6-5 transition. The $J$=7-6 transition appears as two overlapping lines,
while for the transition $J$=8-7 the hyperfine structure is collapsed into a single feature (see Figure \ref{fig:CCC33S}).
All the observed laboratory frequencies, together with those observed in TMC-1
(see Table B.1), were analysed using
the SPFIT program \citep{Pickett1991}, using a Hamiltonian for singlet linear molecules, with the
following form: $H$ = $H_{rot}$ + $H_{nqc}$ where $H_{rot}$ contains rotational and centrifugal
distortion parameters and $H_{nqc}$ the nuclear quadrupole coupling interactions. The coupling
scheme used is \textbf{F}\,=\,\textbf{J}\,+\,\textbf{I}($^{33}$S). The analysis rendered the
experimental constants listed in Table  \ref{constants_CCC33S}.

\begin{figure}
\centering
\includegraphics[angle=0,width=0.5\textwidth]{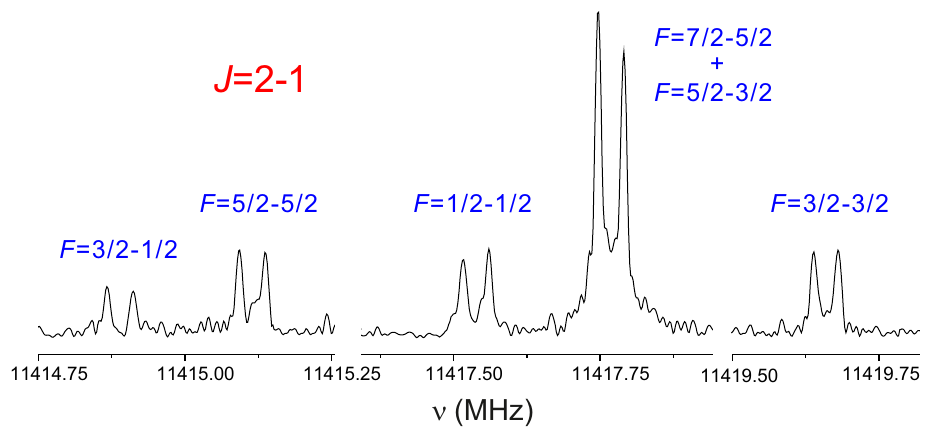}
\caption{Laboratory measurements of a section of the rotational spectrum for the CCC$^{33}$S observed in this work,
showing the five hyperfine components for the $J$=2-1 rotational transition.
The spectra were achieved by 200-shots of accumulation. The coaxial arrangement of
the adiabatic expansion and the resonator axis in the FTMW spectrometer produces an
instrumental Doppler doubling. The resonance frequencies are calculated as the average
of the two Doppler components.} \label{ftmw_cccs}
\end{figure}

\begin{table}
\caption{Observed laboratory transition frequencies for CCC$^{33}$S}
\label{lines_CCC33S}
\centering
\begin{tabular}{cccccc}
\hline
\hline
$J'$ & $F'$ &   $J''$ & $F''$ &   $\nu_{obs}$  &   Obs-Calc \\
\hline
2 & 2 & 1 & 2  &   11419.659 &  -0.001 \\
2 & 3 & 1 & 2  &   11417.768 &   0.002 \\
2 & 4 & 1 & 3  &   11417.768 &   0.002 \\
2 & 1 & 1 & 1  &   11417.538 &  -0.001 \\
2 & 3 & 1 & 3  &   11415.115 &  -0.000 \\
2 & 2 & 1 & 1  &   11414.889 &   0.000 \\
3 & 4 & 2 & 4  &   17123.770 &  -0.001 \\
3 & 3 & 2 & 2  &   17125.766 &   0.000 \\
3 & 2 & 2 & 1  &   17125.766 &   0.000 \\
3 & 4 & 2 & 3  &   17126.424 &   0.002 \\
3 & 5 & 2 & 4  &   17126.424 &   0.002 \\
3 & 3 & 2 & 3  &   17127.658 &  -0.001 \\
3 & 2 & 2 & 2  &   17128.416 &  -0.000 \\
4 & 3 & 3 & 2  &   22834.810 &  -0.000 \\
4 & 5 & 3 & 4  &   22835.120 &   0.002 \\
4 & 6 & 3 & 5  &   22835.120 &   0.002 \\
5 & 4 & 4 & 3  &   28543.632 &  -0.000 \\
5 & 5 & 4 & 4  &   28543.632 &  -0.000 \\
5 & 6 & 4 & 5  &   28543.816 &   0.001 \\
5 & 7 & 4 & 6  &   28543.816 &   0.001 \\
6 & 6 & 5 & 5  &   34252.371 &  -0.003 \\
6 & 5 & 5 & 4  &   34252.371 &  -0.003 \\
6 & 7 & 5 & 6  &   34252.497 &   0.002 \\
6 & 8 & 5 & 7  &   34252.497 &   0.002 \\
\hline
\hline
\end{tabular}
\end{table}

\subsection{HCC$^{34}$S$^+$}\label{cons_HCC34S+}

The HCC$^{34}$S$^+$ isotopic species has not been observed before in the laboratory nor in space. The rotational constant for 
this isotopologue can be estimated from quantum chemical calculations as shown above for other species as CC$^{33}$S. We 
detected four unidentified lines in our survey in the Q-band very close to the predicted frequencies at 31.4 and 42.1 GHz
(see Fig. \ref{fig:HCC34S+} and Table B.1). 
They were attributed to the most intense hyperfine components of the rotational transitions with quantum numbers 
$N_J$=2$_3$-1$_2$ and 3$_4$-2$_3$, respectively. Each transition is split in two intense hyperfine components due to 
the presence of the hydrogen nucleus, which has a non-zero nuclear spin. The frequencies for these four hyperfine 
components were analysed with the SPFIT program \citep{Pickett1991} using a Hamiltonian for a linear molecule in a 
$^3\Sigma$ electronic state. The employed Hamiltonian has been described above. The analysis rendered the experimental 
constants listed in Table \ref{constants_HCC34S+}. This is the first time that this isotopologue of HCCS$^+$ is detected 
in space and provides a definitive identification of this species as previously
claimed by \citet{Cabezas2022a}.

\begin{table}
\caption{Spectroscopic parameters of HCC$^{34}$S$^+$}
\label{constants_HCC34S+}
\centering
\begin{tabular}{lc}
\hline
\hline
\multicolumn{1}{c}{Parameter}  & \multicolumn{1}{c}{HCC$^{34}$S$^+$} \\
\hline
$B$ (MHz)                      &  ~  5889.02214(82)\,$^a$     \\
$D$ (KHz)                      &  ~   [1.2543]\,$^b$       \\
$\lambda$ (MHz)                &  ~   [108970.78]\,$^b$       \\
$\lambda_D$ (kHz)              &  ~    [40.60]\,$^b$        \\
$\gamma$ (MHz)                 &  ~   [$-$41.776]\,$^b$       \\
$b_F$$^{\rm(H)}$ (MHz)         &  ~   $-$45.024(99)           \\
$c$$^{\rm(H)}$ (MHz)           &  ~    [31.663]\,$^b$         \\
$\sigma$ (kHz)                 &  ~        8.0                \\
$N_{lines}$                    &  ~        4                  \\
\hline
\hline
\end{tabular}
\tablefoot{
\tablefoottext{a}{Numbers in parentheses are 1$\sigma$ uncertainties in units of the last digits.}\\
\tablefoottext{b}{Fixed to the value reported by \citet{Cabezas2022a} for HCCS$^{+}$.}\\
}
\end{table}

\subsection{HC$^{33}$S$^+$}\label{cons_HC33S+}

To the best of our knowledge the rotational spectrum for the HC$^{33}$S$^+$ species has not
been observed in the laboratory. The rotational constant for the vibrational ground state of this
isotopologue can be estimated from the
substitution structure provided by \citet{Margules2003} for the HCS$^+$ molecule to be 21160.0$\pm$0.5 MHz.
The uncertainty has been estimated from the observed differences between observed and predicted 
rotational constants for the other isotopologues of HCS$^+$.
Using this value as a starting point we easily measured in our line surveys eight hyperfine components
of the rotational transitions $J$=1-0 and 2-1 of HC$^{33}$S$^+$ (see Fig. \ref{fig:HCS+}). The measured frequencies
are given in Table B.1. We use the same Hamiltonian and identical coupling scheme
than for CCC$^{33}$S. The analysis provides the rotational constant and the nuclear quadrupole coupling
interaction given in Table \ref{constants_HC33S+}. This is the first time that this isotopologue
of HCS$^+$ is detected in space.

\begin{table}
\caption{Spectroscopic parameters of HC$^{33}$S$^+$}
\label{constants_HC33S+}
\centering
\begin{tabular}{lc}
\hline \hline
\multicolumn{1}{c}{Parameter}  & \multicolumn{1}{c}{TMC-1}  \\
\hline
$B$ (MHz)                &  ~  21158.9439(12)\,$^a$   \\
$D$ (kHz)                &  ~   21.17$^b$         \\
$eQq$${(^{33}S)}$ (MHz)  &  ~   13.900(26)        \\
$c$  (kHz)               &  ~   11.6(24) \\
$\sigma$\, (kHz)         &  ~        12.4              \\
$N_{lines}$              &  ~        8               \\
$J_{max}$                &  ~        2             \\ 
\hline
\end{tabular}
\tablefoot{
\tablefoottext{a}{Numbers in parentheses are 1$\sigma$ uncertainties in units of the last digits.}
\tablefoottext{b}{Fixed to the averaged value of $D$ for HCS$^+$ and HC$^{34}$S$^+$.}
}
\end{table}

\subsection{C$_4$S}\label{sec:C4S}
\begin{figure}
\includegraphics[angle=0,width=0.48\textwidth]{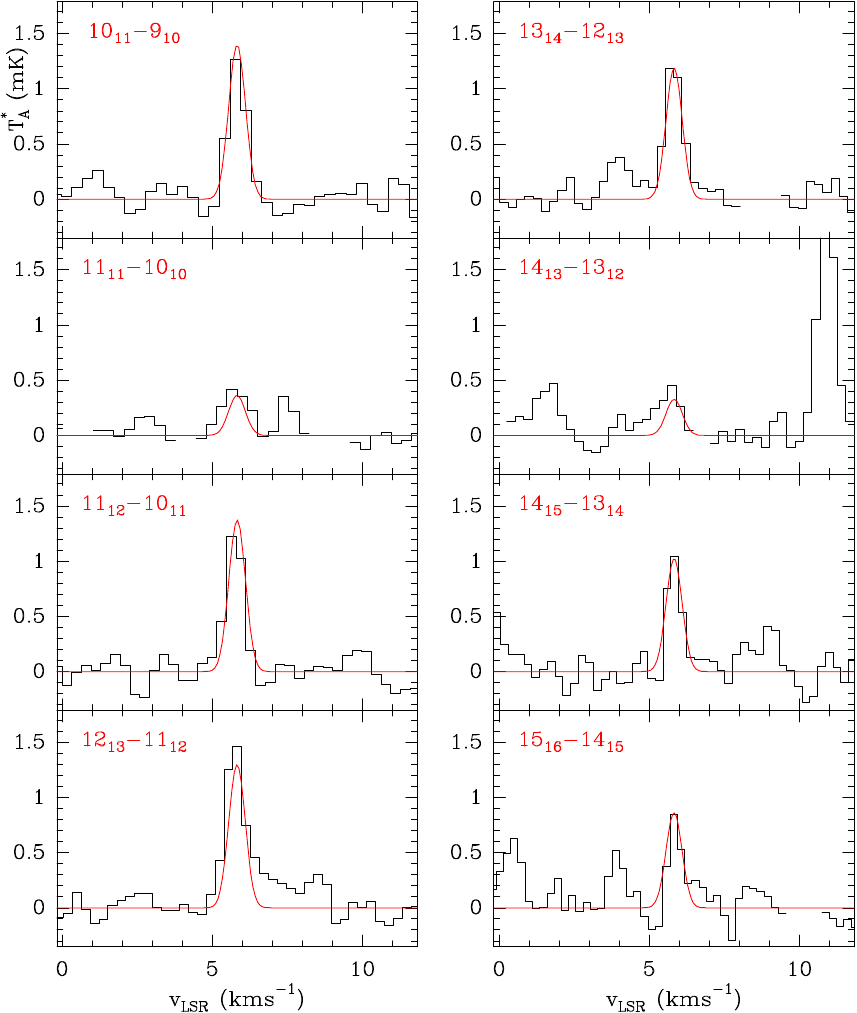}
\caption{Observed transitions of C$_4$S observed towards TMC-1. 
The abscissa 
corresponds to the local standard of rest velocity of the lines adopting the rest frequencies given in Table 
B.1. The 
ordinate corresponds to the antenna temperature corrected for 
atmospheric and telescope losses in mK. 
The derived line parameters of the observed lines are given in Table B.1. 
The synthetic spectra (red line) are derived for $N$=3.8x10$^{10}$cm$^{-2}$ and the LVG analysis
described in section \ref{sec:C4S}.} 
\label{fig:C4S}
\end{figure}

C$_4$S (thiobutatrienylidene) has a ground electronic state $^3$$\Sigma$$^-$, and has been previously detected with the QUIJOTE line survey \citep{Cernicharo2021c}.
This species has been observed in the laboratory by \citet{Hirahara1993} and \citet{Gordon2001} and its dipole moment has been 
derived from ab initio calculations to be 4.03\,D \citep{Pascoli1998}. 
Eight rotational transitions, with $N$=10 up to 15, have been detected and are shown in  Fig. \ref{fig:C4S}. 
The derived line parameters are given in Table B.1.
The predicted frequencies using the rotational constants derived from the available laboratory data show systematic 
differences with respect to the observed lines in the QUIJOTE's domain of up to 70 kHz. 
Hence, we derive improved rotational constants for this species by fitting the standard Hamiltonian of a $^3$$\Sigma$ molecule
to the laboratory data and to the observed transitions in TMC-1. The results are given in
Table \ref{constants_C4S}. 

No collisional rate coefficients are available for this species. In order to have
an estimation of its excitation we have calculated the C$_4$S-He rate coefficients 
from the HC$_3$N-He rate coefficients of \citep{Green1978} using the IOS approximation for a $^3$$\Sigma$ 
molecule \citep{Corey1984,Corey1984b,Corey1986,
Fuente1990}. Although these collisional rate coefficients are rather uncertain, they allow us to reproduce reasonably well the observed intensities for a volume
density of 4$\times$10$^4$ cm$^{-3}$ (see Fig. \ref{fig:C4S}). This value of n(H$_2$) has been considered with caution.
The calculated excitation temperatures between 31 GHz and 50 GHz vary between 6.5 K and 8 K. The column density for this species
is 3.8$\times$10$^{10}$ cm$^{-2}$. A similar value can be obtained adopting LTE for a rotational temperature of 8\,K. The 
CCCS/CCCCS abundance ratio of $\sim$180 is 
consistent with the value derived by \citet{Cernicharo2021c}, but much lower than that of
CCS/CCCS ($\sim$5). The expected line intensities of the isotopologues of C$_4$S are too low to be detected with the present
sensitivity of our line survey.

\begin{table}
\caption{Spectroscopic parameters of C$_4$S}
\label{constants_C4S}
\centering
\begin{tabular}{lcc}
\hline \hline
\multicolumn{1}{c}{Parameter}  & \multicolumn{1}{c}{Laboratory$^b$} & \multicolumn{1}{c}{Lab+TMC-1$^c$}  \\
\hline
$B$ (MHz)                &  ~  1519.20622(14)\,$^a$ &  ~  1519.206116(94)   \\
$D$ (Hz)                 &  ~   48.0(15)      &  ~   46.7(9)       \\
$\lambda$ (MHz)          &  ~   113848(24)    &  ~   113844(15)    \\
$\lambda_D$ (kHz)        &  ~   12.26(12)     &  ~   12.29(11)        \\
$\gamma$ (MHz)           &  ~   -4.15(32)     &  ~   -4.1(2)          \\   
$\sigma$\ (kHz)          &  ~        2.9      &  ~        4.6         \\
$N_{lines}$              &  ~       24        &  ~       32           \\
$J_{max}$                &  ~        8        &  ~       16           \\
\hline
\end{tabular}
\tablefoot{
\tablefoottext{a}{Numbers in parentheses are 1$\sigma$ uncertainties in units of the last digits.}
\tablefoottext{b}{Constants derived from a fit to the laboratory frequencies measured by \citet{Hirahara1993} and \citet{Gordon2001}.}
\tablefoottext{c}{Fit to the laboratory and the TMC-1 data.}
}
\end{table}

\subsection{C$_5$S}\label{sec:C5S}
\begin{figure}
\includegraphics[angle=0,width=0.48\textwidth]{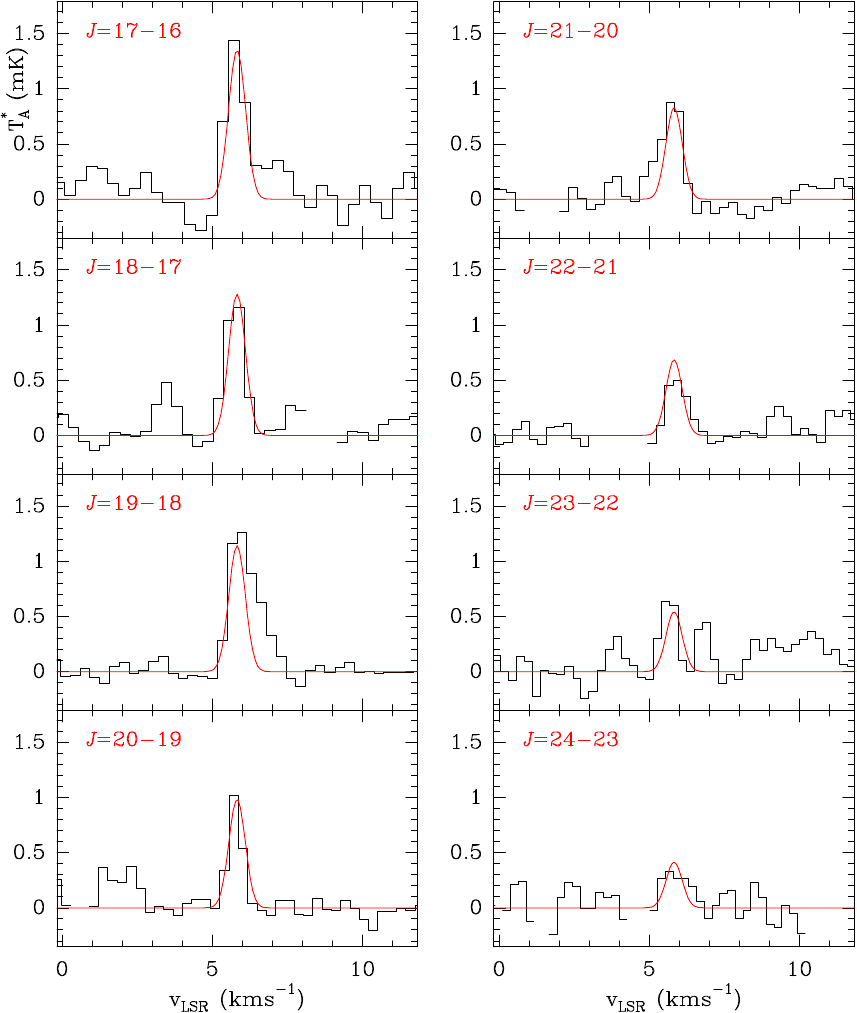}
\caption{Observed transitions of C$_5$S observed towards TMC-1. 
The abscissa 
corresponds to the local standard of rest velocity of the lines adopting the rest frequencies given in Table 
B.1. The 
ordinate corresponds to the antenna temperature corrected for 
atmospheric and telescope losses in mK. 
The derived line parameters of the observed lines are given in Table B.1. 
The synthetic spectra (red line) are derived for $N$=3.0x10$^{10}$cm$^{-2}$ and the LVG analysis
described in section \ref{sec:C5S}.} 
\label{fig:C5S}
\end{figure}

C$_5$S (thiopentatetraenylidene) has a ground electronic state $^1$$\Sigma$$^+$. This molecule
was first tentatively detected towards IRC +10216 by \citet{Bell1993} and confirmed later by \citet{Agundez2014} in the same
source. It has been also previously detected towards TMC-1 \citep{Cernicharo2021c}. The molecule
has been observed in the laboratory up to $J_u$=10 by \citep{Kasai1993,Gordon2001}. Laboratory data for all its
singly substituted isotopologues are also available \citep{Gordon2001}. The dipole moment of the molecule
has been calculated to be 4.65\,D by \citet{Lee1997}. An inspection of the QUIJOTE data permits to detect all lines
of this species between $J_u$=17 and 26. The observed lines are shown in Fig. \ref{fig:C5S} and 
the derived line parameters are given in Table B.1. 
Some systematic frequency differences are found between 
observed and predicted frequencies. Hence, improved rotational constants are provided in Table \ref{constants_C5S} which have been obtained
from a fit to the laboratory and space frequencies using the standard Hamiltonian for a linear molecule.

\begin{table}
\caption{Spectroscopic parameters of C$_5$S}
\label{constants_C5S}
\centering
\begin{tabular}{lcc}
\hline \hline
\multicolumn{1}{c}{Parameter}  & \multicolumn{1}{c}{Laboratory$^b$} & \multicolumn{1}{c}{Lab+TMC-1$^c$}  \\
\hline
$B$ (MHz)                &  ~  922.703433(96)\,$^a$ &  ~  922.703392(46)   \\
$D$ (Hz)                 &  ~   14.08(65)     &  ~   13.80(9)       \\
$\sigma$\ (kHz)          &  ~        1.8      &  ~        8.1         \\
$N_{lines}$              &  ~       13        &  ~       22           \\
$J_{max}$                &  ~       10        &  ~       23           \\
\hline
\end{tabular}
\tablefoot{
\tablefoottext{a}{Numbers in parentheses are 1$\sigma$ uncertainties in units of the last digits.}
\tablefoottext{b}{Fit to the laboratory transitions measured by \citet{Kasai1993} and \citet{Gordon2001}.}
\tablefoottext{c}{Fit to the laboratory and QUIJOTE data.}
}
\end{table}

Collisional rate coefficients for C$_5$S with He have been calculated by \citet{Khadri2020}. Using this data, we derive from a
fit to the observed line profiles a value for the density of H$_2$ of 5$\times$10$^4$ cm$^{-3}$ and a column density
for C$_5$S of 3.0$\times$10$^{10}$ cm$^{-2}$. Similar column densities are obtained for an uniform rotational
temperature of 8.5\,K. Hence, it seems that C$_4$S and C$_5$S have similar abundances.

\onecolumn
\section{Derived line parameters}

\begin{longtable}{lccccccc}
\caption{Observed line parameters for S-bearing species and their isotopologues in this study} \label{line_parameters}\\
\hline \hline
Molecule        &Transition& $F_u-F_l$ &$\nu_{rest}$$^a$  & v$_{LSR}$$^b$   & $\int$ $T_A^*$ dv $^c$ & $\Delta$v$^d$&   T$_A^*$$^e$\\
                &        &         &(MHz)         & km s$^{-1}$ & mK km s$^{-1}$&km s$^{-1}$  &   mK     \\
\hline 
\endfirsthead
\caption{continued.}\\
\hline \hline
Molecule        &Transition& $F_u-F_l$ &$\nu_{rest}$$^a$  & v$_{LSR}$$^b$   & $\int$ $T_A^*$ dv $^c$ & $\Delta$v$^d$&   T$_A^*$$^e$\\
                &        &         &(MHz)         & km s$^{-1}$ & mK km s$^{-1}$&km s$^{-1}$  &   mK     \\
            
\hline
\endhead
\hline
\endfoot
CS$*$             &  1-0   &       & 48990.957$\pm$0.000& 5.74$\pm$0.01&1098.62$\pm$1.73& 0.74$\pm$0.01&1390.60$\pm$0.18\\
                &        &       &                    & 6.27$\pm$0.01&  93.34$\pm$1.81& 0.62$\pm$0.03& 141.68$\pm$0.18\\
                &        &       &                    & 6.54$\pm$0.01&  61.30$\pm$1.51& 0.95$\pm$0.03&  60.85$\pm$0.18\\
                &  2-1   &       & 97980.953$\pm$0.000& 5.70$\pm$0.01&1087.83$\pm$0.97& 0.60$\pm$0.01&1699.60$\pm$2.20\\
                &        &       &                    & 6.25$\pm$0.01& 105.77$\pm$0.04& 0.68$\pm$0.01& 145.83$\pm$2.20\\  
                &        &       &                    & 6.81$\pm$0.01&  13.55$\pm$0.91& 0.67$\pm$0.01&  18.89$\pm$2.20\\  
                &        &       &                    &              &                &              &                \\                             
$^{13}$CS       &  1-0   &       & 46247.564$\pm$0.001& 5.78$\pm$0.01& 151.69$\pm$0.14& 0.70$\pm$0.01& 203.31$\pm$0.17\\
                &  2-1   &       & 92494.271$\pm$0.001& 5.80$\pm$0.01& 121.92$\pm$0.33& 0.60$\pm$0.01& 191.23$\pm$0.58\\
                &        &       &                    &              &                &              &                \\                             
C$^{34}$S       &  1-0   &       & 48206.942$\pm$0.001& 5.82$\pm$0.01& 430.30$\pm$0.21& 0.66$\pm$0.01& 617.05$\pm$0.21\\
                &  2-1   &       & 96412.053$\pm$0.001& 5.83$\pm$0.01& 385.18$\pm$0.30& 0.57$\pm$0.01& 632.74$\pm$0.69\\
                &        &       &                    &              &                &              &                \\                             
$^{13}$C$^{34}$S&  1-0   &       & 45463.412$\pm$0.001& 5.78$\pm$0.02&   6.97$\pm$0.11& 0.72$\pm$0.01&   9.11$\pm$0.14\\
                &  2-1   &       & 90925.997$\pm$0.001& 5.65$\pm$0.04&   7.78$\pm$1.16& 0.55$\pm$0.10&  13.40$\pm$2.22\\
                &        &       &                    &              &                &              &                \\                             
$^{13}$C$^{33}$S&  1-0   &1/2-3/2& 45839.802$\pm$0.004& 5.76$\pm$0.06&   0.55$\pm$0.10& 0.58$\pm$0.13&   0.89$\pm$0.12\\
                &        &5/2-3/2& 45842.444$\pm$0.010& 5.75$\pm$0.06&   0.77$\pm$0.07& 0.77$\pm$0.08&   0.94$\pm$0.10\\
				&        &3/2-3/2& 45845.614$\pm$0.010& 5.66$\pm$0.07&   0.27$\pm$0.05& 0.48$\pm$0.09&   0.54$\pm$0.10\\
                &        &       &                    &              &                &              &                \\                             
C$^{36}$S       &  1-0   &       & 47508.784$\pm$0.001& 5.78$\pm$0.03&   1.92$\pm$0.15& 0.70$\pm$0.06&   2.58$\pm$0.21\\
                &  2-1   &       & 95016.664$\pm$0.001& 5.64$\pm$0.03&   1.26$\pm$0.45& 0.55$\pm$0.00&   2.15$\pm$1.25\\
                &        &       &                    &              &                &              &                \\                             
C$^{33}$S       &   1-0  &1/2-3/2& 48583.262$\pm$0.017& 5.83$\pm$0.01&  19.34$\pm$0.17& 0.63$\pm$0.01&  28.69$\pm$0.17\\
                &        &5/2-3/2& 48585.891$\pm$0.004& 5.81$\pm$0.01&  55.27$\pm$0.15& 0.63$\pm$0.01&  82.18$\pm$0.18\\
                &        &3/2-3/2& 48589.048$\pm$0.014& 5.76$\pm$0.01&  36.70$\pm$0.18& 0.64$\pm$0.01&  53.96$\pm$0.18\\
                &   2-1  &1/2-3/2& 97166.245$\pm$0.030&              &               &               &  $\le$8.40     \\          
                &        &3/2-3/2& 97169.471$\pm$0.013& 5.72$\pm$0.02&   8.18$\pm$1.02&  0.37$\pm$0.06&  20.72$\pm$2.82\\
                &        &5/2-3/2& 97171.801$\pm$0.003& 5.81$\pm$0.01&  57.19$\pm$1.26&  0.58$\pm$0.02&  92.21$\pm$2.82\\
                &        &7/2-5/2& 97171.819$\pm$0.003&              &                &               &                \\                                  
                &        &1/2-1/2& 97172.031$\pm$0.003& 5.68$\pm$0.02&   5.95$\pm$0.90&  0.30$\pm$0.05&  18.75$\pm$2.82\\
                &        &3/2-5/2& 97172.626$\pm$0.004& 5.66$\pm$0.06&   2.63$\pm$0.96&  0.28$\pm$0.14&   8.69$\pm$2.82\\
                &        &5/2-5/2& 97174.956$\pm$0.015& 5.70$\pm$0.03&   6.16$\pm$0.94&  0.36$\pm$0.06&  16.03$\pm$2.82\\
                &        &3/2-1/2& 97175.257$\pm$0.017& 5.71$\pm$0.03&   6.18$\pm$1.01&  0.35$\pm$0.07&  16.42$\pm$2.82\\
                &        &       &                    &              &                &              &                \\                             
\hline
                &        &       &                    &              &                &              &                \\                             
HCS$^+$         & 1-0    &       & 42674.195$\pm$0.001& 5.78$\pm$0.01& 270.30$\pm$0.09& 0.61$\pm$0.01&   415.0$\pm$0.14 \\
                & 2-1    &       & 88347.875$\pm$0.003& 5.80$\pm$0.01& 361.94$\pm$1.30& 0.49$\pm$0.01&   701.5$\pm$2.91 \\
                &        &       &                    &              &                &              &                \\                             
H$^{13}$CS$^+$  & 1-0    &       & 40888.910$\pm$0.002& 5.76$\pm$0.01&   3.18$\pm$0.10& 0.64$\pm$0.02&    4.70$\pm$0.13\\
                & 2-1    &       & 81777.345$\pm$0.004& 5.80$\pm$0.02&   5.41$\pm$0.04& 0.49$\pm$0.03&   10.29$\pm$0.68\\
                &        &       &                    &              &                &              &                \\                             
HC$^{34}$S$^+$  & 1-0    &       & 41983.063$\pm$0.002& 5.77$\pm$0.01&  13.96$\pm$0.09& 0.61$\pm$0.01&   21.56$\pm$0.13\\
                & 2-1    &       & 83965.626$\pm$0.003& 5.76$\pm$0.01&  20.99$\pm$0.43& 0.50$\pm$0.01&   39.83$\pm$0.74\\
                &        &       &                    &              &                &              &                \\                             
HC$^{33}$S$^{+\#}$& 1-0  &1/2-3/2& 42314.299$\pm$0.010& 5.83         &   0.39$\pm$0.09&  0.52$\pm$0.11&   0.71$\pm$0.16\\
                &        &5/2-3/2& 42317.116$\pm$0.010& 5.83         &   1.35$\pm$0.10&  0.58$\pm$0.05&   2.17$\pm$0.16\\
			    &        &3/2-3/2& 42320.562$\pm$0.010& 5.83         &   0.92$\pm$0.10&  0.68$\pm$0.14&   1.28$\pm$0.16\\
                & 2-1    &3/2-3/2& 84632.278$\pm$0.020& 5.83         &   1.21$\pm$0.35&  0.78$\pm$0.14&   1.46$\pm$0.77\\
                &        &5/2-3/2& 84634.817$\pm$0.010& 5.83         &   3.78$\pm$0.37&  0.57$\pm$0.06&   6.49$\pm$0.77\\
	            &        &7/2-5/2& 84634.817$\pm$0.010&              &                &               &                \\
                &        &1/2-1/2& 84635.079$\pm$0.020& 5.83         &   1.21$\pm$0.36&  0.57$\pm$0.17&   2.00$\pm$0.77\\          			 
                &        &5/2-5/2& 84638.265$\pm$0.020& 5.83         &   1.18$\pm$0.48&  0.51$\pm$0.18&   2.18$\pm$0.77\\
                &        &       &                    &              &                &              &                \\                             
\hline
                &        &       &                    &              &                &              &                \\                             
CCS            &$7_6-6_6$&       & 32955.981$\pm$0.007& 5.85$\pm$0.04&   0.78$\pm$0.09&  0.66$\pm$0.08&   1.10$\pm$0.10\\
               &$2_3-1_2$&       & 33751.370$\pm$0.001& 5.78$\pm$0.01&1213.60$\pm$0.11&  0.75$\pm$0.01&1524.50$\pm$0.07\\
			   &$3_3-2_2$&       & 38866.421$\pm$0.001& 5.77$\pm$0.01& 202.32$\pm$0.14&  0.68$\pm$0.01& 281.74$\pm$0.09\\
			   &$8_7-7_7$&       & 42136.104$\pm$0.009& 5.88$\pm$0.04&   0.34$\pm$0.05&  0.62$\pm$0.08&   0.66$\pm$0.09\\
			   &$4_3-3_2$&       & 43981.023$\pm$0.001& 5.79$\pm$0.01& 189.36$\pm$0.13&  0.61$\pm$0.01& 291.54$\pm$0.16\\
			   &$3_4-2_3$&       & 45379.030$\pm$0.001& 5.81$\pm$0.01&1169.50$\pm$0.11&  0.65$\pm$0.01&1690.10$\pm$0.16\\
               &$6_5-5_4$&       & 72323.813$\pm$0.002& 5.80$\pm$0.01& 222.22$\pm$1.87&  0.54$\pm$0.01& 384.26$\pm$3.71\\
               &$6_6-5_5$&       & 77731.723$\pm$0.002& 5.80$\pm$0.01& 160.03$\pm$0.57&  0.59$\pm$0.01& 256.67$\pm$0.95\\
	           &$6_7-5_6$&       & 81505.211$\pm$0.001& 5.83$\pm$0.01& 901.49$\pm$0.41&  0.59$\pm$0.01&1445.30$\pm$0.86\\
	           &$7_6-6_5$&       & 86181.410$\pm$0.002& 5.80$\pm$0.01& 120.89$\pm$1.42&  0.50$\pm$0.01& 229.39$\pm$3.38\\
	           &$7_7-6_6$&       & 90686.381$\pm$0.002& 5.79$\pm$0.01&  85.64$\pm$0.80&  0.52$\pm$0.01& 154.95$\pm$1.87\\
	           &$7_8-6_7$&       & 93870.102$\pm$0.001& 5.83$\pm$0.01& 516.80$\pm$0.41&  0.54$\pm$0.01& 904.86$\pm$0.97\\
	           &$8_7-7_6$&       & 99866.505$\pm$0.002& 5.77$\pm$0.01&  54.12$\pm$0.33&  0.52$\pm$0.01&  98.46$\pm$0.99\\
	           &$8_8-7_7$&       &103640.750$\pm$0.002& 5.78$\pm$0.01&  32.05$\pm$1.45&  0.47$\pm$0.02&  64.52$\pm$3.79\\
	           &$8_9-7_8$&       &106347.744$\pm$0.002& 5.83$\pm$0.01& 241.38$\pm$1.52&  0.52$\pm$0.01& 437.99$\pm$3.87\\
	           &$9_8-8_7$&       &113410.200$\pm$0.002& 5.73$\pm$0.02&  27.90$\pm$2.12&  0.53$\pm$0.05&  49.76$\pm$5.37\\
                &        &       &                    &              &                &              &                \\                             
CC$^{34}$S     &$2_3-1_2$&       & 33111.839$\pm$0.001& 5.76$\pm$0.01& 91.27$\pm$0.09&  0.72$\pm$0.01& 118.35$\pm$0.07\\
               &$3_3-2_2$&       & 38015.235$\pm$0.002& 5.82$\pm$0.01&  7.20$\pm$0.08&  0.67$\pm$0.01&  10.05$\pm$0.10\\
               &$4_3-3_2$&       & 42918.191$\pm$0.002& 5.88$\pm$0.01&  6.84$\pm$0.09&  0.63$\pm$0.01&  10.14$\pm$0.11\\
			   &$3_4-2_3$&       & 44497.599$\pm$0.002& 5.79$\pm$0.01& 87.15$\pm$0.08&  0.62$\pm$0.01& 132.47$\pm$0.13\\
               &$6_6-5_5$&       & 76029.397$\pm$0.003& 5.69$\pm$0.07&  5.05$\pm$1.35&  0.52$\pm$0.15&   9.19$\pm$2.97\\
               &$6_7-5_6$&       & 79827.459$\pm$0.002& 5.82$\pm$0.01& 40.63$\pm$0.40&  0.55$\pm$0.01&  69.46$\pm$0.74\\
               &$7_6-6_5$&       & 84180.570$\pm$0.003& 5.81$\pm$0.03&  3.40$\pm$0.37&  0.53$\pm$0.06&   6.04$\pm$0.55\\
               &$7_7-6_6$&       & 88700.361$\pm$0.003& 5.97$\pm$0.15&  4.57$\pm$1.26&  0.88$\pm$0.29&   4.87$\pm$2.16\\
               &$7_8-6_7$&       & 91913.532$\pm$0.003& 5.83$\pm$0.01& 17.02$\pm$0.65&  0.48$\pm$0.02&  33.33$\pm$1.19\\
               &$8_7-7_6$&       & 97572.235$\pm$0.004& 5.76$\pm$0.10&  2.96$\pm$0.82&  0.64$\pm$0.19&   4.36$\pm$1.79\\
               &$8_8-7_7$&       &101371.047$\pm$0.004&              &               &               & $\le$6.1 \\
               &$8_9-7_8$&       &104109.334$\pm$0.003& 5.89$\pm$0.04&  5.69$\pm$1.23&  0.39$\pm$0.10&  13.67$\pm$3.44\\
                &        &       &                    &              &                &              &                \\                             
CC$^{33}$S$^\#$&$2_3-1_2$&7/2-7/2& 33403.415$\pm$0.005& 5.62$\pm$0.03&  0.84$\pm$0.07&  0.67$\pm$0.06&   1.17$\pm$0.08\\
               &         &5/2-5/2& 33406.288$\pm$0.004& 5.73$\pm$0.06&  0.97$\pm$0.13&  0.55$\pm$0.16&   0.89$\pm$0.08\\
               &         &3/2-3/2& 33407.381$\pm$0.004& 5.72$\pm$0.03&  0.96$\pm$0.10&  0.87$\pm$0.08&   1.05$\pm$0.08\\
               &         &3/2-1/2& 33415.034$\pm$0.004& 5.67$\pm$0.13&  2.26$\pm$0.08&  0.83$\pm$0.03&   2.57$\pm$0.08\\
               &         &5/2-3/2& 33420.244$\pm$0.003& 5.69$\pm$0.02&  2.86$\pm$0.10&  0.71$\pm$0.03&   3.79$\pm$0.09\\
               &         &7/2-5/2& 33425.476$\pm$0.003& 5.72$\pm$0.01&  4.74$\pm$0.10&  0.74$\pm$0.02&   6.01$\pm$0.08\\
               &         &9/2-7/2& 33430.110$\pm$0.005& 5.69$\pm$0.01&  7.15$\pm$0.12&  0.73$\pm$0.02&   9.17$\pm$0.12\\
               &$3_4-2_3$&9/2-9/2& 44901.500$\pm$0.007& 5.80$\pm$0.06&  0.47$\pm$0.08&  0.57$\pm$0.11&   0.78$\pm$0.14\\
               &$       $&7/2-7/2& 44904.692$\pm$0.006& 5.81$\pm$0.03&  0.76$\pm$0.06&  0.54$\pm$0.05&   1.31$\pm$0.13\\
               &$       $&5/2-5/2& 44906.523$\pm$0.007& 5.88$\pm$0.04&  0.61$\pm$0.08&  0.68$\pm$0.10&   0.84$\pm$0.13\\
               &$       $&5/2-3/2& 44919.386$\pm$0.006& 5.83$\pm$0.01&  2.97$\pm$0.12&  0.64$\pm$0.03&   4.34$\pm$0.15\\
               &$       $&7/2-5/2& 44923.880$\pm$0.005& 5.80$\pm$0.01&  3.57$\pm$0.10&  0.57$\pm$0.02&   5.93$\pm$0.14\\
               &$       $&9/2-7/2& 44928.195$\pm$0.006& 5.82$\pm$0.01&  4.69$\pm$0.10&  0.60$\pm$0.02&   7.36$\pm$0.13\\
              &$       $&11/2-9/2& 44932.038$\pm$0.009& 5.81$\pm$0.01&  6.48$\pm$0.07&  0.61$\pm$0.01&   9.92$\pm$0.12\\
                &        &       &                    &              &                &              &                \\                             
$^{13}$CCS     &$2_3-1_2$&5/2-3/2& 32440.193$\pm$0.002& 5.74$\pm$0.02&  1.95$\pm$0.08& 0.78$\pm$0.04&   2.35$\pm$0.07\\                              
                &        &7/2-5/2& 32443.950$\pm$0.002& 5.70$\pm$0.01&  2.80$\pm$0.08& 0.76$\pm$0.03&   3.47$\pm$0.07\\
               &$3_4-2_3$&7/2-5/2& 43574.513$\pm$0.003& 5.73$\pm$0.01&  1.96$\pm$0.08& 0.61$\pm$0.03&   3.03$\pm$0.11\\
                &        &9/2-7/2& 43577.726$\pm$0.003& 5.74$\pm$0.01&  2.79$\pm$0.09& 0.67$\pm$0.03&   3.94$\pm$0.11\\                             
                &        &       &                    &              &                &              &                \\                             
C$^{13}$CS$^\#$&$2_3-1_2$&7/2-5/2& 33613.548$\pm$0.004& 5.83$\pm$0.01& 18.47$\pm$0.10 & 0.73$\pm$0.01 &    23.83$\pm$0.07 \\
              &         & 5/2-3/2& 33615.181$\pm$0.004& 5.82$\pm$0.01& 13.18$\pm$0.11 & 0.74$\pm$0.01 &    16.80$\pm$0.07 \\
              &$3_3-2_2$& 5/2-3/2& 38679.629$\pm$0.010& 5.83         &  1.05$\pm$0.07 & 0.71$\pm$0.05 &     1.38$\pm$0.10 \\
              &         & 7/2-5/2& 38682.766$\pm$0.010& 5.83         &  1.21$\pm$0.06 & 0.63$\pm$0.04 &     1.81$\pm$0.10 \\
              &$4_3-3_2$& 5/2-3/2& 43745.709$\pm$0.010& 5.83         &  0.91$\pm$0.08 & 0.65$\pm$0.08 &     1.31$\pm$0.10 \\
              &         & 7/2-5/2& 43750.477$\pm$0.010& 5.83         &  1.64$\pm$0.09 & 0.73$\pm$0.05 &     2.09$\pm$0.10 \\
              &$3_4-2_3$& 9/2-7/2& 45189.019$\pm$0.007& 5.90$\pm$0.01& 17.35$\pm$0.08 & 0.62$\pm$0.01 &    26.17$\pm$0.13 \\
              &         & 7/2-5/2& 45190.388$\pm$0.007& 5.89$\pm$0.01& 13.25$\pm$0.08 & 0.62$\pm$0.01 &    20.22$\pm$0.13 \\
              &$6_5-5_4$& 9/2-7/2& 71951.803$\pm$0.009&              &               &               & $\le$14.1 \\
              &         &11/2-9/2& 71953.969$\pm$0.009&              &               &               & $\le$14.1 \\
	 		  &$6_6-5_5$&11/2-9/2& 77361.896$\pm$0.008&              &               &               & $\le$3.3 \\
 			 &$       $&13/2-11/2& 77362.613$\pm$0.009&              &               &               & $\le$3.3 \\
             &$6_7-5_6$&15/2-13/2& 81142.453$\pm$0.010& 5.83         &  8.46$\pm$0.43&  0.63$\pm$0.04&  12.69$\pm$0.64\\
             &         &13/2-11/2& 81143.169$\pm$0.010& 5.83         &  7.48$\pm$0.43&  0.70$\pm$0.06&  10.08$\pm$0.64\\
			  &$7_6-6_5$&11/2-9/2& 85744.528$\pm$0.011&              &               &               &       $\le$7.5 \\
			 &$       $&13/2-11/2& 85746.148$\pm$0.011&              &               &               &       $\le$7.5 \\
			 &$7_8-6_7$&17/2-15/2& 93446.744$\pm$0.010& 5.83         &  4.05$\pm$0.39&  0.62$\pm$0.07&   6.10$\pm$0.70\\
			 &         &15/2-13/2& 93447.335$\pm$0.010& 5.83         &  2.16$\pm$0.34&  0.52$\pm$0.09&   3.88$\pm$0.70\\
			 &$8_9-7_8$&19/2-17/2&105863.116$\pm$0.007&              &               &               &       $\le$9.0 \\
			 &$       $&17/2-15/2&105863.579$\pm$0.007&              &               &               &       $\le$9.0 \\
                &        &       &                    &              &                &              &                \\                             
C$^{13}$C$^{34}$S$^\#$
             &$2_3-1_2$& 7/2-5/2 & 32964.623$\pm$0.015& 5.83         &  0.96$\pm$0.09&  0.95$\pm$0.10& 0.55$\pm$0.09\\
             &         & 5/2-3/2 & 32966.233$\pm$0.015& 5.83         &  0.43$\pm$0.08&  0.73$\pm$0.15& 0.96$\pm$0.09\\
             &$3_4-2_3$& 9/2-7/2 & 44295.012$\pm$0.015& 5.83         &  0.29$\pm$0.04&  0.43$\pm$0.08& 0.65$\pm$0.10\\
             &         & 7/2-5/2 & 44296.371$\pm$0.015& 5.83         &  0.37$\pm$0.06&  0.65$\pm$0.14& 0.54$\pm$0.10\\

\hline
                &        &       &                    &              &                &              &                \\                             
CCCS           & 6-5     &       & 34684.368$\pm$0.001& 5.83$\pm$0.01&515.50$\pm$0.11&  0.73$\pm$0.01& 665.02$\pm$0.09 \\
               & 7-6     &       & 40465.014$\pm$0.001& 5.83$\pm$0.01&493.18$\pm$0.08&  0.64$\pm$0.01& 705.12$\pm$0.10 \\
               & 8-7     &       & 46245.623$\pm$0.001& 5.83$\pm$0.01&424.96$\pm$0.13&  0.62$\pm$0.02& 643.99$\pm$0.11 \\
               &13-12    &       & 75147.912$\pm$0.001& 5.82$\pm$0.01&126.03$\pm$1.75&  0.55$\pm$0.01& 215.70$\pm$3.10\\
               &14-13    &       & 80928.182$\pm$0.001& 5.83$\pm$0.01& 63.61$\pm$0.32&  0.59$\pm$0.01& 101.06$\pm$0.67\\
               &15-14    &       & 86708.376$\pm$0.001& 5.92$\pm$0.02& 55.76$\pm$2.22&  0.55$\pm$0.02&  95.49$\pm$3.48$^f$\\
               &16-15    &       & 92488.489$\pm$0.001& 5.84$\pm$0.01& 17.34$\pm$0.36&  0.60$\pm$0.01&  27.18$\pm$0.72\\
               &17-16    &       & 98268.516$\pm$0.002& 5.83$\pm$0.01&  7.32$\pm$0.40&  0.49$\pm$0.15&  14.09$\pm$1.47\\
               &18-17    &       &104048.452$\pm$0.002& 6.03$\pm$0.05&  6.87$\pm$1.43&  0.52$\pm$0.14&  12.40$\pm$2.97\\
               &19-18    &       &109828.291$\pm$0.002&              &               &               &   $\le$13.5    \\
               &         &       &                    &              &               &               &                \\                             
$^{13}$CCCS    &   6-5   &       & 33396.562$\pm$0.001& 5.79$\pm$0.02&  1.98$\pm$0.10&  0.84$\pm$0.05&   2.22$\pm$0.07\\
               &   7-6   &       & 38962.579$\pm$0.001& 5.80$\pm$0.02&  1.64$\pm$0.08&  0.65$\pm$0.04&   2.36$\pm$0.09\\
               &   8-7   &       & 44528.564$\pm$0.001& 5.88$\pm$0.02&  1.48$\pm$0.07&  0.71$\pm$0.04&   1.94$\pm$0.11\\
               &         &       &                    &              &               &               &                \\                             
C$^{13}$CCS    &   6-5   &       & 34336.263$\pm$0.001& 5.81$\pm$0.01&  6.18$\pm$0.09&  0.75$\pm$0.01&   7.78$\pm$0.08\\
               &   7-6   &       & 40058.893$\pm$0.001& 5.84$\pm$0.01&  5.74$\pm$0.07&  0.66$\pm$0.01&   8.26$\pm$0.08\\
               &   8-7   &       & 45781.487$\pm$0.001& 5.84$\pm$0.01&  4.89$\pm$0.11&  0.62$\pm$0.02&   7.43$\pm$0.09\\
               &         &       &                    &              &               &               &                \\                             
CCC$^{34}$S    &   6-5   &       & 33844.246$\pm$0.001& 5.82$\pm$0.01& 21.05$\pm$0.13&  0.71$\pm$0.01&  27.71$\pm$0.08\\
               &   7-6   &       & 39484.876$\pm$0.001& 5.82$\pm$0.01& 20.11$\pm$0.10&  0.65$\pm$0.01&  29.00$\pm$0.09\\
               &   8-7   &       & 45125.470$\pm$0.001& 5.84$\pm$0.01& 18.15$\pm$0.13&  0.62$\pm$0.02&  27.31$\pm$0.13\\
               &  13-12  &       & 73327.721$\pm$0.002& 5.76$\pm$0.08&  4.05$\pm$1.27&  0.53$\pm$0.15&   7.22$\pm$2.37\\
               &         &       &                    &              &               &               &                \\                             
               &         &       &                    &              &               &               &                \\                             
C$^{13}$CC$^{34}$S&6-5   &       & 33491.358$\pm$0.005& 5.92$\pm$0.04&  0.17$\pm$0.06&  0.59$\pm$0.19&   0.28$\pm$0.09\\
               &   7-6   &       & 39073.176$\pm$0.009& 5.82$\pm$0.01&  0.46$\pm$0.08&  0.79$\pm$0.15&   0.55$\pm$0.09\\
               &   8-7   &       & 44654.960$\pm$0.015&              &               &               &    $\le$0.42   \\                             
               &         &       &                    &              &               &               &                \\                             
CCC$^{33}$S$^\#$& 6-5   &11/2-9/2& 34252.374$\pm$0.001& 5.76$\pm$0.02&  1.41$\pm$0.11&  0.69$\pm$0.07&   1.92$\pm$0.07\\
               &        & 9/2-7/2& 34252.374$\pm$0.001&              &               &               &                \\                             			   
               &       &15/2-13/2& 34252.495$\pm$0.001& 5.80$\pm$0.01&  2.25$\pm$0.09&  0.71$\pm$0.04&   2.99$\pm$0.13\\
			   &       &13/2-11/2& 34252.495$\pm$0.001&              &               &               &                \\                             			   
               & 7-6   &13/2-11/2& 39961.062$\pm$0.002& 5.83$\pm$0.04&  1.30$\pm$0.19&  0.55$\pm$0.07&   2.20$\pm$0.09\\
			   &       &11/2- 9/2& 39961.062$\pm$0.002&              &               &               &                \\                             			   
               &       &17/2-15/2& 39961.148$\pm$0.002& 5.83$\pm$0.04&  2.02$\pm$0.19&  0.66$\pm$0.05&   2.89$\pm$0.09\\
			   &       &15/2-13/2& 39961.148$\pm$0.002&              &               &               &                \\                             			   
               & 8-7     &       & 45669.746$\pm$0.010& 5.88$\pm$0.02&  3.45$\pm$0.12&  0.78$\pm$0.03&   4.15$\pm$0.15\\
               &         &       &                    &              &               &               &                \\                             
\hline	   
                &        &       &                    &              &               &               &                \\                             
HCC$^{34}$S$^+$$^\#$&$2_3-1_2$&7/2-5/2&31485.474$\pm$0.010& 5.83     &  0.76$\pm$0.08& 0.88$\pm$0.12 &   0.81$\pm$0.10\\
               &              &5/2-3/2&31490.924$\pm$0.010& 5.83          &  0.52$\pm$0.09& 0.77$\pm$0.15 &   0.64$\pm$0.10\\
               &$3_4-2_3$&9/2-7/2&42180.241$\pm$0.010& 5.83          &  0.54$\pm$0.05& 0.55$\pm$0.05 &   0.94$\pm$0.10\\
               &         &7/2-5/2&42185.089$\pm$0.010& 5.83          &  0.89$\pm$0.09& 0.80$\pm$0.10 &   1.05$\pm$0.10\\
               &         &       &                    &              &               &               &                \\                             
\hline	   
                &        &       &                    &              &               &               &                \\                             
H$_2$CS        &$1_{01}-0_{00}$&       & 34351.430$\pm$0.020& 5.85$\pm$0.01&458.03$\pm$0.15& 0.73$\pm$0.01 & 590.93$\pm$0.08\\
               &$3_{13}-2_{12}$&       &101477.805$\pm$0.001& 5.78$\pm$0.01&582.07$\pm$0.09& 0.54$\pm$0.01 &1016.10$\pm$2.21\\
               &$3_{03}-2_{02}$&       &103040.447$\pm$0.001& 5.79$\pm$0.01&577.08$\pm$1.16& 0.54$\pm$0.01 &1010.80$\pm$2.94\\
               &$3_{12}-2_{11}$&       &104617.027$\pm$0.001& 5.80$\pm$0.01&523.31$\pm$0.09& 0.54$\pm$0.01 & 905.67$\pm$2.30\\
               &         &       &                    &              &               &               &                \\                             
H$_2$C$^{34}$S &$1_{01}-0_{00}$&       & 33765.749$\pm$0.001& 5.74$\pm$0.01& 19.48$\pm$0.12& 0.74$\pm$0.01 &  24.72$\pm$0.08\\ 
               &$3_{13}-2_{12}$&       & 99774.077$\pm$0.001& 5.74$\pm$0.01& 24.17$\pm$0.33& 0.53$\pm$0.03 &  41.37$\pm$0.07\\
               &$3_{03}-2_{02}$&       &101284.314$\pm$0.001& 5.78$\pm$0.01& 23.02$\pm$1.04& 0.53$\pm$0.03 &  40.59$\pm$2.15\\ 
               &$3_{12}-2_{11}$&       &102807.337$\pm$0.001& 5.73$\pm$0.02& 18.78$\pm$1.38& 0.51$\pm$0.04 &  34.30$\pm$3.14\\ 
               &         &       &                    &              &               &               &                \\                             
H$_2$C$^{33}$S &$1_{01}-0_{00}$&3/2-3/2& 34047.023$\pm$0.010& 5.72$\pm$0.02&  1.41$\pm$0.09& 0.72$\pm$0.02 &   1.83$\pm$0.09\\
               &               &5/2-3/2& 34050.010$\pm$0.012& 5.70$\pm$0.02&  2.09$\pm$0.08& 0.75$\pm$0.03 &   2.63$\pm$0.09\\
               &               &1/2-3/2& 34052.355$\pm$0.012& 5.63$\pm$0.04&  0.92$\pm$0.08& 0.86$\pm$0.09 &   1.00$\pm$0.09\\
               &         &       &                    &              &               &               &                \\                             
H$_2$$^{13}$CS &$1_{01}-0_{00}$&       & 33029.940$\pm$0.050& 6.01$\pm$0.01&  5.62$\pm$0.09& 0.76$\pm$0.01 &   7.00$\pm$0.09\\
               &$3_{13}-2_{12}$&       & 97632.178$\pm$0.001& 5.75$\pm$0.04&  5.62$\pm$0.08& 0.54$\pm$0.08 &   9.72$\pm$2.04\\
               &$3_{03}-2_{02}$&       & 99077.813$\pm$0.001& 5.75$\pm$0.02&  6.77$\pm$0.04& 0.54$\pm$0.04 &  11.83$\pm$0.86\\
               &$3_{12}-2_{11}$&       &100534.726$\pm$0.001& 5.70$\pm$0.02&  5.96$\pm$0.06& 0.40$\pm$0.05 &  14.09$\pm$1.40\\
\hline	   
                &        &       &                    &              &               &               &                \\                             
C$_4$S$^\#$  &$10_{11}-9_{10}$ & & 32553.166$\pm$0.010& 5.83         &  0.98$\pm$0.08&  0.71$\pm$0.06&   1.29$\pm$0.10 \\
             &$11_{11}-10_{10}$& & 33422.473$\pm$0.010& 5.83         &  0.48$\pm$0.06&  1.00$\pm$0.14&   0.43$\pm$0.09  \\
             &$11_{12}-10_{11}$& & 35519.772$\pm$0.010& 5.83         &  0.96$\pm$0.08&  0.68$\pm$0.06&   1.32$\pm$0.10 \\
             &$12_{13}-11_{12}$& & 38488.046$\pm$0.010& 5.83         &  1.00$\pm$0.19&  0.69$\pm$0.08&   1.35$\pm$0.10 \\
             &$14_{13}-13_{12}$& & 40509.643$\pm$0.020& 5.83         &  0.30$\pm$0.07&  0.71$\pm$0.18&   0.40$\pm$0.11 \\
             &$13_{14}-12_{13}$& & 41458.064$\pm$0.010& 5.83         &  0.98$\pm$0.09&  0.74$\pm$0.08&   1.25$\pm$0.10 \\
             &$14_{15}-13_{14}$& & 44429.887$\pm$0.010& 5.83         &  0.63$\pm$0.09&  0.54$\pm$0.08&   1.09$\pm$0.12 \\
             &$15_{16}-14_{15}$& & 47403.545$\pm$0.010& 5.83         &  0.47$\pm$0.07&  0.52$\pm$0.10&   0.84$\pm$0.13 \\
                &        &       &                    &              &                &              &                \\                             			
\hline
                &        &       &                    &              &                &              &                \\                             
C$_5$S$^\#$  & 17-16            &  & 31371.650$\pm$0.010 &5.83       & 1.20$\pm$ 0.14 &   0.79$\pm$  0.11 &     1.45$\pm$0.13 \\
             & 18-17            &  & 33217.010$\pm$0.010 &5.83       & 1.00$\pm$ 0.09 &   0.73$\pm$  0.08 &     1.29$\pm$0.09 \\
             & 19-18            &  & 35062.356$\pm$0.010 &5.83       & 0.70$\pm$ 0.02 &   0.60$\pm$  0.10 &     1.09$\pm$0.09 \\
             & 20-19            &  & 36907.703$\pm$0.010 &5.83       & 0.60$\pm$ 0.05 &   0.55$\pm$  0.05 &     1.03$\pm$0.09 \\
             & 21-20            &  & 38753.024$\pm$0.010 &5.83       & 0.43$\pm$ 0.07 &   0.51$\pm$  0.14 &     0.80$\pm$0.11 \\
             & 22-21            &  & 40598.349$\pm$0.010 &5.83       & 0.44$\pm$ 0.08 &   0.77$\pm$  0.15 &     0.53$\pm$0.12 \\
             & 23-22            &  & 42443.710$\pm$0.010 &5.83       & 0.44$\pm$ 0.08 &   0.58$\pm$  0.11 &     0.73$\pm$0.13 \\
             & 24-23            &  & 44289.034$\pm$0.020 &5.83       & 0.22$\pm$ 0.07 &   0.60$\pm$  0.23 &     0.39$\pm$0.13$^f$ \\ 
             & 25-24            &  & 46134.304$\pm$0.017 &5.83       & 0.34$\pm$ 0.10 &   0.51$\pm$  0.12 &     0.64$\pm$0.15 \\
             & 26-25            &  & 47979.593$\pm$0.040 &5.83       & 0.79$\pm$ 0.16 &   0.52$\pm$  0.10 &     1.43$\pm$0.17$^g$ \\ 
             &          &       &                    &              &               &               &                \\                             
\hline
\hline
\end{longtable}
\tablefoot{\\
\tablefoottext{$\$$}{All uncertainties correspond to 1$\sigma$. However, upper limits to any of the parameters correspond to 3$\sigma$ values.}\\
\tablefoottext{*}{Lines have been fitted using three gaussians.}\\
\tablefoottext{a}{Adopted rest frequency (see text).}\\
\tablefoottext{b}{Local standard of rest (LSR) velocity of the line in km\,s$^{-1}$.
If the uncertainty is not given, then the velocity has been fixed to 5.83 km\,s$^{-1}$ 
and the derived frequency has been used to improve the rotational constants of the molecule (see Appendix \ref{sec:lab}).}\\
\tablefoottext{c}{Integrated line intensity in mK\,km\,s$^{-1}$.}\\
\tablefoottext{d}{Linewidth at half intensity derived by fitting a Gaussian function to
the observed line profile (in km\,s$^{-1}$).}\\
\tablefoottext{e}{Antenna temperature in milli Kelvin.}\\
\tablefoottext{f}{Blended with HCO. Line parameters are uncertain.}\\
\tablefoottext{g}{Fully blended with a line of similar intensity from CC$^{13}$CO $J$=5-4. The frequency difference 
between both features is 10 kHz. Several lines of the $^{13}$C isotopologues of CCCO are detected with QUIJOTE. Around
50\% of the observed intensity can be attributed to CC$^{13}$CO.}\\
\tablefoottext{\#}{Molecule for which new rotational constants are provided 
(see Appendix \ref{sec:lab}).}\\
}
\end{appendix} 


\begin{thebibliography}{}
\bibitem[Ag{\'u}ndez et al.(2008)]{Agundez2008} Ag{\'u}ndez, M., Fonfr\'ia, J.~P., Cernicharo, J., et al. 2008, \aap, 479, 493
\bibitem[Ag{\'u}ndez et al.(2014)]{Agundez2014} Ag{\'u}ndez, M., Cernicharo, J., \& Gu{\'e}lin, M.\ 2014, \aap, 570, A45 
\bibitem[Ag{\'u}ndez et al.(2015)]{Agundez2015} Ag{\'u}ndez, M., Cernicharo, J., de Vicente, P., et al.\ 2015, \aap, 579, L10
\bibitem[Ag{\'u}ndez et al.(2019)]{Agundez2019} Ag{\'u}ndez, M., Marcelino, N., Cernicharo, J., et al.\ 2019, \aap, 625, A147
\bibitem[Ag{\'u}ndez et al.(2022)]{Agundez2022} Ag{\'u}ndez, M., Cabezas, C., Marcelino, N. et al. 2022, \aap, 659, L9 
\bibitem[Ag{\'u}ndez et al.(2023)]{Agundez2023} Ag{\'u}ndez, M., Marcelino, N., Tercero, B., et al. 2023, \aap, 677, A106
\bibitem[Agundez et al.(2025)]{Agundez2024} Agundez, M., Molpeceres, G., Cabezas, C., et al.\ 2025, \aap, 693, L20
\bibitem[Ahrens \& Winnewisser (1999)]{Ahrens1999}Ahrens, V. \& Winnewisser, G. 1999, Z. Naturforsch, 54a, 131
\bibitem[Anders \& Grevesse(1989)]{Anders1989} Anders, E. \& Grevesse, N.\ 1989, \gca, 53, 197 
\bibitem[Bell et al.(1993)]{Bell1993} Bell, M.~B., Avery, L.~W., \& Feldman, P.~A.\ 1993, \apjl, 417, L37 
\bibitem[Bogey et al. (1981)]{Bogey1981}Bogey, M., Demuynch, C. \& Destombes, J.L. 1981, Chem. Phys. Lett., 81, 256
\bibitem[Bogey et al. (1982)]{Bogey1982}Bogey, M., Demuynch, C. \& Destombes, J.L. 1982, \jms, 95, 35
\bibitem[Bogey et al. (1984)]{Bogey1984}Bogey, M., Demuynch, C., Destombes, J.L., \& Lemoine, B. 1984, \jms, 107, 417
\bibitem[Botschwina \& Sebald (1985)]{Botschwina1985}Botchswina, P. \& Sebald, P. 1985 \jms, 110, 1
\bibitem[Burkholder et al. (1987)]{Burkholder1987}Burkholder, J.~B., Lovejoy, E.~R., Hammer, P.~D., \& Howard, C.~J. 1987, \jms, 124, 450
\bibitem[Cabezas et al.(2016)]{Cabezas2016} Cabezas, C., Guillemin, J.-C., Endo, Y. 2016, J. Chem. Phys., 145, 184304.
\bibitem[Cabezas et al.(2021)]{Cabezas2021} Cabezas, C., Roueff, E., Tercero, B., et al.\ 2021, \aap, 650, L15
\bibitem[Cabezas et al.(2022a)]{Cabezas2022a} Cabezas, C., Fuentetaja, R., Roueff, E., et al.\ 2022a, \aap, 657, L5 
\bibitem[Cabezas et al.(2022b)]{Cabezas2022b} Cabezas, C., Ag{\'u}ndez, M., Marcelino, N., et al.\ 2022b, \aap, 657, L4
\bibitem[Cabezas et al.(2024)]{Cabezas2024} Cabezas, C., Ag{\'u}ndez, M., Endo, Y., et al.\ 2024, \aap, 686, L3
\bibitem[Cernicharo(1985)]{Cernicharo1985} Cernicharo, J. 1985, Internal IRAM report (Granada: IRAM)
\bibitem[Cernicharo \& Gu\'elin (1987)]{Cernicharo1987} Cernicharo, J. \& Gu\'elin, M. 1987, \aap, 176, 299 
\bibitem[Cernicharo et al. (1987)]{Cernicharo1987}Cernicharo, J., Gu\'elin, M., Hein, H. Kahane, C. 1987b, \aap, 181, L9
\bibitem[Cernicharo et al. (2012a)]{Cernicharo2012a} Cernicharo, J., Marcelino, N., Rouef, E., et al. 2012a, \apj, 759, L43
\bibitem[Cernicharo (2012b)]{Cernicharo2012b} Cernicharo, J., 2012b, in ECLA 2011: Proc. of the European Conference on Laboratory Astrophysics,
EAS Publications Series, 2012, Ed.: C. Stehl, C. Joblin, \& L. d'Hendecourt (Cambridge: Cambridge Univ. Press),
251; \texttt{https://nanocosmos.iff.csic.es/?page$\_$id=1619}
\bibitem[Cernicharo et al.(2020)]{Cernicharo2020} Cernicharo, J., Marcelino, N., Ag\'undez, et al. 2020, \aap, 642, L8 
\bibitem[Cernicharo et al.(2021a)]{Cernicharo2021a} Cernicharo, J., Ag\'undez, M., Kaiser, R. et al. 2021a, \aap, 652, L9 
\bibitem[Cernicharo et al.(2021b)]{Cernicharo2021b} Cernicharo, J., Ag\'undez, M., Cabezas, C., et al.\ 2021b, \aap, 649, L15 
\bibitem[Cernicharo et al.(2021c)]{Cernicharo2021c} Cernicharo, J., Ag{\'u}ndez, M., Kaiser, R.~I., et al.\ 2021c, \aap, 655, L1 
\bibitem[Cernicharo et al.(2021d)]{Cernicharo2021d} Cernicharo, J., Cabezas, C., Ag\'undez, M., et al. 2021d, \aap, 648, L3 
\bibitem[Cernicharo et al.(2021e)]{Cernicharo2021e} Cernicharo, J., Cabezas, C., Endo, Y., et al. 2021e, \aap, 650, L14
\bibitem[Cernicharo et al.(2021f)]{Cernicharo2021f} Cernicharo, J., Cabezas, C., Endo, Y., et al. 2021f, \aap, 646, L3 
\bibitem[Cernicharo et al. (2023)]{Cernicharo2023} Cernicharo, J., Tercero, B., Marcelino, N., et al. 2023, \aap, 674, L4
\bibitem[Cernicharo et al. (2024a)]{Cernicharo2024a}Cernicharo, J., Tercero, B., Cabezas, C. et al. 2024a, \aap, 682, L13
\bibitem[Cernicharo et al.(2024b)]{Cernicharo2024b} Cernicharo, J., Ag{\'u}ndez, M., Cabezas, C., et al.\ 2024c, \aap, 682, L4
\bibitem[Chin et al.(1996)]{Chin1996} Chin, Y.-N., Henkel, C., Whiteoak, J. B., et al. 1996, \aap, 305, 960
\bibitem[Colzi et al.(2020)]{Colzi2020} Colzi, L., Sipilä, O., Roueff, E., et al. 2020, \aap, 640, A51
\bibitem[Corey (1984)]{Corey1984}Corey, G.~C. 1984, \jcp, 81, 2678
\bibitem[Corey \& McCourt (1984)]{Corey1984b}Corey, G.~C. \& McCourt, F.R. 1984, \jcp, 84, 2723
\bibitem[Corey et al. (1986)]{Corey1986}Corey, G.~C. Alexander, M.~H. \& Schaefer, J. 1986, \jcp, 85, 2726
\bibitem[Denis-Alpizar et al.(2018)]{Denis-alpizar2018} Denis-Alpizar, O., Stoecklin, T., Guilloteau, S., et al.\ 2018, \mnras, 478, 1811
\bibitem[Denis-Alpizar et al.(2022)]{Denis-alpizar2022} Denis-Alpizar, O., Quintas-S\'anchez, E. \& Dawes. R. 2022, \mnras, 512, 5546
\bibitem[Dunning(1989)]{Dunning1989} Dunning, T. H., 1989, J. Chem. Phys. 90, 1007
\bibitem[Endo et al.(1994)]{Endo1994} Endo, Y., Kohguchi, H., Ohshima, Y. 1994, Faraday Discuss., 97, 341.
\bibitem[Faure et al. (2016)]{Faure2016}Faure, A., Lique, F. \& Wiesenfeld, L. 2016, MNRAS, 460, 2103
\bibitem[Feh\`er et al. (2016)]{Feher2016}Feh\`er, O., T\'oth, L.V., Ward-Thompson, D. et al. 2016, \aap, 590, A75
\bibitem[Flower \& Lique(2015)]{Flower2015} Flower, D.~R. \& Lique, F.\ 2015, \mnras, 446, 1750.
\bibitem[Foss\'e et al.(2001)]{Fosse2001} Foss\'e, D., Cernicharo, J., Gerin, M., Cox, P. 2001, \apj, 552, 168
\bibitem[Frisch et al.(2016)]{Frisch2016} Frisch, M. J., Trucks, G. W., Schlegel, H. B., et al. 2016, Gaussian~16 Revision A.03
\bibitem[Fuente et al.(1990)]{Fuente1990} Fuente, A., Cernicharo, J., Barcia, A. \& G\'omez-G\'onzalez, J. 1990, \aap, 231, 151
\bibitem[Fuente et al.(2019)]{Fuente2019} Fuente, A., Navarro, D. G., Caselli, P. et al. 2019, \aap, 624, A105
\bibitem[Fuente et al.(2023)]{Fuente2023} Fuente, A., Rivi\`ere-Marichalar, P., Beitia-Antero, L. et al. 2023, \aap, 670, A114
\bibitem[Fuentetaja et al.(2022)]{Fuentetaja2022} Fuentetaja, R., Ag{\'u}ndez, M., Cabezas, C., et al.\ 2022, \aap, 667, L4
\bibitem[Furuya et al.(2011)]{Furuya2011} Furuya, K., Aikawa, Y., Sakai, N., et al.\ 2011, \apj, 731, 38.
\bibitem[Godard Palluet \& Lique(2023)]{Godard2023} Godard Palluet, A. \& Lique., F 2023, J. Chem. Phys., 158, 044303
bibitem[Godard Palluet \& Lique(2024)]{Godard2024} Godard Palluet, A. \& Lique, F.\ 2024, \mnras, 527, 6702
\bibitem[Godard Palluet \& Lique(2024)]{Godard2024} Godard Palluet, A. \& Lique, F.\ 2024, \mnras, 527, 6702
\bibitem[Gordon et al.(2001)]{Gordon2001} Gordon, V.~D., McCarthy, M.~C., Apponi, A.~J., et al.\ 2001, \apjs, 134, 311
\bibitem[Goldreich \& Kwan (1974)]{Goldreich1974}Goldreich, P. \& Kwan, J. 1974, \apj, 189, 441
\bibitem[Gottlieb et al. (2003)]{Gottlieb2003}Gottlieb, C.~A., Myers, P.~C. \& Thaddeus, P. 2003, \apj, 588, 655
\bibitem[Gudeman et al. (1981)]{Gudeman1981}Gudeman, C. S., Haese, N. N., Piltch, N. D. \& Woods, R. C. 1981, \apj, 246, L47
\bibitem[Green \& Chapman(1978)]{Green1978} Green, S. \& Chapman, S.\ 1978, \apjs, 37, 169
\bibitem[Green (1991)]{Green1991} Green, S. 1991, \apjs, 76, 979
\bibitem[Hirahara et al.(1993)]{Hirahara1993} Hirahara, Y., Ohshima, Y., \& Endo, Y.\ 1993, \apjl, 408, L113
\bibitem[Ikeda et al.(1997)]{Ikeda1997}Ikeda, M., Sekimoto, Y. \& Yamamoto, S. 1997, Journal of Molecular Spectroscopy, 185, 21
\bibitem[Kaifu et al.(1987)]{Kaifu1987} Kaifu, N., Suzuki, H., Ohishi, M., et al.\ 1987, \apjl, 317, L111 
\bibitem[Kaifu et al.(2004)]{Kaifu2004} Kaifu, N., Ohishi, M., Kawaguchi, K., et al. 2004, PASJ, 56, 69
\bibitem[Kasai et al. (1993)]{Kasai1993}Kasai, Y., Obi, K., Ohshima, Y. et al. 1993, \apj, 410, L45
\bibitem[Khadri et al.(2020)]{Khadri2020} Khadri, F., Chefai, A., \& Hammami, K.\ 2020, \mnras, 498, 5159 
\bibitem[Kim \& Yamamoto(2003)]{Kim2003} Kim, E. \& Yamamoto, S.\ 2003, Journal of Molecular Spectroscopy, 219, 296
\bibitem[Kirk et al. (2013)]{Kirk2013}Kirk, J. M., Ward-Thompson, D., Palmeirim, P. et al. 2013, MNRAS, 432, 1424
\bibitem[Lee (1997)]{Lee1997} Lee, S., 1997, Chem. Phys., 268, 69
\bibitem[Loison et al.(2020)]{Loison2020} Loison, J.-C., Wakelam, V., Gratier, P., et al.\ 2020, \mnras, 498, 4663. 
\bibitem[Lovas et al. (1992)]{Lovas1992} Lovas, F.~J., Suenram, R.~D., Ogata, T. \& Yamamoto, S. 1992, \apj, 399, 325
\bibitem[Lucas \& Liszt (1998)]{Lucas1998}Lucas, R. \& Liszt, H. 1998, \aap, 337, 246
\bibitem[Margul\`es et al. (2003)]{Margules2003}Margul\`es, L., Lewen, F., Winnewisser, G. et al. 2003, Phys. Chem. Chem. Phys., 5, 2770
\bibitem[McGuire et al.(2018)]{McGuire2018} McGuire, B. A., Martin-Drumel, M.-A., Lee, K. L. K., et al. 2018b, PCCP, 20, 13870
\bibitem[Milam et al. (2005)]{Milam2005}Milam, S.~N., Savage, C., Brewster, M.~A., et al. 2005, \apj, 634, 1126
\bibitem[M{\o}ller \& Plesset(1934)]{Moller1934} M{\o}ller, C., \& Plesset, M. S. 1934, \prev, 46, 618
\bibitem[M\"uller et al.(2005)]{Muller2005} M\"uller, H.~S.~P., Schl\"oder, F., Stutzki, J., Winnewisser, G. 2005, \jmst, 742, 215
\bibitem[Navarro-Almaida et al. (2020)]{Navarro2020} Navarro-Almaida, D., Le Gal, R., Fuente, A. et al. 2020, \aap, 637, A39
\bibitem[Navarro-Almaida et al.(2023)]{Navarro2023} Navarro-Almaida, D., Bop, C.~T., Lique, F., et al.\ 2023, \aap, 670, A110
\bibitem[Ohshima \& Endo  (1992)]{Ohshima1992}Ohshima, Y. \& Endo, Y. 1992, \jms, 153, 627
\bibitem[Pardo et al.(2001)]{Pardo2001} Pardo, J.~R., Cernicharo, J., Serabyn, E. 2001, IEEE Trans. Antennas and Propagation, 49, 12
\bibitem[Pascoli \& Lavendy(1998)]{Pascoli1998} Pascoli, G. \& Lavendy, H.\ 1998, International Journal of Mass Spectrometry, 181, 11.
\bibitem[Pickett(1991)]{Pickett1991} Pickett, H. M. 1991, \jms, 148, 371
\bibitem[Pickett et al.(1998)]{Pickett1998} Pickett, H.~M., Poynter, R.~L., Cohen, E.~A., et al. 1998, J. Quant. Spectrosc. Radiat. Transfer, 60, 883
\bibitem[Ram et al. (1995)]{Ram1995}Ram, R.~S., Bernath, P.~F. \& Davies, S. P. 1995, \jms, 173, 146
\bibitem[Ritchey et al. (2011)]{Ritchey2011} Ritchey, A. M., Federman, S. R., \& Lambert, D. L. 2011, \apj, 728, 36
\bibitem[Roueff et al. (2015)]{Roueff2015} Roueff, E., Loison, J.-C., \& Hickson, K. M. 2015, \aap, 576, A99
\bibitem[Sahnoun et al.(2020)]{Sahnoun2020} Sahnoun, E., Ben Khalifa, M., Khadri, F., et al.\ 2020, \apss, 365, 183
\bibitem[Saito et al.(1987)]{Saito1987} Saito, S., Kawaguchi, K., Yamamoto, S., et al.\ 1987, \apjl, 317, L115
\bibitem[Sakai et al. (2007)]{Sakai2007} Sakai, N., Ikeda, M., Morita, M. et al. 2007, \apj, 663, 1174
\bibitem[Sakai et al. (2013)]{Sakai2013} Sakai, N., Takano, S., Sakai, S., Shiba, S., Sumiyoshi, Y., Endo, Y.,  \& Yamamoto, S., 2013, J. Phys. Chem., A117, 9831
\bibitem[Sheffer et al. (2007)]{Sheffer2007} Sheffer, Y., Rogers, M., Federman, S. R., et al. 2007, \apj, 667, 1002
\bibitem[Stahl et al.(2008)]{Stahl2008} Stahl, O., Casassus, S., Wilson, T. 2008, \aap, 477, 865
\bibitem[Suenram et al. (1994)]{Suenram1994} Suenram, R.~D., Lovas, F.~J., et al. 1994, \apj, 429, L89 
\bibitem[Suzuki et al. (1984)]{Suzuki1984}Suzuki, H., Kaifu, N., Miyaji, T. et al. 1984, \apj, 282, 197
\bibitem[Takano et al. (1998)]{Takano1998}Takano, S., Masuda, A., Hirahara, Y. et al. 1998, \aap, 329, 1156
\bibitem[Talbi (2018)]{Talbi2018} Talbi, D., 2018, Australian Journal of Chemistry 71, 311-314.
\bibitem[Tang \&  Saito (1995)]{Tang1995}Tang, J \& Saito, S. 1995, \pccp, 5, 2770
\bibitem[Tercero et al. (2021)]{Tercero2021} Tercero, F., L\'opez-P\'erez, J. A., Gallego, et al., 2021, \aap, 645, A37
\bibitem[Tercero et al. (2024)]{Tercero2024}Tercero, B., Marcelino, N., Ag\'undez, M. et al. 2024, \aap, 682, L12
\bibitem[Thaddeus et al. (1981)]{Thaddeus1981}Thaddeus, P., Gu\'elin, M. \& Linke, R.A. 1981, \apj, 246, L41
\bibitem[Todd (1977)]{Todd1977}Todd, T.~T. 1977, \jms, 66, 162
\bibitem[Todd \& Olson (1979)]{Todd1979}Todd, T.~T. \& Olson W.~B. 1979, \jms, 74, 190
\bibitem[Troscompt et al. (2009)]{Troscompt2009}Troscompt, N., Faure, A., Wiesenfeld, L. et al. 2009, \aap, 493, 687
\bibitem[Uehara et al. (2015)]{Uehara2015}Uehara, H., Horiai, K. \& Sakamoto, Y. 2015, \jms, 313, 19
\bibitem[Wilson(1999)]{Wilson1999} Wilson, T.~L.\ 1999, Reports on Progress in Physics, 62, 143
\bibitem[Winkel et al. (1984)]{Winkel1984}Winkel, R.~J., Davies, S.~P., Pecumer R. \& Brault, J.~W. 1984, Can. J. Phys., 62, 1414
\bibitem[Winnewisser \& Cook (1968)]{Winnewisser1968} Winnewisser, G. \& Cook, R.~L. 1968, J. Mol. Spectrosc., 28, 266
\bibitem[Yamamoto et al.(1987)]{Yamamoto1987} Yamamoto, S., Saito, S., Kawaguchi, K., et al.\ 1987, \apjl, 317, L119
\bibitem[Yamamoto et al. (1990)]{Yamamoto1990}Yamamoto, S., Saito, S., Kawaguchi, K., et al., 1990, \apj, 361, 318
\bibitem[Yan et al. (2023)]{Yan2023}Yan, Y. T., Henkel, C., Kobayashi, C. et al. 2023, \aap, 670, A98

\end{thebibliography}
\end{document}